\documentclass[11pt]{article}
\usepackage[dvips]{graphicx}
\usepackage{amssymb}
\usepackage{amsmath}
\usepackage{epsfig} 
\usepackage[]{caption}
\usepackage{verbatim}
\usepackage{amsfonts}
\usepackage{amsthm}
\usepackage{enumerate}
\usepackage{float}
\usepackage{morefloats}
\usepackage{authblk}

\title{Intrinsic group behaviour: dependence of pedestrian dyad dynamics on principal social and personal features}

\author[1]{Francesco Zanlungo\thanks{zanlungo@atr.jp}}
\author[1,2]{Zeynep Y{\"u}cel}
\author[1,3]{Dra\v{z}en Br\v{s}\v{c}i\'c}
\author[1]{Takayuki Kanda}
\author[1]{Norihiro Hagita}
\affil[1]{ATR International, Kyoto, Japan}
\affil[2]{Research fellow, JSPS, Tokyo, Japan}
\affil[3]{University of Rijeka, Croatia}

\begin{document}  
\maketitle
\section{Abstract}
{\footnotesize
In recent years, researchers in pedestrian behaviour and crowd modelling have become more and more interested in the behaviour of walking social groups, since these groups
represent an important portion of pedestrian crowds, and present peculiar dynamical features. It is anyway clear that, being group dynamics determined by human social behaviour,
it probably depends on properties such as the purpose of the pedestrians, their personal relation, their gender, age, and body size.
We may call these the ``intrinsic properties'' of the group (opposed to extrinsic ones such as crowd density or environmental features). In this work we quantitatively analyse the dynamical properties of
pedestrian dyads (distance, spatial formation and velocity) by analysing a large data set of automatically tracked pedestrian trajectories in an unconstrained ``ecological'' setting (a shopping mall), whose relational group properties have been analysed by three different human coders. We observed that females walk slower and closer than males, that workers walk faster, at a larger distance 
and more abreast than leisure oriented people, and that inter group relation has a strong effect on group structure, with couples walking very close and abreast, colleagues walking at a larger distance, and friends walking more abreast than family members. Pedestrian height (obtained automatically through our tracking system) influences velocity and abreast distance, both growing functions of the average group height.
Results regarding pedestrian age show as expected that elderly people walk slowly, while active age adults walk at the maximum velocity. 
Groups with children have a strong
tendency to walk in a non abreast formation, with a large distance (despite a low abreast distance).
A cross-analysis of the interplay between these intrinsic features, taking in account also the effect of extrinsic crowd density,
confirms these major effects but reveals also a richer structure. An interesting and unexpected result, for example, is that the velocity of groups with children {\it increases} with
density, at least in the low-medium density range found under normal conditions in shopping malls. Children also appear to behave differently according to the gender of the parent.}
\newpage 
{\footnotesize \tableofcontents}
\newpage
\section{Introduction}
Urban crowds are characterised by the presence of a large number of social groups. The ratio between individual pedestrians and pedestrians moving in groups
may change considerably between different environments and at different times of the day \cite{M,M2,OOC}, but it is in general never negligible, with groups representing up to 85\% of the walking population \cite{schultz2,mou2}.
Despite this empirical evidence about the importance of groups, the standard approach in microscopic (agent-based) pedestrian modelling has been for long time to assume that the crowd is composed of individuals,
moving without any preferential ties to other pedestrians. This is an extremely strong simplification of the system, although it was obviously understandable as a first approach to the problem.
Nevertheless, it is intuitive that groups behave in a specific way (they move together and close) and their presence should clearly influence the dynamics of the crowd. 
Not taking in account the group component of crowds may have an impact on the planning of buildings and emergency evacuation plans. For example,
\cite{kumagai} reports that around 48\% of people that evacuated by foot the city of Sendai (Miyagi, Japan) during the 2011 Tohoku great earthquake and tsunami did it by moving in groups, with 
a probable effect on evacuation times (for the smaller city of Kamaishi, in Iwate prefecture, the figure was as high as 71\%). Furthermore an understanding of pedestrian behaviour
is essential for robots and automatic navigation vehicles (such as wheelchairs or delivery carts) that will become arguably common in future pedestrian areas \cite{Kanda,shiomi,Luis}.

Indeed, in recent years, a few studies concerning empirical observations and mathematical modelling of the groups' characteristic configuration and velocity 
have been introduced \cite{M,schultz2,mou2,costa,zan3,dyads,vizzarri,koster,koster2,wei,kara,zhang,seyfried,gocrofe,zhao,kruchten,wangside,huang}. In a recent series of papers \cite{M,M2,M3}, we focused on the development of a mathematical model to describe group interaction, and in specific
the group spatial structure and velocity. The model proposed in \cite{M} introduced a non Newtonian \cite{Turchetti} potential for group interaction on the basis of few and intuitive ideas 
about social interaction in pedestrian groups, and its predictions for group size, structure and velocity are in agreement with the observed natural behaviour of pedestrians.
In \cite{M2} we studied a large data set of pedestrian trajectories to see how an {\it extrinsic}, i.e.,
environmental, property such as crowd density influences the dynamics of groups, and in \cite{M3} we introduced a mathematical model to explain
such a crowd density effect on groups \footnote{Density is probably only one of the environmental properties affecting group dynamics, a second one being the environment architectural features such as
corridor width \cite{mped14}.}.

Nevertheless, we may expect that a social behaviour such as walking in groups depends also on {\it intrinsic} properties of the groups. It is known by studies with subjects
that   age, gender and height affect walking speed \cite{velh}, but here we are interested on how group behaviour is affected by the  nature of the group itself: not only by the
characteristics of the individuals that compose it, but also by the relation between them, which is
expected to have a strong impact on group dynamics (see \cite{costa,wei,bode,sexual,sivers,feng,Muller} and our preliminary study \cite{mped16}).
To study natural human behaviour we use a large {\it ecological} (i.e. obtained by observing unconstrained
pedestrians in their natural environment, see \cite{costa}) data set to describe how the group spatial structure, size and velocity of {\it dyads} (two people groups) change based on the following intrinsic properties of
groups:
\begin{enumerate}
\item {\bf purpose} of movement,
\item {\bf relation} between the members,
\item {\bf gender} of the members,
\item {\bf age} of the members,
\item {\bf height} of the members.
\end{enumerate}
Being the data set based on unconstrained trajectories of unknown pedestrians, such features are necessarily (with the exclusion of pedestrian height, obtained automatically through our tracking system
\cite{Drz}) {\it apparent}, i.e. based on the judgement of human coders, and thus an analysis of their reliability is performed. Furthermore, being social behaviour cultural dependent,
the results are probably influenced by the place in which data were collected (a shopping mall in Osaka, western Japan). Nevertheless, they provide a useful insight on how these intrinsic
features affect in a quantitative way the behaviour of dyads.
\section{Data set}
The pedestrian group data base used for this work is based on the freely available set \cite{data}, introduced by \cite{M2}. This set is again based on a very large pedestrian trajectory set \cite{OOC},
collected in a $\approx 900$ m$^2$ area of the Asia and Pacific Trade Center (ATC), a multi-purpose building located in the Osaka (Japan) port area. 
For the purpose of this work, in order to avoid taking in consideration the effect of architectural features of the environment \cite{mped14}, 
such as its width, we use data only from the corridor area as defined in \cite{M2}.\\
The trajectories have been automatically tracked using 3D range sensors and the algorithm introduced in \cite{Drz}, which provides, along with the pedestrian position on the plane, the height of their head,
for more than 800 hours during a one year time span. At the same time, we  video recorded the tracking area using 16 different cameras. A subset of the video recordings were used by a human coder
to identify the
pedestrian social groups reported in data set \cite{data}.
\subsection{An ecological data set}
The data set concerns the natural behaviour of pedestrians, i.e. the pedestrians were behaving in an unconstrained way, and observed in their natural environment\footnote{With the consent of local authorities
  and building managers. Posters explaining that an experiment concerning pedestrian tracking was being hold were present in the environment.}. 
Collecting data in the pedestrians' natural environment obviously presents some technical problems and an overall lower quality in tracking data (higher tracking noise),
but it is an approach with growing popularity \cite{austria,corb1}, that allows for removing possible influence on pedestrian behaviour due to performing experiments in laboratories, i.e. artificial
environments, using selected subjects following the experimenters' instructions.
This is extremely important for this study, since we may hardly believe that
social pedestrian group behaviour could be observed in such controlled laboratory experiments \cite{costa}.

The pedestrians in this data set are all {\it socially interacting}, i.e. they were, on the basis of conversation and gaze clues, coded as not only moving together, but also performing some
kind of social interaction \cite{M,M2}.
\subsection{Group composition coding}
In order to obtain the ``ground truth'' for the inter-group composition and social relation, we proceeded similarly to our previous works \cite{M,M2}, 
 and asked three different human coders to observe the video recordings corresponding to the data set  \cite{data} and analyse the group composition, and in detail
to code, when possible,
\begin{enumerate}
\item the apparent purpose of the group's visit to the area ({\bf work} or {\bf leisure}),
\item the apparent gender of their members,
\item their apparent relation ({\bf colleagues}, {\bf friends}, {\bf family} or {\bf couples}\footnote{The Japanese term used, {\it koibito}, could be translated also as ``lovers'', and
suggests the idea of a relatively young, unmarried couple.}),
\item their apparent age (in decades, such as 0-9, 10-19, etc.)
\end{enumerate}
While one coder examined data from five different days (three working days and two holidays), corresponding to 1168 different socially interacting dyads, the other two examined only one day (283 dyads).
The coders are not specialised in pedestrian studies, are not aware of our mathematical models of pedestrian behaviour\footnote{One of the coders that analysed data from a single day is a 
non-technical member of our lab, but she did not take part in the development of mathematical modelling or quantitative data analysis.},
and did not have access to our quantitative measurements of position and velocity. They thus relied only on visual features such as clothing, gestures, behaviour and gazing \cite{knapp,kleine,arg} to identify the groups' 
social roles and composition\footnote{Distance and velocity are obviously features of the pedestrians' behaviour, but the coders had access only to visual clues concerning these properties,
and not to quantitative measurements. Furthermore, they were not given instructions such as ``friends walk closer than colleagues'' or similar. They were simply told to
use the available visual clues to code the social roles and composition.}.
\subsection{Coders' agreement}
The coding process is obviously strongly dependent on the subjective evaluation of the coder. Nevertheless,
the 283 dyads examined by all coders may be used to examine their agreement, and provide thus some information about the reliability of their coding.
To this end, we use in appendix \ref{coderel} two different approaches. On one hand, we use the standard approach used in social sciences of directly comparing the results of the coding,
through statistical indicators such as Cohen's and Fleiss's kappa, or Krippendorf's alpha (appendix \ref{coderag}). On the other hand, we also use an approach more rooted in the ``hard'' sciences, and treat the different 
codings as independent experiments, and quantitatively and quantitatively compare the findings (appendix \ref{codercomp}).
\subsection{Trajectories}
\label{traj}
While our tracking system provides us with pedestrian positions and velocities at time intervals $\delta t$ in the order of tens of milliseconds, 
we average pedestrian positions over time intervals 
$\Delta t=0.5$ s, to reduce 
the effect of measurement noise and the influence of pedestrian gait. As a result, we obtain pedestrian positions at discrete times $k$, as 
\begin{equation}
\mathbf{x}(k \Delta t)=(x(k \Delta t),y(k \Delta t),z(k \Delta t)),
\end{equation}
where $z$ gives the height of the top of the pedestrian head,
and  define pedestrian velocities in 2D as 
\begin{equation}
\mathbf{v}(k \Delta t)=\left[(x(k \Delta t)-x((k-1) \Delta t))/\Delta t,(y(k \Delta t)-y((k-1) \Delta t))/\Delta t\right].
\end{equation}
Following \cite{M,M2}, only data points with both the average group velocity $V$ (eq. \ref{eV}) and all individual velocities $v_i$ larger 0.5 m/s, and with all pedestrian positions falling inside a square with side 2.5 m centred on the group centre, were used\footnote{These thresholds were again based on our analysis of probability distribution functions of group positions in \cite{M,M2} and pedestrian
velocities in \cite{OOC}.}.
\subsubsection{Pedestrian Height}
Pedestrian height measurement is obviously subject to oscillations (see \cite{Drz}). A major problem with height tracking is that there are situations in which the head is hidden or poorly tracked,
and the pedestrian height is wrongly assigned as the height of the shoulders. To avoid this problem, for each pedestrian we first compute the median height over the whole trajectory, and then 
define the pedestrian height as the average $\overline{z}$ of all $z$ measurements above the median\footnote{As
discussed in  \cite{Drz} and \cite{OOC}, tracking errors in which the tracking ID is misassigned to a different pedestrian, or, in the worst of cases, to an object,
are possible, in particular in crowded environments. Such errors obviously affect the measurement of the pedestrian height, but since our pedestrians move in group, 
the large majority of these errors may be identified by noticing
that the pedestrians corresponding to the IDs in the group are not moving together anymore. For this reason, when computing the median of $z$ we use only data points 
for which the distance between pedestrians in the group was less than 4 meters. This is a conservative
threshold justified by our findings in \cite{M,M2} suggesting that interacting groups have extremely low probability of having a distance larger than 2 meters.}.

By being smaller and thus more easily occluded, the tracking of children is more difficult than the tracking of adults\footnote{For example, a couple of children resulted to have 
a height in the 160-170 cm range, due to a confusion between the parent and children position.}. Since our statistical analysis identified a few interesting results related
to children, we visually analysed the video recordings to verify that this problem did not affect significantly our findings.
\subsection{Density}
As we report in \cite{M2}, group velocity and spatial configuration depend on crowd density. In this paper we follow again our main analysis of \cite{M2} and compute pedestrian density with a
good spatial resolution (more than a good time resolution) as time averages over 300 seconds in a $L=0.5$ meters square area. More details, along with a discussion of possible density definitions,
may be found in \cite{M2} (refer also to \cite{VoroZhang} and \cite{corb2} for possible alternative definitions of density).
\section{Quantitative observables}
\begin{figure}[ht!]
\begin{center}
\includegraphics[width=0.7\linewidth]{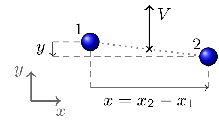} 
\caption{Group observables.}
\label{f0}
\end{center}
\end{figure}

Based on our analysis performed in works \cite{M,M2,M3}, we define the following quantitative observables for the dynamics of a pedestrian dyad (Fig \ref{f0}):
\begin{enumerate}
\item  The {\it group velocity} $V$, defined as
\begin{equation}
\label{eV}
  V=|\mathbf{V}| \qquad \mathbf{V}=(\mathbf{v}_1+\mathbf{v}_2)/2,
\end{equation}
$\mathbf{v}_i$ being the velocities of the two pedestrians in an arbitrary reference frame co-moving with the environment (i.e. in which the velocity of walls and other architectural features is zero).
\item The {\it pedestrian distance} or {\it group size} $r$, defined as
\begin{equation}
  r=|\mathbf{r}| \qquad \mathbf{r}=\mathbf{r}_1-\mathbf{r}_2,
\end{equation}
$\mathbf{r}_i$ being the positions of the two pedestrians in the above reference frame.
\item The {\it group abreast distance} or {\it abreast extension} $x$, that may be defined as follows. First we identify a unit vector in the direction of $\mathbf{V}$
\begin{equation}
\hat{\mathbf{g}}=\mathbf{V}/V.
\end{equation}
For each pedestrian we compute the clockwise angle $\theta_i$ between $\hat{\mathbf{g}}$ and $\mathbf{r}_i$, and define the projection of each $\mathbf{r}_i$ orthogonal to
the velocity
\begin{equation}
x_i = r_i \sin \theta_i.
\end{equation}
If necessary, we rename the pedestrians so that $x_1\leq x_2$ and finally define
\begin{equation}
x = x_2-x_1 \geq 0.
\end{equation}
\item  We can also define the group extension in the direction of motion,
\begin{equation}
y_i = r_i \cos \theta_i,\qquad y = y_2-y_1. 
\end{equation}
This is a signed quantity, a property that was particularly useful in our previous works. Nevertheless, for the purpose of this paper, it resulted more useful to analyse
the {\it group depth} 
\begin{equation}
|y|.
\end{equation}
\end{enumerate}
As described in detail in appendix \ref{detstat}, to which the reader should refer for technical details, for each observable $O$ and relation or composition category $k$ 
we provide the number of groups $N^k_g$, the category average $<O>_k$, standard deviation $\sigma_k$ and standard error $\varepsilon_k$, all based on the
analysis of groups that contributed with at least 10 usable data points, and reported in tables as
\begin{equation}
<O>_k \pm \varepsilon_k \;\;(\sigma).
\end{equation}
Furthermore we provide an analysis of the overall observable probability distribution function, and some parameters to estimate the differences between categories (ANOVA $p$ values,
effect size $\delta$, coefficient of determination $R^2$). The cross-analysis regarding the common effect of different ``factors'' (i.e., purpose, relation, gender, age and height) may be found 
in appendix \ref{accounting}. 
\section{The effect of purpose}
\label{purpose}
\subsection{Overall statistical analysis}
The results related to the purpose dependence of all observables
concerning the 1088 dyads whose purpose was coded (and that provided enough
data to be analysed) are shown in table \ref{table1a} (refer to appendix \ref{detstat} for an explanation of all terms).
\begin{table}[!ht]
\scriptsize
\caption{Observable dependence on purpose for dyads. Lengths in millimetres, times in seconds.}
\label{table1a}
\begin{center}
\begin{tabular} {|c|c|c|c|c|c|}
\hline
Purpose &  $N^k_g$ &  $V$ &    $r$ & $x$  & $|y|$ \\
\hline
Leisure & 716 & 1118 $\pm$ 7.3   ($\sigma$=195) &815 $\pm$ 9.5  ($\sigma$=253) &628 $\pm$ 6.1 ($\sigma$=162) &383 $\pm$ 12 ($\sigma$=334)    \\ 
\hline
Work & 372 & 1271 $\pm$ 8.2   ($\sigma$=158) &845 $\pm$ 12  ($\sigma$=228) &713 $\pm$ 8 ($\sigma$=154) &332 $\pm$ 15 ($\sigma$=289)    \\ 
\hline
$F_{1,1086}$ & & 169 & 3.75 & 69.4 & 6.25\\
\hline
$p$ & & $<10^{-8}$ & 0.053 & $<10^{-8}$ & 0.0126\\
\hline
$R^2$ & & 0.135 & 0.00344 & 0.0601 & 0.00572\\
\hline
$\delta$ & & 0.832 & 0.124 & 0.533 & 0.16\\
\hline
\end{tabular}
\end{center}
\end{table}
We have thus a very strong and significant evidence that pedestrians that visited the environment for working walk at an higher velocity and with a larger abreast distance, as shown
by the comparison of averages and standard errors, and by the corresponding high $F$ and $\delta$ and low $p$ values (see appendix \ref{detstat} for definitions and meaning
of these quantities). We have also a difference in distance and ``group depth'', although its  significance is less strong.
\subsection{Probability distribution functions}
We can get further insight about the differences in behaviour between workers and leisure oriented people by studying explicitly the probability distribution functions for the observables $V$, $r$, $x$
and $|y|$,
which are shown respectively in figures \ref{f1d}, \ref{f1a}, \ref{f1b} and \ref{f1c}, and whose statistical analysis is reported in appendix \ref{overpur} (refer again to appendix \ref{detstat} for the 
difference between the analysis reported in the main text an the one of appendix \ref{overpur}).

\begin{figure}[ht!]
\begin{center}
\includegraphics[width=0.7\linewidth]{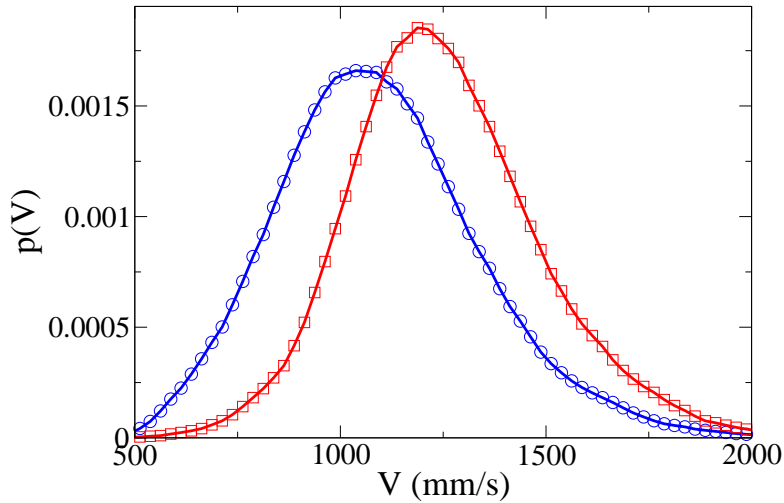} 
\caption{Pdf of the $V$ observable in leisure (blue, centre of bin identified by circles) and work (red, squares) oriented dyads. All pdfs in this work are shown after having been smoothed with
a moving average filter.}
\label{f1d}
\end{center}
\end{figure}

\begin{figure}[ht!]
\begin{center}
\includegraphics[width=0.7\linewidth]{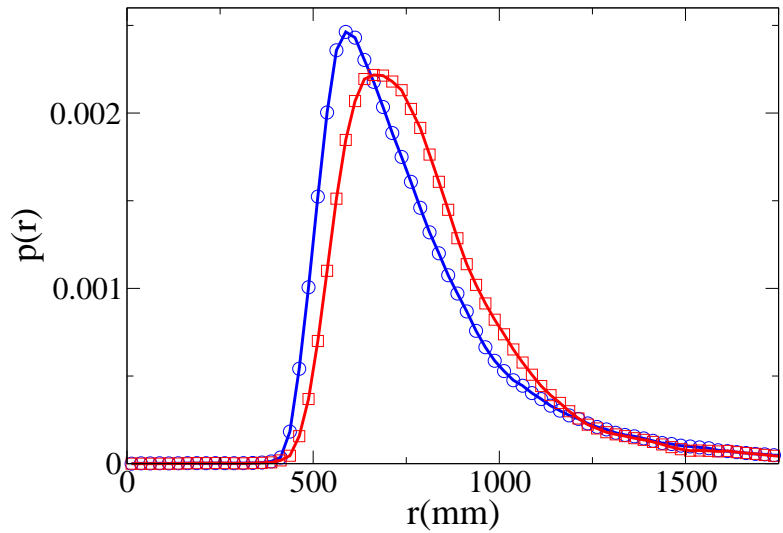} 
\caption{Pdf of the $r$ observable in leisure (blue, circles) and work (red, squares) oriented dyads.}
\label{f1a}
\end{center}
\end{figure}

\begin{figure}[ht!]
\begin{center}
\includegraphics[width=0.7\linewidth]{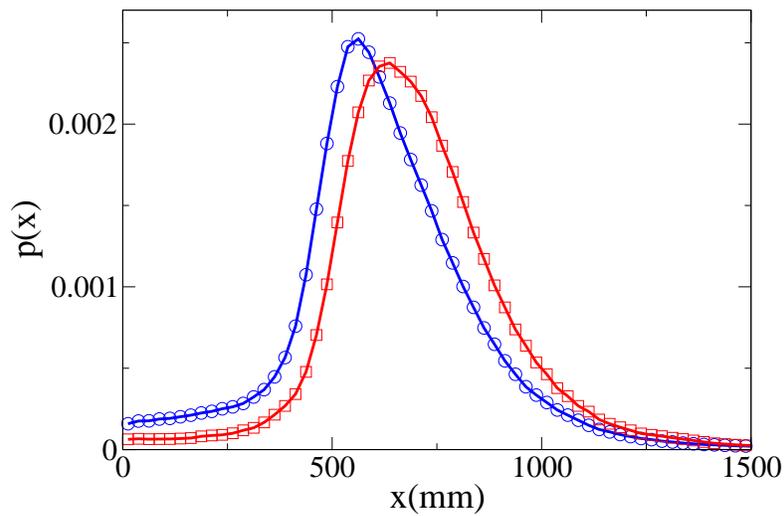} 
\caption{Pdf of the $x$ observable in leisure (blue, circles) and work (red, squares) oriented dyads.}
\label{f1b}
\end{center}
\end{figure}

\begin{figure}[ht!]
\begin{center}
\includegraphics[width=0.7\linewidth]{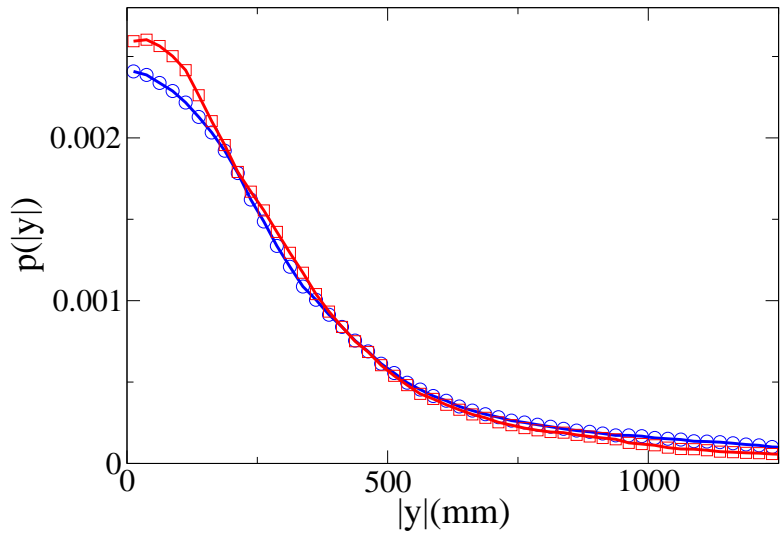} 
\caption{Pdf of the $|y|$ observable in leisure (blue, circles) and work (red, squares) oriented dyads.}
\label{f1c}
\end{center}
\end{figure}

These pdfs provide an easy interpretation of the data. The $x$ and $V$ peaks and tails are displaced to higher values for workers. The $r$ peak is also displaced to an higher value, but leisure distribution has
a fatter tail. Correspondingly, the $|y|$ distribution is slightly more spread in leisure oriented pedestrians. Furthermore, the $x$ distribution presents a considerably higher value for low $x$ values in the
``leisure'' case.
These latter consideration show that while ``workers'' walk strongly abreast, the ``leisure'' dyads are less ordered.
\subsection{Further analysis}
In appendix \ref{furtherpur} we further analyse these results, to understand the effect on them of age, gender, density and height, while in appendix  \ref{coderpur} we verify that
 the major findings are confirmed by all coders. We may see that in general, even when age, gender, density and height are kept fixed, the results exposed above are confirmed
 by this further analysis. 
\section{The effect of relation}
Groups may also be analysed according to the relation between their members (colleagues, couples, friends, family). There is obviously a strong overlap between the ``colleagues'' category and the ``work'' one
analysed above (and obviously between ``leisure'' and the three categories couples, friends and families), but since they are are conceptually different (colleagues could visit the shopping mall for
lunch, or for shopping outside of working time), we will provide an independent analysis\footnote{Although in the cross-analysis of appendix \ref{accounting} we usually drop the analysis of
purpose and focus on relation.}.
\subsection{Overall statistical analysis}
The results related to the relation dependence for all observables
concerning the 1018 dyads whose purpose was coded (and that provided enough
data to be analysed) are shown in table \ref{table2a}.
\begin{table}[!ht]
\scriptsize
\caption{Observable dependence on purpose for dyads. Lengths in millimetres, times in seconds.}
\label{table2a}
\begin{center}
\begin{tabular} {|c|c|c|c|c|c|}
\hline
Relation&  $N^k_g$ &  $V$ &    $r$ & $x$  & $|y|$ \\
\hline
Colleagues & 358 & 1274 $\pm$ 8.3   ($\sigma$=157) &851 $\pm$ 12  ($\sigma$=231) &718 $\pm$ 8.3 ($\sigma$=157) &334 $\pm$ 15 ($\sigma$=292)    \\ 
\hline
Couples & 96 & 1099 $\pm$ 17   ($\sigma$=169) &714 $\pm$ 22  ($\sigma$=219) &600 $\pm$ 15 ($\sigma$=150) &291 $\pm$ 24 ($\sigma$=231)    \\ 
\hline
Families & 246 & 1094 $\pm$ 13   ($\sigma$=197) &863 $\pm$ 19  ($\sigma$=302) &583 $\pm$ 11 ($\sigma$=171) &498 $\pm$ 25 ($\sigma$=391)    \\ 
\hline
Friends & 318 & 1138 $\pm$ 11   ($\sigma$=200) &792 $\pm$ 11  ($\sigma$=199) &662 $\pm$ 7.5 ($\sigma$=134) &314 $\pm$ 15 ($\sigma$=268)    \\ 
\hline
$F_{3,1014}$ & & 60.7 & 12.2 & 42.3 & 21.4\\
\hline
$p$ & & $<10^{-8}$ & 7.39$\cdot 10^{-8}$ & $<10^{-8}$ & $<10^{-8}$\\
\hline
$R^2$ & & 0.152 & 0.0349 & 0.111 & 0.0595\\
\hline
$\delta$ & & 1.03 & 0.529 & 0.828 & 0.587\\
\hline
\end{tabular}
\end{center}
\end{table}

We may see that, as expected from the previous analysis, there is a considerable difference between the velocity of colleagues and the velocity of the other groups. Friends appear to be faster
than couples or families, although the difference is limited to 2-3 standard errors. We may also see that couples walk at the closest distance, followed by friends, colleagues and then families.
On the other hand, families walk at the shortest abreast distance, although at a value basically equivalent to that of couples. The abreast distance of friends is significantly larger, and
the one of colleagues assumes the greatest value. The ``depth'' $|y|$ assumes the smallest value in couples, followed by friends and workers, and the, by a large margin, highest value in
families.
\subsection{Probability distribution functions}
\begin{figure}[ht!]
\begin{center}
\includegraphics[width=0.7\linewidth]{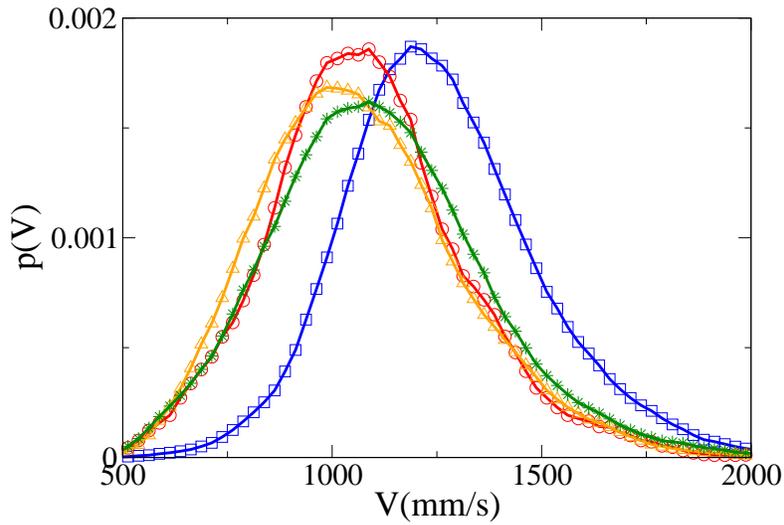} 
\caption{Pdf of the $V$ observable in dyads with colleague (blue, squares), couple (red, circles), family (orange, triangles) and friend (green, stars) relation.}
\label{f2d}
\end{center}
\end{figure}

\begin{figure}[ht!]
\begin{center}
\includegraphics[width=0.7\linewidth]{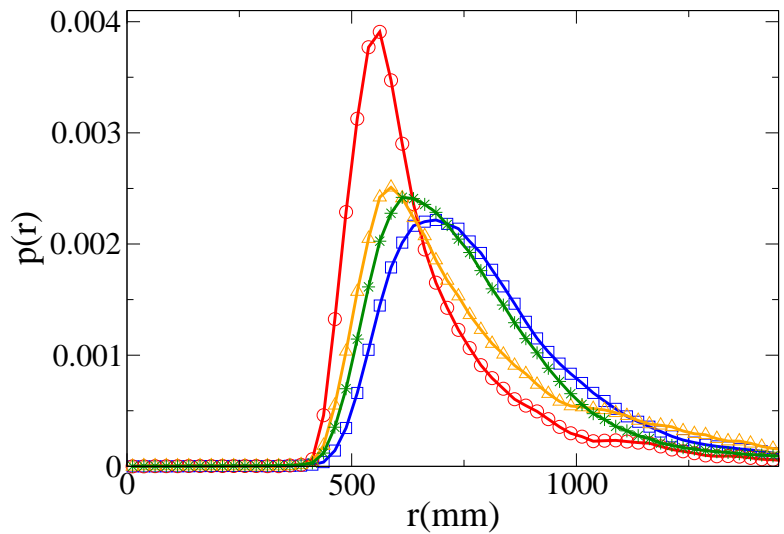} 
\caption{Pdf of the $r$ observable in dyads with colleague (blue, squares), couple (red, circles), family (orange, triangles) and friend (green, stars) relation.}
\label{f2a}
\end{center}
\end{figure}

\begin{figure}[ht!]
\begin{center}
\includegraphics[width=0.7\linewidth]{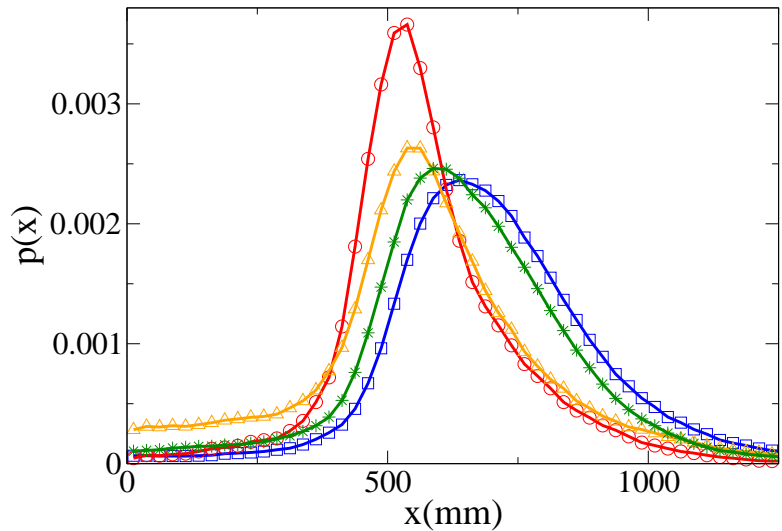} 
\caption{Pdf of the $x$ observable in dyads with colleague (blue, squares), couple (red, circles), family (orange, triangles) and friend (green, stars) relation.}
\label{f2b}
\end{center}
\end{figure}

\begin{figure}[ht!]
\begin{center}
\includegraphics[width=0.7\linewidth]{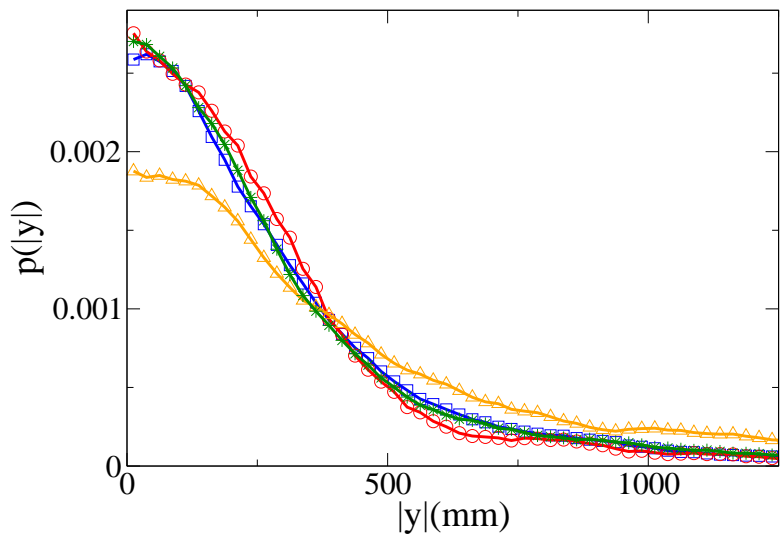} 
\caption{Pdf of the $|y|$ observable in dyads with colleague (blue, squares), couple (red, circles), family (orange, triangles) and friend (green, stars) relation.}
\label{f2c}
\end{center}
\end{figure}

These results may be completely understood only by analysing the probability distribution functions, which are shown in  figures \ref{f2d}, \ref{f2a}, \ref{f2b} and \ref{f2c} for,
respectively, $V$, $r$, $x$ and $|y|$ (the statistical analysis of these distributions is reported in section \ref{overrel}).

The pdfs provide again an easy interpretation of the data. The $V$ distributions for friends, families and couples are quite similar, while the one for colleagues is clearly different (displaced to
higher values). This suggests that ``relation'' influences velocity in a limited way, with respect to ``purpose''.
The peaks of both $r$ and $x$ distributions assume the minimum value for couples, followed by families, friends and colleagues. The distributions for families present the following peculiar properties:
the $r$ distribution has a fat tail (causing the high average value), the $x$ distribution assumes large values for small $x$, and the $|y|$ distribution is more spread (on the other hand,
$|y|$ distributions are very similar in the other categories).

We may thus conclude that ``relation'' has an influence on distance, with couples walking at the closest distance, followed by families, friends and colleagues. At the same time, families walk in a less
ordered formation (less abreast). This behaviour is probably mainly due to children (see also section \ref{ageef}), and influences the results of the previous section (``leisure'' oriented dyads walking less abreast).
\subsection{Further analysis}
In appendix \ref{furtherrel} we further analyse these results, to understand the effect on them of age, gender, density and height, while in appendix  \ref{coderrel} we verify if
 the major findings are confirmed by all coders. The major trends exposed above are all  confirmed by this further analysis. In particular, the tendency of families to have a wider $|y|$ distribution
may be diminished but does not disappear when we keep fixed gender, age or height, showing that it is probably not only due to children.
\section{The effect of gender}
\subsection{Overall statistical analysis}
The results related to the relation dependence for all observables
concerning the 1089 dyads whose gender was coded (and that provided enough
data to be analysed) are shown in table \ref{table3a}.
\begin{table}[!ht]
\scriptsize
\caption{Observable dependence on gender for dyads. Lengths in millimetres, times in seconds.}
\label{table3a}
\begin{center}
\begin{tabular} {|c|c|c|c|c|c|}
\hline
Gender&  $N^k_g$ &  $V$ &    $r$ & $x$  & $|y|$ \\
\hline
Two females & 252 & 1102 $\pm$ 12   ($\sigma$=193) &790 $\pm$ 14  ($\sigma$=227) &647 $\pm$ 7.8 ($\sigma$=123) &321 $\pm$ 20 ($\sigma$=311)    \\ 
\hline
Mixed & 371 & 1111 $\pm$ 9.5   ($\sigma$=183) &824 $\pm$ 14  ($\sigma$=273) &613 $\pm$ 9 ($\sigma$=174) &416 $\pm$ 18 ($\sigma$=350)    \\ 
\hline
Two males & 466 & 1254 $\pm$ 8.3   ($\sigma$=178) &846 $\pm$ 11  ($\sigma$=228) &699 $\pm$ 7.7 ($\sigma$=166) &349 $\pm$ 14 ($\sigma$=293)    \\ 
\hline
$F_{2,1086}$ & & 84.6 & 4.37 & 30.7 & 7.69\\
\hline
$p$ & & $<10^{-8}$ & 0.0129 & $<10^{-8}$ & 0.000484\\
\hline
$R^2$ & & 0.135 & 0.00798 & 0.0535 & 0.014\\
\hline
$\delta$ & & 0.825 & 0.248 & 0.51 & 0.282\\
\hline
\end{tabular}
\end{center}
\end{table}
\subsection{Probability distribution functions}
\begin{figure}[ht!]
\begin{center}
\includegraphics[width=0.7\linewidth]{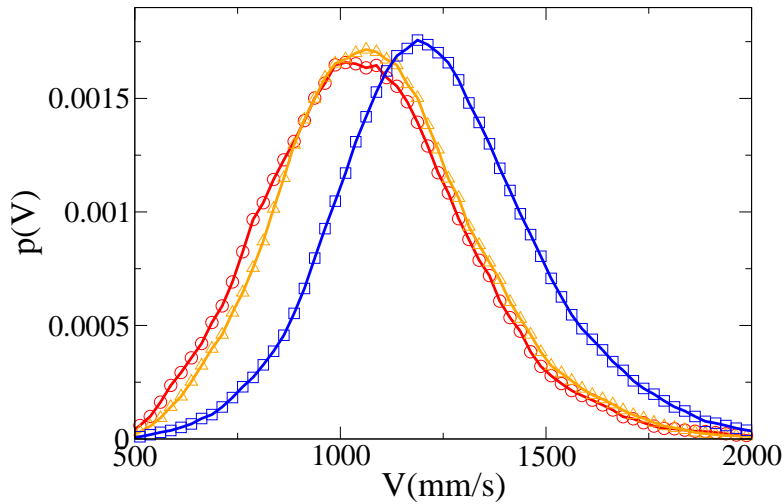} 
\caption{Pdf of the $V$ observable in dyads with two females (red, circles), mixed (orange, triangles) and two males (blue, squares).}
\label{f3d}
\end{center}
\end{figure}

\begin{figure}[ht!]
\begin{center}
\includegraphics[width=0.7\linewidth]{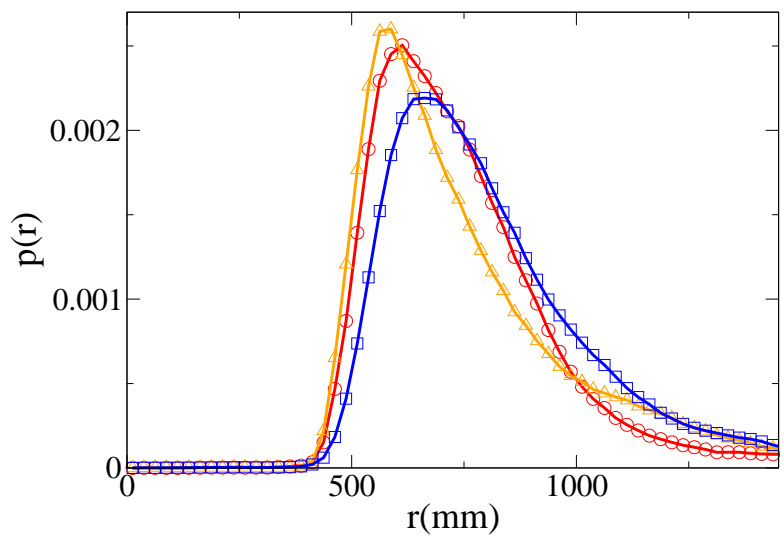} 
\caption{Pdf of the $r$ observable in dyads with two females (red, circles), mixed (orange, triangles) and two males (blue, squares).}
\label{f3a}
\end{center}
\end{figure}

\begin{figure}[ht!]
\begin{center}
\includegraphics[width=0.7\linewidth]{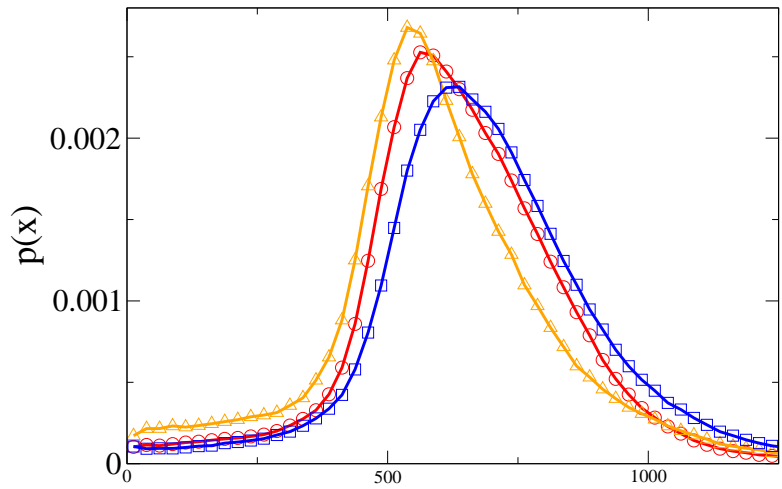} 
\caption{Pdf of the $x$ observable in dyads with two females (red, circles), mixed (orange, triangles) and two males (blue, squares).}
\label{f3b}
\end{center}
\end{figure}

\begin{figure}[ht!]
\begin{center}
\includegraphics[width=0.7\linewidth]{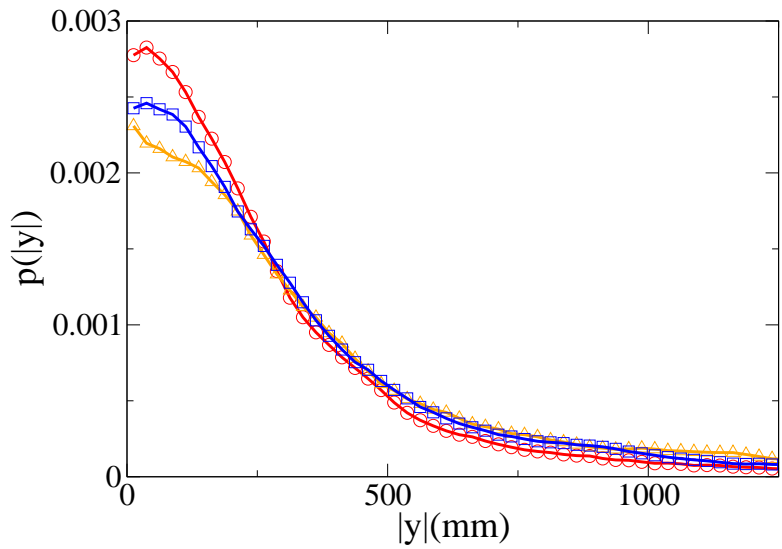} 
\caption{Pdf of the $|y|$ observable in dyads with two females (red, circles), mixed (orange, triangles) and two males (blue, squares).}
\label{f3c}
\end{center}
\end{figure}

Although we may easily see that the differences between the distributions are statistically significant (with stronger differences in the $V$ and $x$ distributions),
it is again useful, in order to understand these results, to analyse the probability distribution functions, which are shown in figures \ref{f3d}, \ref{f3a}, \ref{f3b} and \ref{f3c} for, respectively,
the $V$, $r$, $x$ and $|y|$ observables (the statistical analysis of these distributions is reported in section \ref{overgen})

The difference between the females and males distributions is very clear, with both peaks and tails in the velocity and (abreast or absolute) distance distributions displaced to higher results.
Regarding the $|y|$ distribution, we may see that the male distribution is more spread than the female one, and thus females have a stronger tendency to walk abreast.
The mixed dyads absolute and abreast distance distribution are characterised by low values for the peaks and fat tails, in particular for the absolute value distribution. The $x$ distribution
presents relatively high values at low $x$, and correspondingly
the $|y|$ distribution is very spread
(tendency not to walk abreast). The mixed dyads velocity distribution is interestingly very similar to the female one. 
\subsection{Further analysis}
The peculiarity of the mixed distributions may be better understood by taking in consideration
the other effects, in particular those related to relation, as shown in appendix \ref{furthergen}. Coder reliability is analysed in appendix \ref{codergen}.
\section{The effect of age}
\label{ageef}
To study the dependence of the $r$, $x$, $|y|$ and $V$ observable on age, we used three different approaches, namely to study how these observable change depending on {\it average},
{\it maximum} and {\it minimum} group age. The latter analysis appears to be the most interesting one, since it allows us to spot the presence of children, and we limit ourselves to it in the main text.
Results corresponding to the dependence on average and maximum age are found in appendix \ref{maxavage}. 
\subsection{Overall statistical analysis}
\begin{figure}[ht!]
\begin{center}
\includegraphics[width=0.7\linewidth]{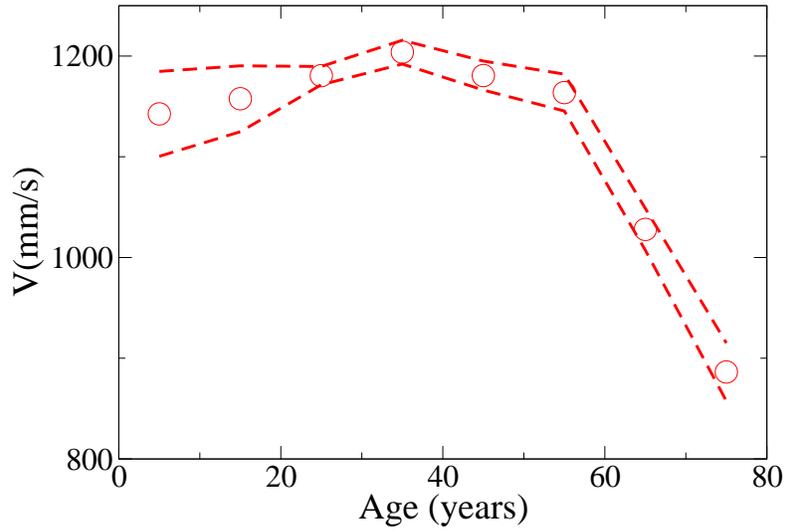} 
\caption{$V$ dependence on minimum age. Dashed lines provide standard error confidence intervals. The point at 75 years corresponds to the ``70 years or more'' slot.}
\label{f4b}
\end{center}
\end{figure}

\begin{figure}[ht!]
\begin{center}
\includegraphics[width=0.7\linewidth]{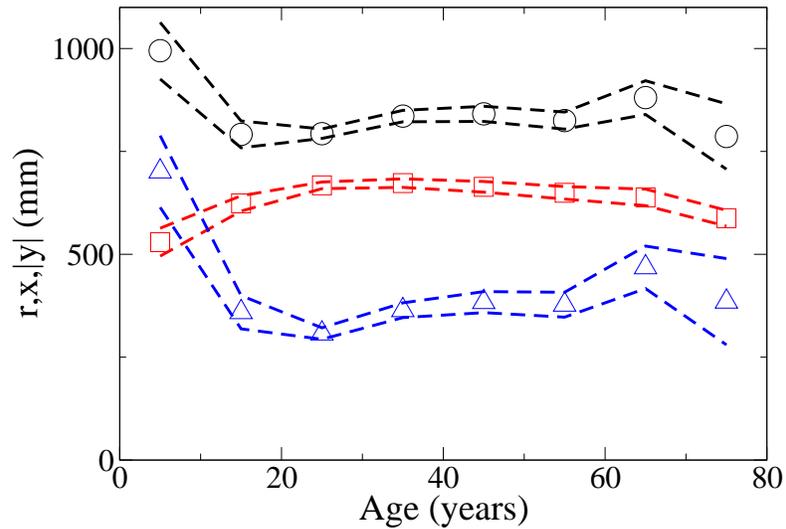} 
\caption{$r$, $x$ and $|y|$ dependence on minimum age. Black circles: $r$; red squares: $x$; blue triangles: $|y|$. Dashed lines provide standard error confidence intervals. The point at 75 years corresponds to the ``70 years or more'' slot.}
\label{f4b2}
\end{center}
\end{figure}

Table \ref{table4a} and figures \ref{f4b}, \ref{f4b2} show the minimum age dependence of all observables (based on the analysis of 1089 dyads). Although differences between distributions are statistically significant, both velocity and distance observables are mostly constant for 
groups whose minimum age is in the 20-60 years range. We nevertheless find that the group depth $|y|$ (the observable characterising thus the tendency of pedestrians not to walk abreast) assumes 
a very high value in groups with children, a minimum in the 20-29 years range, and then grows with age. On the other hand, abreast distance $x$ is relatively low for groups with children (as we will see
below, $x$ grows with body size\footnote{Body size could also influence the $x$ distance in elderly people, due to the shorter height of elderly people in the Japanese population \cite{heightdis}.}). Velocity is mostly constant below 60 years, but drops for elderly groups.

\begin{table}[!ht]
\scriptsize
\caption{Observable dependence on minimum age for dyads. Lengths in millimetres, times in seconds.}
\label{table4a}
\begin{center}
\begin{tabular} {|c|c|c|c|c|c|}
\hline
Minimum age&  $N^k_g$ &  $V$ &    $r$ & $x$  & $|y|$ \\
\hline
0-9 years & 31 & 1143 $\pm$ 42   ($\sigma$=235) &995 $\pm$ 69  ($\sigma$=383) &529 $\pm$ 34 ($\sigma$=189) &701 $\pm$ 87 ($\sigma$=485)    \\ 
\hline
10-19 years & 63 & 1158 $\pm$ 33   ($\sigma$=259) &791 $\pm$ 33  ($\sigma$=259) &624 $\pm$ 19 ($\sigma$=148) &359 $\pm$ 40 ($\sigma$=320)    \\ 
\hline
20-29 years & 364 & 1181 $\pm$ 9.1   ($\sigma$=173) &793 $\pm$ 11  ($\sigma$=218) &668 $\pm$ 8.1 ($\sigma$=154) &307 $\pm$ 14 ($\sigma$=264)    \\ 
\hline
30-39 years & 292 & 1204 $\pm$ 12   ($\sigma$=202) &836 $\pm$ 14  ($\sigma$=238) &673 $\pm$ 10 ($\sigma$=176) &364 $\pm$ 18 ($\sigma$=307)    \\ 
\hline
40-49 years & 149 & 1181 $\pm$ 14   ($\sigma$=176) &841 $\pm$ 18  ($\sigma$=224) &664 $\pm$ 13 ($\sigma$=158) &384 $\pm$ 26 ($\sigma$=311)    \\ 
\hline
50-59 years & 111 & 1164 $\pm$ 18   ($\sigma$=193) &825 $\pm$ 21  ($\sigma$=219) &649 $\pm$ 15 ($\sigma$=160) &378 $\pm$ 30 ($\sigma$=318)    \\ 
\hline
60-69 years & 67 & 1028 $\pm$ 21   ($\sigma$=170) &881 $\pm$ 41  ($\sigma$=335) &638 $\pm$ 20 ($\sigma$=164) &468 $\pm$ 52 ($\sigma$=422)    \\ 
\hline
$\geq$ 70 years & 12 & 886 $\pm$ 29   ($\sigma$=99.8) &786 $\pm$ 79  ($\sigma$=275) &588 $\pm$ 19 ($\sigma$=66.6) &385 $\pm$ 100 ($\sigma$=363)    \\ 
\hline
$F_{7,1081}$ & & 10.7 & 3.96 & 4.23 & 8.02\\
\hline
$p$ & & $<10^{-8}$ & 0.000282 & 0.000128 & $<10^{-8}$\\
\hline
$R^2$ & & 0.065 & 0.025 & 0.0267 & 0.0494\\
\hline
$\delta$ & & 1.6 & 0.583 & 0.808 & 1.37\\
\hline
\end{tabular}
\end{center}
\end{table}

\subsection{Probability distribution functions}
\begin{figure}[ht!]
\begin{center}
\includegraphics[width=0.7\linewidth]{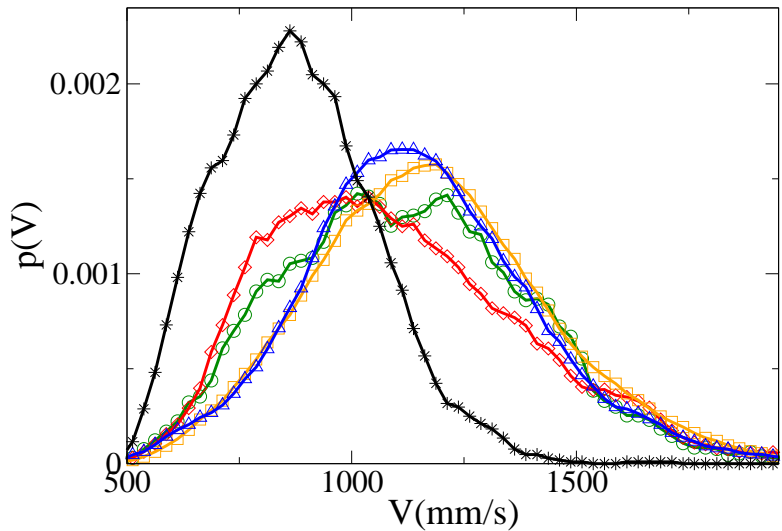} 
\caption{Probability distribution function for $V$. (Minimum) age in the 0-9 years range: green and circles; in the 10-19 range: red and diamonds, in the 30-39 range: orange and squares;
 in the 50-59 range: blue, and triangles; in the over 70 range: black and stars.}
\label{f4g}
\end{center}
\end{figure}

\begin{figure}[ht!]
\begin{center}
\includegraphics[width=0.7\linewidth]{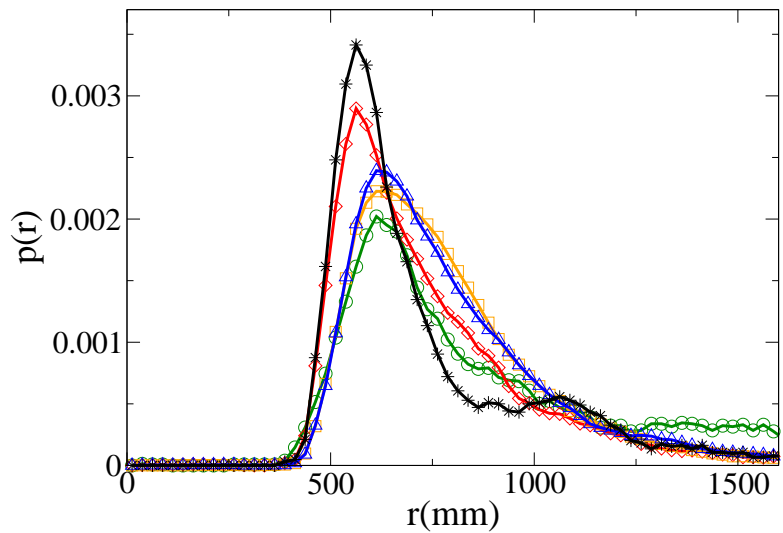} 
\caption{Probability distribution function for $r$. (Minimum) age in the 0-9 years range: green and circles; in the 10-19 range: red and diamonds, in the 30-39 range: orange and squares;
 in the 50-59 range: blue, and triangles; in the over 70 range: black and stars.}
\label{f4d}
\end{center}
\end{figure}

\begin{figure}[ht!]
\begin{center}
\includegraphics[width=0.7\linewidth]{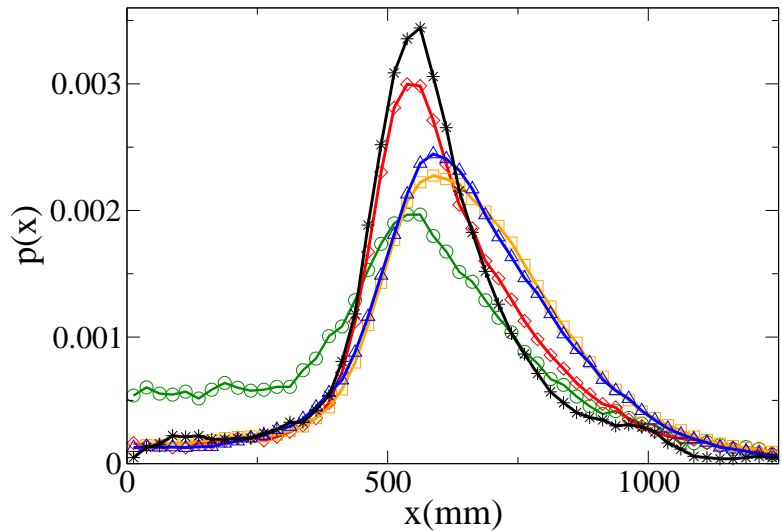} 
\caption{Probability distribution function for $x$. (Minimum) age in the 0-9 years range: green and circles; in the 10-19 range: red and diamonds, in the 30-39 range: orange and squares;
 in the 50-59 range: blue, and triangles; in the over 70 range: black and stars.}
\label{f4e}
\end{center}
\end{figure}

\begin{figure}[ht!]
\begin{center}
\includegraphics[width=0.7\linewidth]{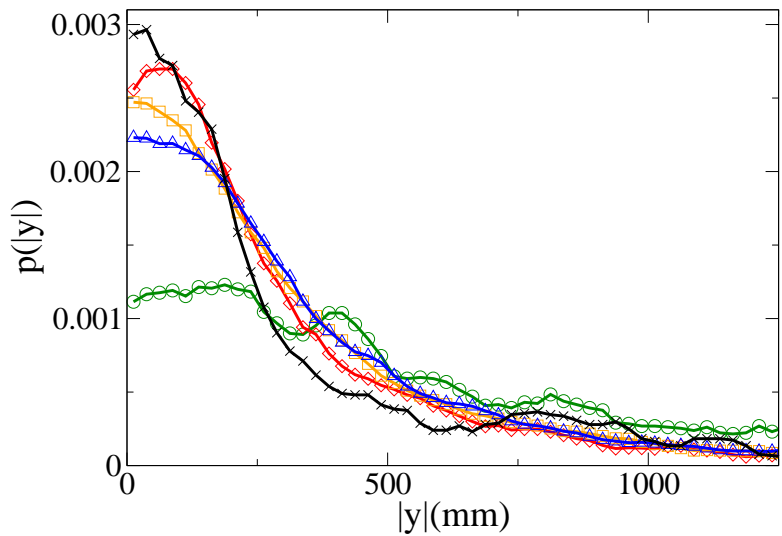} 
\caption{Probability distribution function for $|y|$. (Minimum) age in the 0-9 years range: green and circles; in the 10-19 range: red and diamonds, in the 30-39 range: orange and squares;
 in the 50-59 range: blue, and triangles; in the over 70 range: black and stars.}
\label{f4f}
\end{center}
\end{figure}

The probability functions for different observables in different age ranges are shown in figures \ref{f4d}, \ref{f4e}, \ref{f4f} and \ref{f4g} respectively
for observables $V$, $r$, $x$ and $|y|$, and their statistical analysis is presented in section
\ref{overage}.
We may easily see from the large tail of the $r$ distribution, the high values for the $x$ distribution, the spread of the $|y|$ distribution, that the presence of a child causes the
group not to walk very abreast. The abreast distance peak is higher in ``working age people''with respect to young and elderly dyads.
Elderly people have a very narrow peak in the $|y|$ distribution, but also a fat tail. Velocity
 in the 0-19 age range assumes lower peaks than in the
 20-59 range, but has a large spread, while in elderly people it assumes clearly lower values\footnote{As stated in section \ref{traj}, the tracking of short people (and thus of children) 
is more difficult, and thus the tracked position could be affected by higher sensor noise, although our time filter (see again section \ref{traj}) should remove this problem.
We thus examined a portion of the videos corresponding to group with children, and noticed that children have indeed
 an erratic behaviour that leads them to sudden accelerations and non-abreast formations. We thus believe that the large spread of observables for dyads with children is due to actual pedestrian 
behaviour.}

\subsection{Further analysis}
In appendix \ref{furtherage} we analyse possible effects due to density, relation, gender and age. Interesting results reported in the appendix suggest a tendency of families {\it not} to walk abreast
even when formed only by adults\footnote{This could be related to a visual bias of coders, that code mixed dyads as families when not walking abreast, and couples when walking abreast.},
and differences in groups with children based on gender (probably affected by the gender of the parent). Coder reliability is analysed in appendix \ref{coderage}.

A further interesting result is that, as shown in figure \ref{fastchild} (based on the analysis of appendix \ref{accounting}), dyads with children {\it walk faster at higher density}, in contrast with the usual pedestrian behaviour.

\begin{figure}[ht!]
\begin{center}
\includegraphics[width=0.7\linewidth]{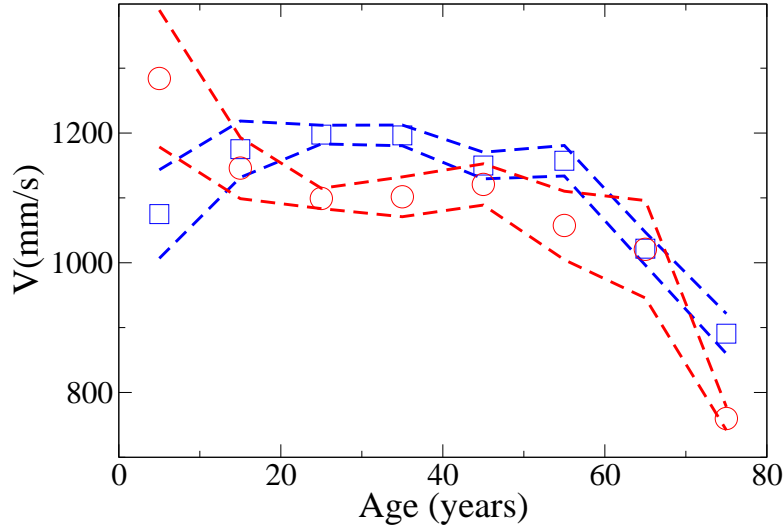} 
\caption{{\bf Dependence of $V$ on minimum age at different densities.} Blue squares: $0 \leq \rho \leq 0.05$ ped/m$^2$ range; red circles: $0.15 \leq \rho \leq 0.2$ ped/m$^2$ range. The point at 75 years corresponds to the ``70 years or more'' slot. Dashed lines show standard error confidence intervals.}
\label{fastchild}
\end{center}
\end{figure}
\section{The effect of height}
Height is the only pedestrian feature that is not the result of coding, since it is automatically tracked by our system \cite{Drz}. We again considered (see appendix \ref{heicomp}) average, minimum and maximum height. The three indicators give similar results, and in the following we use minimum height to better identify the presence of children.
\subsection{Overall statistical analysis}
\begin{figure}[ht!]
\begin{center}
\includegraphics[width=0.7\linewidth]{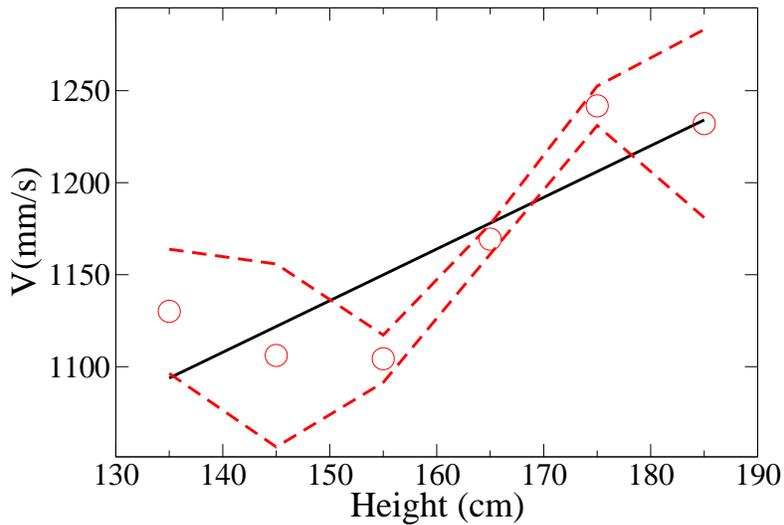} 
\caption{$V$ dependence on minimum height.  Data points shown by red circles. Continuous black line: linear best fit with $V=\alpha+\beta h$, $\alpha$=715 mm/s, $\beta$=2.81 s$^{-1}$. Dashed lines provide standard error confidence intervals, the point at 135 cm corresponds to the ``less than 140 cm'' slot, the one at 185 cm to the ``more than 180 cm'' slot.}
\label{f5d}
\end{center}
\end{figure}

\begin{figure}[ht!]
\begin{center}
\includegraphics[width=0.7\linewidth]{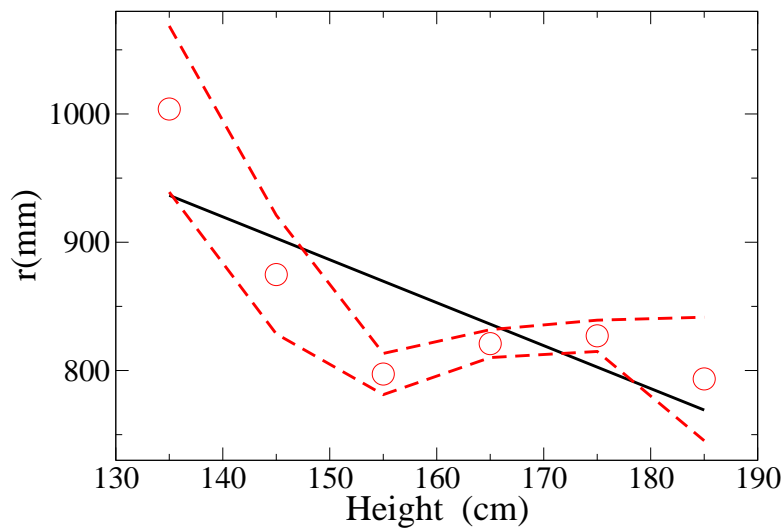} 
\caption{$r$ dependence on minimum height. Data points shown by red circles. Continuous black line: linear best fit with $r=\alpha+\beta h$, $\alpha$=1390 mm, $\beta$=-3.35.
Dashed lines provide standard error confidence intervals, the point at 135 cm corresponds to the ``less than 140 cm'' slot, the one at 185 cm to the ``more than 180 cm'' slot.}
\label{f5d2}
\end{center}
\end{figure}

\begin{figure}[ht!]
\begin{center}
\includegraphics[width=0.7\linewidth]{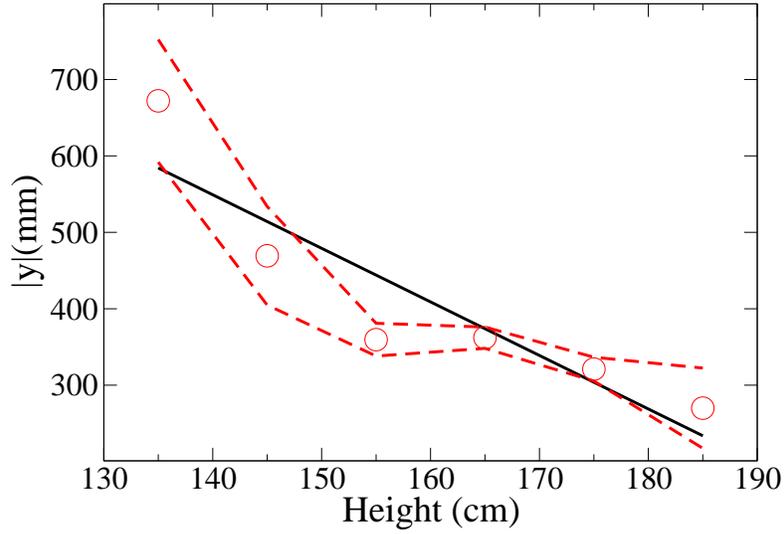} 
\caption{ $|y|$ dependence on minimum height. Data points shown by red circles. Continuous black line: linear best fit with $|y|=\alpha+\beta h$, $\alpha$=1530 mm, $\beta$=-7.01.
Dashed lines provide standard error confidence intervals, the point at 135 cm corresponds to the ``less than 140 cm'' slot, the one at 185 cm to the ``more than 180 cm'' slot.}
\label{f5b}
\end{center}
\end{figure}

\begin{figure}[ht!]
\begin{center}
\includegraphics[width=0.7\linewidth]{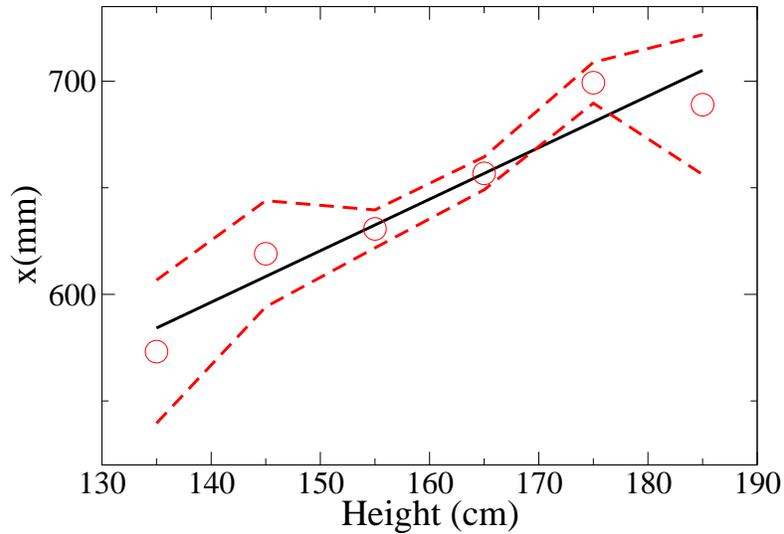} 
\caption{$x$ dependence on minimum height. Data points shown by red circles. Continuous black line: linear best fit with $x=\alpha+\beta h$, $\alpha$=258 mm, $\beta$=2.42.
Dashed lines provide standard error confidence intervals, the point at 135 cm corresponds to the ``less than 140 cm'' slot, the one at 185 cm to the ``more than 180 cm'' slot.}
\label{f5b2}
\end{center}
\end{figure}

The dependence of all observables on minimum height (based on 1089 dyads) is shown in table \ref{table_h_min}. We have significant statistical difference for all observables, but the interpretation of the results is not
straightforward, due to the peculiar behaviour of dyads including short people (most probably children). As shown in Figure \ref{f5d} velocity grows (as expected, see for example \cite{velh} and \cite{OOC}), with height, but dyads with a very short individual represent an exception (children move fast despite the short height).
In figure \ref{f5d2} we may see that distance is mostly independent of height above 150 cm,
but assumes a very high value for dyads including short pedestrians. Figure \ref{f5b} shows the height dependence of $|y|$,
which results to be a decreasing function, although a comparison 
with a linear fit shows that dyads including people under 140 cm walk with a particularly spread (non abreast) $|y|$ distribution, while above 150 cm the
group depth is almost constant. The $x$ observable, on the other hand, appears to grow mostly in a linear way (figure \ref{f5b2}).
This could lead us to think that abreast distance depends only on body size. Nevertheless, while there is probably a strong dependence of abreast distance on height, this linear
dependence is also due to the balance between the non-linear male and female behaviour, as shown in figure \ref{fastchild2}, based on the analysis of appendix \ref{accounting}. Furthermore, as we will see below when studying the 
distribution probability functions of the $x$ observable, the growth in $x$ with height is a combination of a increase of peak position and decrease of people walking in non abreast formation
(figure \ref{f5f}).

\begin{table}[!ht]
\scriptsize
\caption{Observable dependence on minimum height for dyads. Lengths in millimetres, times in seconds.}
\label{table_h_min}
\begin{center}
\begin{tabular} {|c|c|c|c|c|c|}
\hline
Minimum height&  $N^k_g$ &  $V$ &    $r$ & $x$  & $|y|$ \\
\hline
$<$ 140 cm & 39 & 1130 $\pm$ 34   ($\sigma$=211) &1004 $\pm$ 65  ($\sigma$=404) &573 $\pm$ 34 ($\sigma$=210) &672 $\pm$ 80 ($\sigma$=501)    \\ 
\hline
140-150 cm & 39 & 1106 $\pm$ 50   ($\sigma$=311) &875 $\pm$ 46  ($\sigma$=289) &619 $\pm$ 25 ($\sigma$=156) &469 $\pm$ 64 ($\sigma$=403)    \\ 
\hline
150-160 cm & 234 & 1104 $\pm$ 13   ($\sigma$=197) &797 $\pm$ 16  ($\sigma$=246) &631 $\pm$ 8.9 ($\sigma$=136) &360 $\pm$ 21 ($\sigma$=328)    \\ 
\hline
160-170 cm & 498 & 1169 $\pm$ 8.1   ($\sigma$=182) &821 $\pm$ 11  ($\sigma$=243) &657 $\pm$ 7.7 ($\sigma$=172) &362 $\pm$ 14 ($\sigma$=311)    \\ 
\hline
170-180 cm & 262 & 1242 $\pm$ 11   ($\sigma$=173) &827 $\pm$ 12  ($\sigma$=197) &699 $\pm$ 9.6 ($\sigma$=155) &321 $\pm$ 16 ($\sigma$=251)    \\ 
\hline
$>$ 180 cm & 17 & 1232 $\pm$ 51   ($\sigma$=211) &793 $\pm$ 48  ($\sigma$=198) &689 $\pm$ 33 ($\sigma$=135) &270 $\pm$ 53 ($\sigma$=217)    \\ 
\hline
$F_{5,1083}$ & & 14.5 & 5.25 & 7.45 & 9.69\\
\hline
$p$ & & $<10^{-8}$ & 9.03$\cdot 10^{-5}$ & 6.9$\cdot 10^{-7}$ & $<10^{-8}$\\
\hline
$R^2$ & & 0.0626 & 0.0237 & 0.0333 & 0.0428\\
\hline
$\delta$ & & 0.744 & 0.591 & 0.773 & 0.922\\
\hline
\end{tabular}
\end{center}
\end{table}

\begin{figure}[ht!]
\begin{center}
\includegraphics[width=0.7\linewidth]{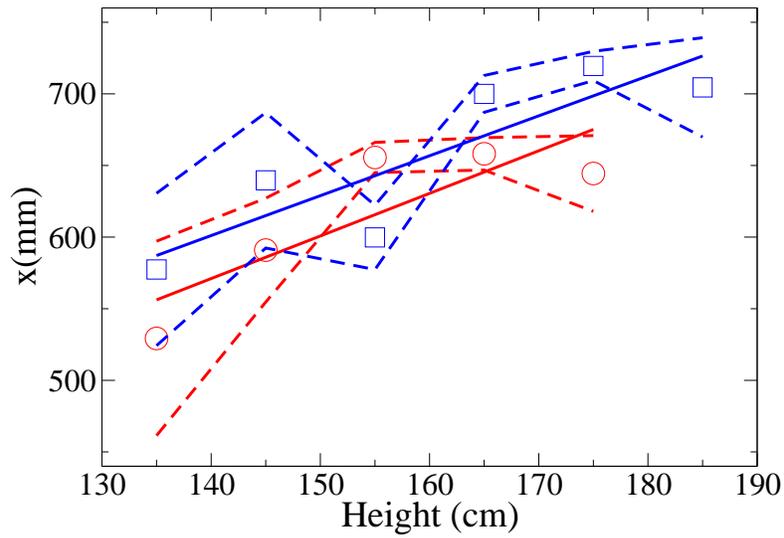} 
\caption{$x$ dependence on minimum height for different genders. Red circles: two females; blue squares: two males (continuous lines:
linear fits $x=\alpha+\beta h$, $\alpha_{\text{female}}$=154 mm, $\beta_{\text{female}}$=2.98; $\alpha_{\text{male}}$=211 mm, $\beta_{\text{male}}$=2.79). The points at 135 and 185 cm represent the ``less than 
140'' and ``more than 180'' cm slots).
Dashed lines show confidence intervals.}
\label{fastchild2}
\end{center}
\end{figure}

\subsection{Probability distribution functions}
\begin{figure}[ht!]
\begin{center}
\includegraphics[width=0.7\linewidth]{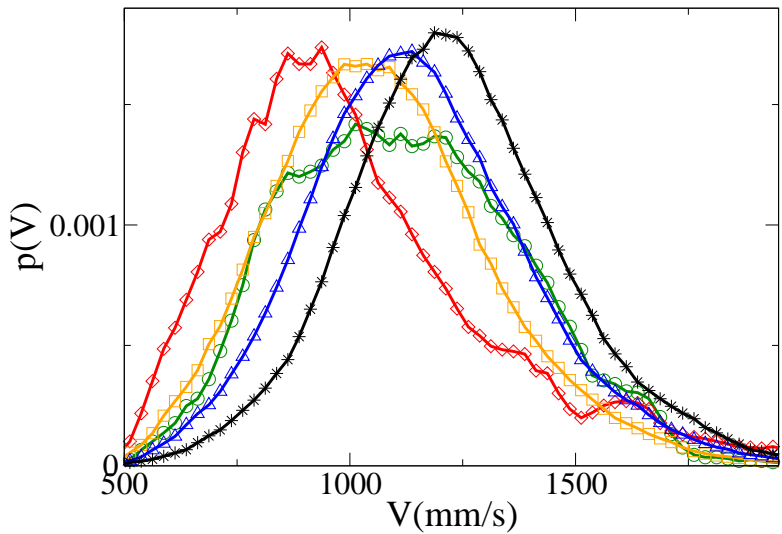} 
\caption{Probability distribution function for $V$. (Minimum) height in the 0-140 cm range: green circles; in the 140-150 range: red diamonds; in the 150-160 range: orange squares;
in the 160-170 range: blue triangles;
in the 170-180 range: black stars. }
\label{f5h}
\end{center}
\end{figure}

\begin{figure}[ht!]
\begin{center}
\includegraphics[width=0.7\linewidth]{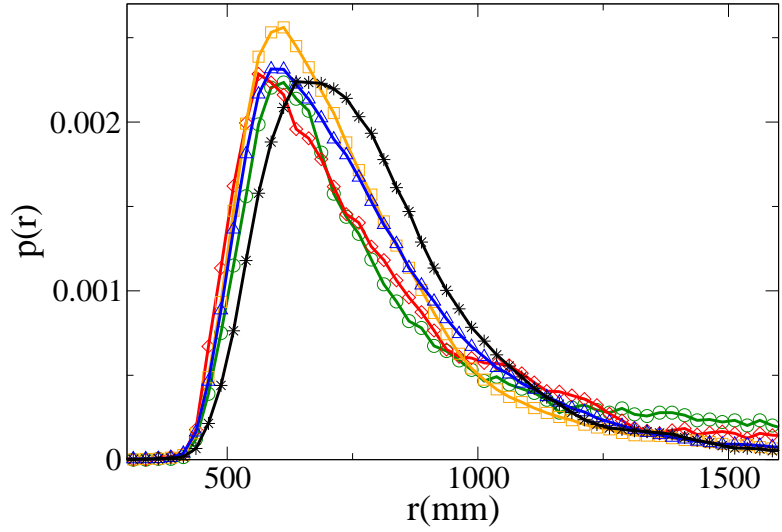} 
\caption{Probability distribution function for $r$. (Minimum) height in the 0-140 cm range: green circles; in the 140-150 range: red diamonds; in the 150-160 range: orange squares; 
in the 160-170 range: blue triangles;
in the 170-180 range: black stars.}
\label{f5e}
\end{center}
\end{figure}

\begin{figure}[ht!]
\begin{center}
\includegraphics[width=0.7\linewidth]{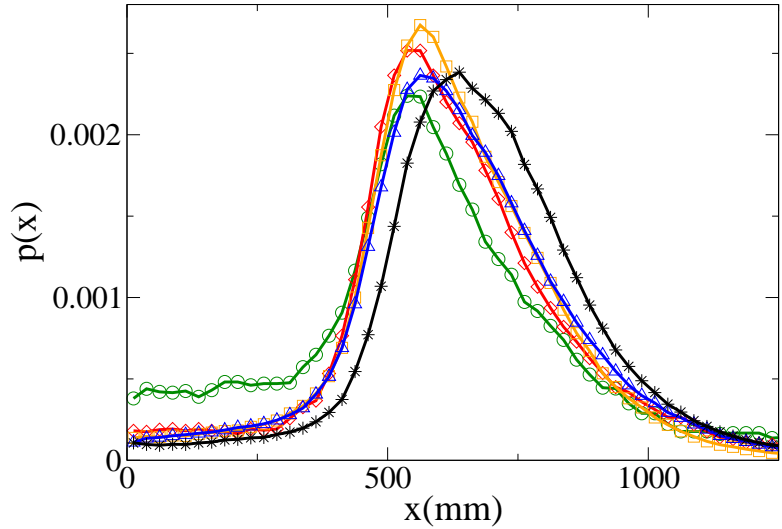} 
\caption{Probability distribution function for $x$. (Minimum) height in the 0-140 cm range: green circles; in the 140-150 range: red diamonds; in the 150-160 range: orange squares; 
in the 160-170 range: blue triangles;
in the 170-180 range: black stars.}
\label{f5f}
\end{center}
\end{figure}

\begin{figure}[ht!]
\begin{center}
\includegraphics[width=0.7\linewidth]{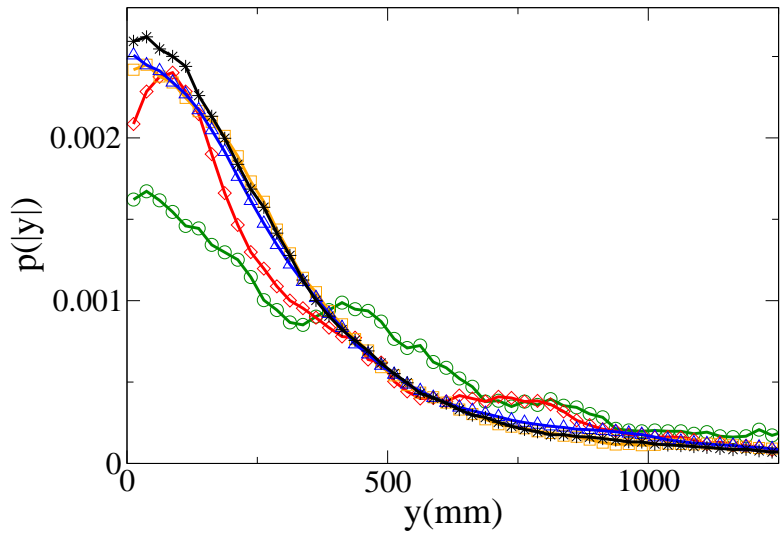} 
\caption{Probability distribution function for $|y|$. (Minimum) height in the 0-140 cm range: green circles; in the 140-150 range: red diamonds; in the 150-160 range: orange squares; 
in the 160-170 range: blue triangles;
in the 170-180 range: black stars.}
\label{f5g}
\end{center}
\end{figure}

The probability functions for different observables in different minimum height ranges are shown in figures \ref{f5h} \ref{f5e}, \ref{f5f} and \ref{f5g}, respectively for
the $V$, $r$, $x$ and $|y|$ observables, and their statistical analysis is shown in section
\ref{overh}.

We see that the abreast distance distributions
are displaced to the right with growing height, with a corresponding decrease in the values assumed around zero (particularly high in the 0-140 cm distribution,
probably due to children behaviour). Similarly, the $|y|$ distribution becomes narrower with growing height, and presents a very different behaviour in the shortest 
height slot. The  absolute distance distributions are displaced to the right with growing height, but the very fat tail for the  0-140 cm distribution causes the average value to
have a more complex dependence on height. The $V$ distribution shows a clear displacement to the right with growing height,
both in peaks and tails, although the 0-140 cm distribution has again a peculiar behaviour due to its very pronounced 
width.

\subsection{Further analysis}
In appendix \ref{furtherhei} we analyse the validity of these results on height dependence when we consider other effects such as age, relation, gender and density, and verify that,
although sometimes diminished, height related results are present also when analysing groups with fixed age, relation and gender.

\section{Discussion and conclusion}
\subsection{Summary of our findings}
By analysing how pedestrian dyad behaviour depends on the group's ``intrinsic properties'', namely the characteristics of its members and the relation between them, we observed that females dyads are 
slower and walk closer than males, that workers walk faster, at a larger distance 
and more abreast than leisure oriented people, and that inter-group relation has a strong effect on group structure, with couples walking very close and abreast, colleagues walking at a larger distance, and friends walking more abreast than family members.
Pedestrian height influences velocity and abreast distance, observables that grow with the average or minimum group height. We also found that
 elderly people walk slowly, while active age adults walk at the maximum velocity. Dyads with children have a strong
tendency to walk in a non abreast formation, with a large distance but a shorter abreast distance. 

In the supplementary materials appendices, we analysed how these features affect each
other, and we verified that the effects of the different features are present, even though sometimes diminished, even when the other features are kept fixed (e.g., when we compare
colleagues of different gender, and the like). The cross-analysis of the interplay between these features revealed also a richer structure. 
Interesting results are, for example, that the velocity of dyads with children appears to increase with
density (at least in the low-medium density range), and that children behaviour appears to be influenced by the gender of the parent.

In this work we focused on ``group features'' more than ``individual features'', i.e. we did not explicitly address questions such as the age or height difference, and similar. We may nevertheless
infer from our results some information about how group members with different height, age and gender ``compromise'' on group dynamics. We may see, for example, in appendix \ref{heicomp} that
average age gives, for the $V$ and $x$ observables that are growing function of height, a result in between those obtained for minimum and maximum height. Height is a physical and not social feature,
and it appears that, after averaging over all social features, the chosen velocity and abreast distance are the averages of those preferred by the individuals. Gender and age appear, on the other hand,
to have a deeper impact on social interactions. In mixed groups, males appear to adapt to female velocity when we average over relations, but when we analyse for secondary effects in 
appendix \ref{accounting}, we see that this is true for couples and families, but it does not apply to friends or colleagues. Similarly, while couples walk closer than male or female same sex dyads, 
mixed colleague groups walk farther than same sex dyads of both genders. In a similar way, due also to the peculiar behaviour of children, it is impossible to find a simple ``compromise'' rule for age 
related behaviour. More information could be inferred by an analysis taking in explicit account age differences, that we reserve for a future work.

The exact figures found in this work may depend strongly on the environment in which they have been recorded, and vary not only with density, but also with other macroscopic crowd dynamics features
(uni-directional flow, bi-directional flow, multi-directional flow, presence or not of standing pedestrians, etc.) as well as architectural features of the environment (open space or large corridor
or narrow corridor, etc.). For this reason, attempts to verify our findings in different environments should be directed not at specific quantitative figures (e.g. male dyads walk at 1.25 m/s and females at 1.1 meters per second)
but at qualitative patterns (e.g. males walk faster than females in a statistically significant way, with a difference in velocity comparable to the standard deviation in distributions).
It would be in particular very interesting to compare our findings with different cultural settings, since it may be expected that social group behaviour is strongly dependent
on culture, so that at least some of the patterns could change when similar data collection experiments are performed outside of (western) Japan.
\subsection{Future work}
A possible extension of this work regard the analysis of three people group behaviour. Furthermore, as stated above, in this work we limited ourselves to group properties
and not individual properties
(e.g., we verified if a group was mixed, but we did not study the specific position of the male or female). After a revision of the coding procedure, we could analyse if, according to gender, age or
height differences, roles such as ``leader'' or ``follower'' emerge. Finally, a mathematical modelling following \cite{M} and \cite{M3} could be performed.
\subsection{Possible technological impact}
Besides the obvious applications to pedestrian simulations, with possible influence in building and events planning, disaster prevention,
and even in entertainment industries such as movies and video games, we are particularly interested in applications in the field of 
robotics and more in general slow vehicles with automatic navigation capabilities deployed in pedestrian facilities, such as delivery vehicles
or automatic wheelchairs and carts. Such vehicles will arguably become more common in the future, and in order to navigate safely inside human crowds, 
and to move together with other humans ``as in a group'', they will need and understanding of pedestrian and group behaviour.

More specifically, a ``companion'' robot or an automatic wheelchair will need
\begin{enumerate}
\item to be able to recognise pedestrian groups, using an automatic recognition algorithm \cite{Z,new}
\item to be able to predict their behaviour, both in order to be able to safely avoid them and to perform a socially acceptable behaviour \cite{shiomi}
\item to be able to move together with other humans, and behave as a member of a group \cite{Luis}
\end{enumerate}
For all these applications it is extremely important to understand deeply how pedestrians actually behave and we plan to use 
these findings to improve our previous algorithms and systems as part of the development of a platform for autonomous personal mobility systems.
\section{Acknowledgements}
This research is partly supported by the Ministry of Internal Affairs and Communications (MIC), Japan,
{\it research and development project on autonomous personal mobility including robots}, 
by CREST, JST and JSPS KAKENHI Grand Number 16J40223.
\clearpage
\appendix
\section{Statistical analysis of observables}
\label{detstat}
In this work we are interested in describing how pedestrian group behaviour is influenced by some {\it intrinsic features}, such as purpose, relation, gender, age or height.
Each feature (or factor) may be divided in $k$ categories (e.g., in the case of relation $k=4$ and the categories are colleagues, couples, family and friends). Each group
is coded as belonging to a specific category, so that each category has $N_g^k$ groups. As described in section \ref{traj}, for each group $i \in N_g^k$
we can measure the value of observable $o$ every 500 ms.
We may call these measurements $o^k_{i,j}$ with $j=1,\hdots,n^k_i$ (i.e. we have $n^k_i$ measurements, or events, corresponding to group $i$ in category $k$).

We believe that the largest amount of quantitative information regarding the dependence of group behaviour on intrinsic features is included in
the overall probability distributions functions concerning all $N^k=\sum_{i \in N^k_g} n^k_i$ measurements of a given observable, as shown for example in figure \ref{f1d}, since from the 
analysis of these figures we can understand what is the probability of having a given value for each observable in each category.

It is nevertheless useful to extract some quantitative information, such as average values and standard deviations, from these distributions. Furthermore, although the purpose of this
paper is not to provide a ``$p$ value statistical independence label'' to each feature, to compare such average values it is customary and useful
 to compute, along with other statistical indicators such as effect size and determination coefficient, the standard error of each distribution
 and to perform the related analysis of variance (ANOVA). The computation of these latter statistical quantities is nevertheless
based on an assumption of statistical independence of the data, an assumption that clearly does not hold for all our $N^k$ observations\footnote{As an extreme case,
we can imagine that for a given $k$ we were following a single group ($N^k_g=1$) for one hour ($n^k_1=7200$). We will have then, if we ignore measurement noise, a perfect information
regarding the behaviour of that group in that hour and, under the strong assumption of time independence in the group behaviour, a good statistics about the behaviour {\it of that particular group}.
We still do not have any information about how group behaviour changes between groups in the category, since that information depends on the number of groups analysed, $N^k_g$. Furthermore,
since in general we track a given group only for the few seconds it needs to cross the corridor, the observations $o_{i,j}$ at fixed $i$ are also strongly time correlated.}.
\subsection{Average values, standard deviations and standard errors}
We thus proceed in the following way, justified by having a similar number of observation for each group\footnote{An average of 49 observations with a standard deviation of 22 over 1168 groups.
We nevertheless exclude from the following analysis groups that provided less than 10 observation points.}.
For each observable $o$ we compute the average over group $i$
\begin{equation}
O^k_i=\left(\sum_{j=1}^{n_i^k} o^k_{i,j}\right)/n^k_i,
\end{equation}
and then provide its average value in the category $k$ as
\begin{equation}
<O>_k \pm \varepsilon_k,
\end{equation}
where $<O>$ and the standard error $\varepsilon$ are given by
\begin{equation}
<O>_k=\left(\sum_{i=1}^{N_g^k} O^k_i\right)/N_g^k,
\end{equation}
\begin{equation}
\label{standard}
\varepsilon_k=\sigma_k/\sqrt{N_g^k},
\end{equation}
and the standard deviation is
\begin{equation}
\sigma_k=\sqrt{\left(\sum_{i=1}^{N_g^k} (O^k_i)^2\right)/N_g^k-<O>_k^2}.
\end{equation}

As a rule of thumb, we may say that $o$ assumes a different value between categories $k$ and $j$ if
\begin{equation}
\label{thumb}
|<O>_k-<O>_j| \gg 2 \max(\varepsilon_k,\varepsilon_k).
\end{equation}
\subsection{Analysis of variance}
This rule of thumb is obviously related to the ANOVA analysis reported in the text. The ANOVA analysis proceeds as follows. We define $n^c$ as the number of categories for a given feature,
\begin{equation}
N=\sum_{k=1}^{n^c} N_g^k,
\end{equation}
as the total number of groups, and the overall average of the observable as
\begin{equation}
<O>=\left(\sum_{k=1}^{n^c}<O>_k N_g^k\right)/N.
\end{equation}
We then define the distance between $<O>$ and $<O>_k$ as
\begin{equation}
d_k=<O>-<O>_k,
\end{equation} 
and the degrees of freedom
\begin{equation}
\gamma_1=n^c-1,\qquad \gamma_2=N-n^c.
\end{equation} 
The $F$ factor is then defined as
\begin{equation}
\label{effe}
F=\left(\gamma_2 \sum_{k=1}^{n^c} d_k^2 N_g^k\right)/\left(\gamma_1 \sum_{k=1}^{n^c} \sigma_k^2 N_g^k\right).
\end{equation}
This result is reported in our tables as $F_{\gamma_1,\gamma_2}$, along with the celebrated
$p$ value, that provides the probability, under the hypothesis of independence of data, that the difference between the distributions is due to chance \cite{Ash}
\begin{equation}
p=1-\int_0^F f_{\gamma_1,\gamma_1}(x) dx.
\end{equation}
The $f$ distribution has to be computed numerically \cite{recipes}, but a value $F \gg 1$ assures a small $p$ value.

Let us see how this relates to the rule of thumb for standard errors. Let us assume we have two categories with the same number of groups for category
\begin{equation}
N_g^1=N_g^2=N_g.
\end{equation}
We clearly have
\begin{equation}
<O>=\left(<O>_1+<O>_2\right)/2,
\end{equation}
\begin{equation}
|d_1|=|d_2|=|<O>_1-<O>_2|/2,
\end{equation}
and
\begin{equation}
F=(N_g-1)|<O>_1-<O>_2|^2/\left( \sigma_1^2+\sigma_2^2 \right).
\end{equation}
Using\footnote{The actual definition of the standard error uses $\sqrt{N_g-1}$ but the numbers shown in the tables use the approximate definition $\sqrt{N_g}$. For $N_g\approx 100$ or more, as it is
usually the case in this work, the difference is at most 5\%.}
\begin{equation}
\sigma_i^2/(N_g-1) \approx \varepsilon_i^2,
\end{equation}
we get the expression
\begin{equation}
F\approx |<O>_1-<O>_2|^2/\left(\varepsilon_1^2+\varepsilon_2^2\right)>|<O>_1-<O>_2|^2/\left(2 \max(\varepsilon_1,\varepsilon_2)\right)^2,
\end{equation}
so that the rule of thumb eq. \ref{thumb} corresponds to have an high $F$ value and thus a low $p$ value.
\subsection{Coefficient of determination}
Eq \ref{effe} says that the $F$ factor is high if the $\sigma_k$ are smaller than the $d_k$, i.e. if the variation inside the categories are smaller than outside the category, 
and if the total number of observation is high. Due to the large number of data points, the $F$ values in appendix \ref{overstat} (where we use all the observable measurement
instead of group averages) are always very high, and the corresponding
$p$ values very low, but the hypothesis of statistical independence of data underlying the usual interpretation of
$p$ is obviously not valid.  There are nevertheless some statistical estimators that do not depend 
dramatically on the number of observations, and that will thus have a similar value either if performed using all the data points or if performed using only group averages.

One such estimator is the coefficient of determination 
\begin{equation}
R^2=1-\left(\sum_{i,k} (o^k_i-<O>_k)^2\right)/\left(\sum_{i,k} (o^k_i-<O>)^2\right),
\end{equation}
which can also be computed as from the $F$ factor as
\begin{equation}
R^2=\left(F \gamma_1\right)/\left(F \gamma_1 +\gamma_2\right),
\end{equation}
and provides an estimate of how much of the variance in the data is ``explained'' by the category averages.
\subsection{Effect size}
The $R^2$ coefficient may attain low values if two or more category distribution functions are very similar, as it usually the case in our work. 
To point out the presence of at least one distribution that is
clearly different from the others we may use the following definition of the effect size $\delta$. We first define \cite{efsiz}
\begin{equation}
\delta_{k,l}=(<O>_k-<O>_l)/\overline{\sigma}, \qquad \overline{\sigma}=\sqrt{\left((\tilde{n}_k-1)\sigma_k^2+(\tilde{n}_l-1)\sigma_l^2\right)/\left(\tilde{n}_k+\tilde{n}_l-2\right)},
\end{equation}
where $\tilde{n}_k$, $\tilde{n}_l$ are the number of points used for computing the averages and standard deviations\footnote{I.e., $\tilde{n_k}=N_g^k$ if we are using group averages, 
$\tilde{n_k}=N^k$ if we are using overall distributions.}, and then we consider the maximum pairwise effect size
\begin{equation}
\delta=\max_{k,l}|\delta_{k,l}|.
\end{equation}
While a $p$ value tells us about the significance of the statistical difference between two distributions, the difference may be often so small that if can be verified only if a large amount of
data are collected. But if we have also $\delta\approx 1$, then the two distributions are different enough to be distinguished also using a relatively reduced amount of data.
\subsection{Multi-factor cross analysis}
We refrain from applying the machinery of two way or $n$ way ANOVA to our data, since our ecological data set is extremely unbalanced, and it is unbalanced for the very reason
that our ``factors'' are not independent variables\footnote{For example, since the average height of females is two standard deviations lower than the male one \cite{heightdis}, the high range height groups will
be entirely composed of males, not to mention more extreme cases, such as the conditional probability of having a children in a group of colleagues, which is arguably zero.}. 

It is nevertheless useful to analyse the interplay between the different features, and we do that in section \ref{accounting} by performing a statistical analysis similar to the one described
above of a given feature $A$ while keeping fixed the value of another feature $B$ to a category $\overline{k}$.\footnote{For the fixed category feature $B$, we use also the external feature of
pedestrian crowd density. Since the same group may contribute to different densities, when operating at a fixed density we use for group averages all groups that contribute with at least
5 data points (instead of the usual 10) to the observable distribution for that density value.} Sometimes this analysis is performed on a reduced number of groups, and thus the corresponding 
$p$ value may be high. This does not imply that the analysis is valueless, at least in our opinion, since it
provides new information. The $F$ and $p$ values are, in this situation, useful to compare different observables on the given condition. 
As an example, table \ref{table2c1} tells us that $x$ has a stronger variation between relation categories for fixed gender than $r$, and so on.
Furthermore, in these situations, an analysis of statistical indicators that do not depend critically on the number of observations, such as the effect size, is particularly valuable.
\clearpage
\section{Statistical analysis of overall probability distributions}
\label{overstat}
\subsection{Purpose}
\label{overpur}
Table \ref{overt1} provides a statistical analysis of the overall probability distributions for the purpose categories.
\begin{table}[!ht]
\scriptsize
\caption{Statistical analysis of the overall probability distributions for the purpose categories. Lengths in millimetres, times in seconds.}
\label{overt1}
\begin{center}
\begin{tabular} {|c|c|c|c|c|c|}
\hline
Purpose &  $N^k$ &  $V$ &    $r$ & $x$  & $|y|$ \\
\hline
Leisure & 38501 & 1096 $\pm$ 1.3   ($\sigma$=251) &799 $\pm$ 1.6  ($\sigma$=309) &630 $\pm$ 1.2  ($\sigma$=236) &360 $\pm$ 2 ($\sigma$=388)    \\ 
\hline
Work & 18936 & 1257 $\pm$ 1.7   ($\sigma$=235) &834 $\pm$ 2.1  ($\sigma$=287) &714 $\pm$ 1.6  ($\sigma$=227) &315 $\pm$ 2.5 ($\sigma$=341)    \\ 
\hline
$F_{1,57435}$ & & 5400 & 169 & 1640 & 184\\
\hline
$p$ & & $<10^{-8}$ & $<10^{-8}$ & $<10^{-8}$ & $<10^{-8}$\\
\hline
$R^2$ & & 0.0859 & 0.00293 & 0.0278 & 0.00319\\
\hline
$\delta$ & & 0.652 & 0.115 & 0.36 & 0.12\\
\hline
\end{tabular}
\end{center}
\end{table}
\subsection{Relation}
\label{overrel}
Table \ref{overt2} provides a statistical analysis of the overall probability distributions for the relation categories.
\begin{table}[!ht]
\scriptsize
\caption{Statistical analysis of the overall probability distributions for the relation categories. Lengths in millimetres, times in seconds.}
\label{overt2}
\begin{center}
\begin{tabular} {|c|c|c|c|c|c|}
\hline
Relation&  $N^k$ &  $V$ &    $r$ & $x$  & $|y|$ \\
\hline
Colleagues & 18172 & 1262 $\pm$ 1.7   ($\sigma$=234) &840 $\pm$ 2.2  ($\sigma$=290) &720 $\pm$ 1.7  ($\sigma$=229) &317 $\pm$ 2.6 ($\sigma$=344)    \\ 
\hline
Couples & 5273 & 1085 $\pm$ 3.2   ($\sigma$=231) &699 $\pm$ 3.7  ($\sigma$=271) &584 $\pm$ 2.6  ($\sigma$=188) &290 $\pm$ 4.4 ($\sigma$=318)    \\ 
\hline
Families & 12596 & 1072 $\pm$ 2.2   ($\sigma$=246) &834 $\pm$ 3.2  ($\sigma$=357) &592 $\pm$ 2.3  ($\sigma$=260) &452 $\pm$ 4 ($\sigma$=447)    \\ 
\hline
Friends & 17634 & 1113 $\pm$ 2   ($\sigma$=260) &788 $\pm$ 2  ($\sigma$=265) &659 $\pm$ 1.6  ($\sigma$=214) &312 $\pm$ 2.5 ($\sigma$=338)    \\ 
\hline
$F_{3,53671}$ & & 1940 & 362 & 975 & 485\\
\hline
$p$ & & $<10^{-8}$ & $<10^{-8}$ & $<10^{-8}$ & $<10^{-8}$\\
\hline
$R^2$ & & 0.0978 & 0.0198 & 0.0517 & 0.0264\\
\hline
$\delta$ & & 0.795 & 0.493 & 0.614 & 0.392\\
\hline
\end{tabular}
\end{center}
\end{table}
\subsection{Gender}
\label{overgen}
Table \ref{overt3} provides a statistical analysis of the overall probability distributions for the relation categories.
\begin{table}[!ht]
\scriptsize
\caption{Statistical analysis of the overall probability distributions for the gender categories. Lengths in millimetres, times in seconds.}
\label{overt3}
\begin{center}
\begin{tabular} {|c|c|c|c|c|c|}
\hline
Gender&  $N^k$ &  $V$ &    $r$ & $x$  & $|y|$ \\
\hline
Two females & 14688 & 1075 $\pm$ 2.1   ($\sigma$=251) &773 $\pm$ 2.2  ($\sigma$=268) &647 $\pm$ 1.7  ($\sigma$=202) &302 $\pm$ 2.9 ($\sigma$=346)    \\ 
\hline
Mixed & 19311 & 1098 $\pm$ 1.7   ($\sigma$=239) &803 $\pm$ 2.4  ($\sigma$=334) &614 $\pm$ 1.8  ($\sigma$=248) &388 $\pm$ 3 ($\sigma$=411)    \\ 
\hline
Two males & 23516 & 1237 $\pm$ 1.6   ($\sigma$=249) &839 $\pm$ 1.9  ($\sigma$=292) &702 $\pm$ 1.6  ($\sigma$=239) &337 $\pm$ 2.3 ($\sigma$=355)    \\ 
\hline
$F_{2,57512}$ & & 2570 & 225 & 791 & 232\\
\hline
$p$ & & $<10^{-8}$ & $<10^{-8}$ & $<10^{-8}$ & $<10^{-8}$\\
\hline
$R^2$ & & 0.0822 & 0.00778 & 0.0268 & 0.008\\
\hline
$\delta$ & & 0.647 & 0.233 & 0.365 & 0.224\\
\hline
\end{tabular}
\end{center}
\end{table}
\subsection{Age}
\label{overage}
Table \ref{overt4} provides a statistical analysis of the overall probability distributions for the minimum age ranges.
\begin{table}[!ht]
\scriptsize
\caption{Statistical analysis of the overall probability distributions for the minimum
  age ranges. Lengths in millimetres, times in seconds.}
\label{overt4}
\begin{center}
\begin{tabular} {|c|c|c|c|c|c|}
\hline
Minimum age&  $N^k$ &  $V$ &    $r$ & $x$  & $|y|$ \\
\hline
0-9 years & 1041 & 1127 $\pm$ 8.4   ($\sigma$=272) &983 $\pm$ 15  ($\sigma$=480) &573 $\pm$ 9.5  ($\sigma$=306) &663 $\pm$ 18 ($\sigma$=580)    \\ 
\hline
10-19 years & 3443 & 1110 $\pm$ 5.2   ($\sigma$=303) &767 $\pm$ 5.1  ($\sigma$=298) &626 $\pm$ 3.8  ($\sigma$=222) &322 $\pm$ 6.2 ($\sigma$=364)    \\ 
\hline
20-29 years & 18679 & 1167 $\pm$ 1.8   ($\sigma$=240) &788 $\pm$ 2.1  ($\sigma$=289) &665 $\pm$ 1.6  ($\sigma$=223) &301 $\pm$ 2.6 ($\sigma$=349)    \\ 
\hline
30-39 years & 15552 & 1179 $\pm$ 2.1   ($\sigma$=264) &816 $\pm$ 2.4  ($\sigma$=294) &667 $\pm$ 2  ($\sigma$=248) &343 $\pm$ 2.9 ($\sigma$=357)    \\ 
\hline
40-49 years & 7974 & 1167 $\pm$ 2.7   ($\sigma$=242) &838 $\pm$ 3.3  ($\sigma$=296) &668 $\pm$ 2.7  ($\sigma$=243) &374 $\pm$ 4.2 ($\sigma$=378)    \\ 
\hline
50-59 years & 6025 & 1153 $\pm$ 3.3   ($\sigma$=253) &812 $\pm$ 3.7  ($\sigma$=284) &653 $\pm$ 2.9  ($\sigma$=223) &358 $\pm$ 4.7 ($\sigma$=367)    \\ 
\hline
60-69 years & 3969 & 1001 $\pm$ 3.5   ($\sigma$=219) &836 $\pm$ 5.4  ($\sigma$=340) &643 $\pm$ 3.8  ($\sigma$=242) &409 $\pm$ 6.7 ($\sigma$=419)    \\ 
\hline
$\geq$ 70 years & 832 & 877 $\pm$ 6   ($\sigma$=172) &793 $\pm$ 13  ($\sigma$=363) &599 $\pm$ 7.8  ($\sigma$=224) &383 $\pm$ 16 ($\sigma$=453)    \\ 
\hline
$F_{7,57507}$ & & 400 & 89.1 & 46.7 & 175\\
\hline
$p$ & & $<10^{-8}$ & $<10^{-8}$ & $<10^{-8}$ & $<10^{-8}$\\
\hline
$R^2$ & & 0.0464 & 0.0107 & 0.00566 & 0.0208\\
\hline
$\delta$ & & 1.16 & 0.619 & 0.382 & 0.991\\
\hline
\end{tabular}
\end{center}
\end{table}
\subsection{Height}
\label{overh}
Table \ref{overt5} provides a statistical analysis of the overall probability distributions for the minimum height ranges.
\begin{table}[!ht]
\scriptsize
\caption{Statistical analysis of the overall probability distributions for the minimum
  height ranges. Lengths in millimetres, times in seconds.}
\label{overt5}
\begin{center}
\begin{tabular} {|c|c|c|c|c|c|}
\hline
Minimum height&  $N^k$ &  $V$ &    $r$ & $x$  & $|y|$ \\
\hline
$<$ 140 cm & 1579 & 1127 $\pm$ 6.9   ($\sigma$=274) &942 $\pm$ 11  ($\sigma$=457) &605 $\pm$ 7.6  ($\sigma$=300) &578 $\pm$ 14 ($\sigma$=553)    \\ 
\hline
140-150 cm & 2206 & 1032 $\pm$ 6.7   ($\sigma$=315) &855 $\pm$ 8  ($\sigma$=374) &644 $\pm$ 5.3  ($\sigma$=248) &420 $\pm$ 10 ($\sigma$=468)    \\ 
\hline
150-160 cm & 13064 & 1076 $\pm$ 2.2   ($\sigma$=251) &779 $\pm$ 2.5  ($\sigma$=281) &628 $\pm$ 1.8  ($\sigma$=209) &337 $\pm$ 3.2 ($\sigma$=365)    \\ 
\hline
160-170 cm & 26345 & 1151 $\pm$ 1.5   ($\sigma$=245) &810 $\pm$ 1.9  ($\sigma$=306) &655 $\pm$ 1.5  ($\sigma$=243) &348 $\pm$ 2.3 ($\sigma$=374)    \\ 
\hline
170-180 cm & 13497 & 1234 $\pm$ 2.1   ($\sigma$=243) &819 $\pm$ 2.3  ($\sigma$=269) &700 $\pm$ 2  ($\sigma$=232) &309 $\pm$ 2.8 ($\sigma$=323)    \\ 
\hline
$>$ 180 cm & 824 & 1224 $\pm$ 9.3   ($\sigma$=268) &823 $\pm$ 11  ($\sigma$=325) &686 $\pm$ 8.1  ($\sigma$=234) &309 $\pm$ 14 ($\sigma$=404)    \\ 
\hline
$F_{5,57509}$ & & 648 & 102 & 149 & 171\\
\hline
$p$ & & $<10^{-8}$ & $<10^{-8}$ & $<10^{-8}$ & $<10^{-8}$\\
\hline
$R^2$ & & 0.0533 & 0.00875 & 0.0128 & 0.0146\\
\hline
$\delta$ & & 0.796 & 0.534 & 0.398 & 0.532\\
\hline
\end{tabular}
\end{center}
\end{table}
\clearpage
\section{Accounting for other effects}
\label{accounting}
\subsection{Secondary effects and purpose}
\label{furtherpur}
\subsubsection{Density}
Work-oriented dyads are more frequently found during working days, in which the environment presents a lower density (and thus higher velocity and inter-group pedestrian
distance, \cite{M2}). It is thus important to analyse the results of section \ref{purpose} when they are divided for density ranges, for example by comparing results in the 
$0 \leq \rho <0.05$ pedestrian per square meter range
with those in the $0.15 \leq \rho <0.2$ range\footnote{Groups may contribute to different density ranges, see \cite{M2} for details.}. The results are reported in table \ref{table1b1}
and \ref{table1b2}, showing
that the differences in $V$ and $x$ remain significant regardless of density. The difference in $|y|$ becomes significant at high density, while at very low density is not significant
(while the opposite happens to $r$).

\begin{table}[!ht]
\scriptsize
\caption{Observable dependence on purpose for dyads in the $0 \leq \rho \leq 0.05$ pedestrian per square meter density range. Lengths in millimetres, times in seconds.}
\label{table1b1}
\begin{center}
\begin{tabular} {|c|c|c|c|c|c|}
\hline
Purpose &  $N^k_g$ &  $V$ &    $r$ & $x$  & $|y|$ \\
\hline
Leisure & 426 & 1113 $\pm$ 10   ($\sigma$=215) &851 $\pm$ 15  ($\sigma$=303) &658 $\pm$ 9 ($\sigma$=186) &400 $\pm$ 18 ($\sigma$=373)    \\ 
\hline
Work & 209 & 1274 $\pm$ 12   ($\sigma$=169) &924 $\pm$ 20  ($\sigma$=296) &741 $\pm$ 14 ($\sigma$=203) &409 $\pm$ 26 ($\sigma$=373)    \\ 
\hline
$F_{1,633}$ & & 89.6 & 8.17 & 25.7 & 0.0807\\
\hline
$p$ & & $<10^{-8}$ & 0.0044 & 5.36$\cdot 10^{-7}$ & 0.776\\
\hline
$R^2$ & & 0.124 & 0.0127 & 0.039 & 0.000128\\
\hline
$\delta$ & & 0.8 & 0.242 & 0.428 & 0.024\\
\hline
\end{tabular}
\end{center}
\end{table}

\begin{table}[!ht]
\scriptsize
\caption{Observable dependence on purpose for dyads in the $0.15 \leq \rho \leq 0.2$ pedestrian per square meter density range. Lengths in millimetres, times in seconds.}
\label{table1b2}
\begin{center}
\begin{tabular} {|c|c|c|c|c|c|}
\hline
Purpose &  $N^k_g$ &  $V$ &    $r$ & $x$  & $|y|$ \\
\hline
Leisure & 145 & 1084 $\pm$ 14   ($\sigma$=170) &764 $\pm$ 17  ($\sigma$=209) &560 $\pm$ 13 ($\sigma$=158) &390 $\pm$ 26 ($\sigma$=308)    \\ 
\hline
Work & 22 & 1229 $\pm$ 27   ($\sigma$=125) &754 $\pm$ 26  ($\sigma$=123) &673 $\pm$ 25 ($\sigma$=117) &237 $\pm$ 40 ($\sigma$=186)    \\ 
\hline
$F_{1,165}$ & & 14.7 & 0.0513 & 10.2 & 5.05\\
\hline
$p$ & & 0.000182 & 0.821 & 0.00167 & 0.026\\
\hline
$R^2$ & & 0.0817 & 0.000311 & 0.0583 & 0.0297\\
\hline
$\delta$ & & 0.881 & 0.0521 & 0.735 & 0.516\\
\hline
\end{tabular}
\end{center}
\end{table}
In table \ref{table1b3} and \ref{table1b4} we report, respectively, $p$ and $\delta$ values for purpose corresponding to each observable and density range, 
showing that the $V$, $x$ and $|y|$ distributions are different  
in a statistically significant way at different density ranges, although the effect on $|y|$ grows with density. Differences in $r$ are significant only at the lowest density range.
\begin{table}[!ht]
\scriptsize
\caption{$p$ values for purpose corresponding to velocity and distance observables at different density ranges.}
\label{table1b3}
\begin{center}
\begin{tabular} {|c|c|c|c|c|}
\hline
Density&  $V$ &    $r$ & $x$  & $|y|$ \\
\hline
0-0.05 ped/m$^2$&  $<10^{-8}$ & 0.0044 & 5.36$\cdot 10^{-7}$ & 0.776\\
\hline
0.05-0.1 ped/m$^2$&  $<10^{-8}$ & 0.682 & $<10^{-8}$ & 0.000517\\
\hline
0.1-0.15 ped/m$^2$&  $<10^{-8}$ & 0.221 & $<10^{-8}$ & 1.32$\cdot 10^{-6}$\\
\hline
0.15-0.2 ped/m$^2$&  0.000182 & 0.821 & 0.00167 & 0.026\\
\hline
\end{tabular}
\end{center}
\end{table}
\begin{table}[!ht]
\scriptsize
\caption{$\delta$ values for purpose corresponding to velocity and distance observables at different density ranges.}
\label{table1b4}
\begin{center}
\begin{tabular} {|c|c|c|c|c|}
\hline
Density&   $V$ &    $r$ & $x$  & $|y|$ \\
\hline
0-0.05 ped/m$^2$ &  0.8 & 0.242 & 0.428 & 0.024\\
\hline
0.05-0.1 ped/m$^2$ &  0.914 & 0.0292 & 0.515 & 0.248\\
\hline
0.1-0.15 ped/m$^2$ &  0.812 & 0.117 & 0.627 & 0.467\\
\hline
0.15-0.2 ped/m$^2$ &  0.881 & 0.0521 & 0.735 & 0.516\\
\hline
\end{tabular}
\end{center}
\end{table}
\subsubsection{Gender}
The work and leisure populations are strongly biased regarding gender. In tables  \ref{table1c1}, \ref{table1c2} and  \ref{table1c3}
we show the results for the work and leisure observables when
limited to, respectively, female, mixed and male dyads.  While velocity is still significantly different also when gender is fixed, absolute distance in men, and all distance observables
in females are not significantly different. We may thus conclude that differences between workers and leisure oriented people are present
regardless of gender, but are magnified by the gender difference in the two populations.

\begin{table}[!ht]
\scriptsize
\caption{Observable dependence on purpose for 2 female dyads. Lengths in millimetres, times in seconds.}
\label{table1c1}
\begin{center}
\begin{tabular} {|c|c|c|c|c|c|}
\hline
Purpose &  $N^k_g$ &  $V$ &    $r$ & $x$  & $|y|$ \\
\hline
Leisure & 222 & 1092 $\pm$ 13   ($\sigma$=194) &794 $\pm$ 16  ($\sigma$=235) &644 $\pm$ 8.4 ($\sigma$=125) &328 $\pm$ 22 ($\sigma$=322)    \\ 
\hline
Work & 29 & 1184 $\pm$ 30   ($\sigma$=162) &755 $\pm$ 28  ($\sigma$=150) &663 $\pm$ 19 ($\sigma$=101) &274 $\pm$ 38 ($\sigma$=204)    \\ 
\hline
$F_{1,249}$ & & 5.97 & 0.733 & 0.615 & 0.777\\
\hline
$p$ & & 0.0153 & 0.393 & 0.433 & 0.379\\
\hline
$R^2$ & & 0.0234 & 0.00293 & 0.00247 & 0.00311\\
\hline
$\delta$ & & 0.484 & 0.17 & 0.155 & 0.175\\
\hline
\end{tabular}
\end{center}
\end{table}

\begin{table}[!ht]
\scriptsize
\caption{Observable dependence on purpose for mixed gender dyads. Lengths in millimetres, times in seconds.}
\label{table1c2}
\begin{center}
\begin{tabular} {|c|c|c|c|c|c|}
\hline
Purpose &  $N^k_g$ &  $V$ &    $r$ & $x$  & $|y|$ \\
\hline
Leisure & 330 & 1097 $\pm$ 9.9   ($\sigma$=180) &814 $\pm$ 15  ($\sigma$=267) &602 $\pm$ 9.6 ($\sigma$=174) &415 $\pm$ 19 ($\sigma$=341)    \\ 
\hline
Work & 41 & 1226 $\pm$ 26   ($\sigma$=167) &902 $\pm$ 48  ($\sigma$=308) &698 $\pm$ 24 ($\sigma$=152) &420 $\pm$ 65 ($\sigma$=419)    \\ 
\hline
$F_{1,369}$ & & 18.9 & 3.79 & 11.3 & 0.00662\\
\hline
$p$ & & 1.77$\cdot 10^{-5}$ & 0.0524 & 0.000849 & 0.935\\
\hline
$R^2$ & & 0.0488 & 0.0102 & 0.0298 & 1.79$\cdot 10^{-5}$\\
\hline
$\delta$ & & 0.722 & 0.323 & 0.558 & 0.0135\\
\hline
\end{tabular}
\end{center}
\end{table}

\begin{table}[!ht]
\scriptsize
\caption{Observable dependence on purpose for 2 male dyads. Lengths in millimetres, times in seconds.}
\label{table1c3}
\begin{center}
\begin{tabular} {|c|c|c|c|c|c|}
\hline
Purpose &  $N^k_g$ &  $V$ &    $r$ & $x$  & $|y|$ \\
\hline
Leisure & 164 & 1196 $\pm$ 16   ($\sigma$=207) &846 $\pm$ 19  ($\sigma$=246) &660 $\pm$ 14 ($\sigma$=173) &392 $\pm$ 25 ($\sigma$=325)    \\ 
\hline
Work & 302 & 1285 $\pm$ 8.7   ($\sigma$=152) &846 $\pm$ 13  ($\sigma$=218) &720 $\pm$ 9 ($\sigma$=157) &325 $\pm$ 16 ($\sigma$=271)    \\ 
\hline
$F_{1,464}$ & & 28.1 & 0.000251 & 14.4 & 5.5\\
\hline
$p$ & & 1.83$\cdot 10^{-7}$ & 0.987 & 0.000165 & 0.0195\\
\hline
$R^2$ & & 0.057 & 5.42$\cdot 10^{-7}$ & 0.0301 & 0.0117\\
\hline
$\delta$ & & 0.515 & 0.00154 & 0.369 & 0.228\\
\hline
\end{tabular}
\end{center}
\end{table}
\subsubsection{Age}
In tables  (\ref{table1d1}) and (\ref{table1d2}) we show the results for the work and leisure observables when
limited to groups of a given average age. The results suggest that differences may be present at any age (in particular concerning $V$), but are definitely more
strong for more mature walkers.

\begin{table}[!ht]
\scriptsize
\caption{Observable dependence on purpose for dyads with average age in the 20-29 years range. Lengths in millimetres, times in seconds.}
\label{table1d1}
\begin{center}
\begin{tabular} {|c|c|c|c|c|c|}
\hline
Purpose &  $N^k_g$ &  $V$ &    $r$ & $x$  & $|y|$ \\
\hline
Leisure & 292 & 1164 $\pm$ 10   ($\sigma$=177) &798 $\pm$ 14  ($\sigma$=242) &656 $\pm$ 9.7 ($\sigma$=165) &326 $\pm$ 17 ($\sigma$=290)    \\ 
\hline
Work & 78 & 1242 $\pm$ 19   ($\sigma$=166) &775 $\pm$ 17  ($\sigma$=152) &684 $\pm$ 12 ($\sigma$=108) &266 $\pm$ 22 ($\sigma$=197)    \\ 
\hline
$F_{1,368}$ & & 12.2 & 0.608 & 1.91 & 2.93\\
\hline
$p$ & & 0.000536 & 0.436 & 0.168 & 0.088\\
\hline
$R^2$ & & 0.0321 & 0.00165 & 0.00515 & 0.00789\\
\hline
$\delta$ & & 0.446 & 0.0996 & 0.176 & 0.219\\
\hline
\end{tabular}
\end{center}
\end{table}
\begin{table}[!ht]
\scriptsize
\caption{Observable dependence on purpose for dyads with average age in the 50-59 years range. Lengths in millimetres, times in seconds.}
\label{table1d2}
\begin{center}
\begin{tabular} {|c|c|c|c|c|c|}
\hline
Purpose &  $N^k_g$ &  $V$ &    $r$ & $x$  & $|y|$ \\
\hline
Leisure & 61 & 1053 $\pm$ 21   ($\sigma$=164) &808 $\pm$ 32  ($\sigma$=247) &601 $\pm$ 20 ($\sigma$=155) &404 $\pm$ 46 ($\sigma$=356)    \\ 
\hline
Work & 53 & 1276 $\pm$ 21   ($\sigma$=153) &845 $\pm$ 24  ($\sigma$=173) &706 $\pm$ 20 ($\sigma$=144) &345 $\pm$ 36 ($\sigma$=261)    \\ 
\hline
$F_{1,112}$ & & 54.8 & 0.808 & 13.7 & 0.966\\
\hline
$p$ & & $<10^{-8}$ & 0.371 & 0.000328 & 0.328\\
\hline
$R^2$ & & 0.329 & 0.00716 & 0.109 & 0.00855\\
\hline
$\delta$ & & 1.4 & 0.17 & 0.702 & 0.186\\
\hline
\end{tabular}
\end{center}
\end{table}
In table \ref{table1d3} and \ref{table1d4} we report, respectively, $p$ and $\delta$ values for purpose corresponding to each observable and average age range, showing
again that differences have a tendency to grow with age.
\begin{table}[!ht]
\scriptsize
\caption{$p$ values for purpose corresponding to velocity and distance observables at different average age ranges.}
\label{table1d3}
\begin{center}
\begin{tabular} {|c|c|c|c|c|}
\hline
Average age&   $V$ &    $r$ & $x$  & $|y|$ \\
\hline
20-29 years&  0.000536 & 0.436 & 0.168 & 0.088\\
\hline
30-39 years&  $<10^{-8}$ & 0.0689 & 6.34$\cdot 10^{-8}$ & 0.144\\
\hline
40-49 years&  $<10^{-8}$ & 0.12 & 4.78$\cdot 10^{-6}$ & 0.264\\
\hline
50-59 years&  $<10^{-8}$ & 0.371 & 0.000328 & 0.328\\
\hline
60-69 years&  0.0233 & 0.221 & 0.48 & 0.463\\
\hline
\end{tabular}
\end{center}
\end{table}
\begin{table}[!ht]
\scriptsize
\caption{$\delta$ values for purpose corresponding to velocity and distance observables at different average age ranges.}
\label{table1d4}
\begin{center}
\begin{tabular} {|c|c|c|c|c|}
\hline
Average age&   $V$ &    $r$ & $x$  & $|y|$ \\
\hline
20-29 years &  0.446 & 0.0996 & 0.176 & 0.219\\
\hline
30-39 years &  0.994 & 0.224 & 0.682 & 0.18\\
\hline
40-49 years &  0.97 & 0.226 & 0.68 & 0.162\\
\hline
50-59 years &  1.4 & 0.17 & 0.702 & 0.186\\
\hline
60-69 years &  1.21 & 0.649 & 0.373 & 0.389\\
\hline
\end{tabular}
\end{center}
\end{table}
\subsubsection{Height}
In tables  (\ref{table1e1}) and (\ref{table1e2}) we show the results for the work and leisure observables when
limited to groups of a given average height, and in tables \ref{table1e3} and \ref{table1e4} we report, respectively, $p$ and $\delta$ 
values for purpose corresponding to each observable and average height range.
Differences appear to grow with height, probably affected also by the gender distributions.

\begin{table}[!ht]
\scriptsize
\caption{Observable dependence on purpose for dyads with average height in the 150-160 cm range. Lengths in millimetres, times in seconds.}
\label{table1e1}
\begin{center}
\begin{tabular} {|c|c|c|c|c|c|}
\hline
Purpose &  $N^k_g$ &  $V$ &    $r$ & $x$  & $|y|$ \\
\hline
Leisure & 108 & 1107 $\pm$ 25   ($\sigma$=260) &821 $\pm$ 26  ($\sigma$=268) &629 $\pm$ 15 ($\sigma$=153) &389 $\pm$ 34 ($\sigma$=352)    \\ 
\hline
Work & 10 & 1152 $\pm$ 51   ($\sigma$=160) &709 $\pm$ 27  ($\sigma$=86) &631 $\pm$ 28 ($\sigma$=88.5) &264 $\pm$ 32 ($\sigma$=101)    \\ 
\hline
$F_{1,116}$ & & 0.283 & 1.71 & 0.00249 & 1.24\\
\hline
$p$ & & 0.596 & 0.194 & 0.96 & 0.268\\
\hline
$R^2$ & & 0.00243 & 0.0145 & 2.15$\cdot 10^{-5}$ & 0.0106\\
\hline
$\delta$ & & 0.177 & 0.434 & 0.0166 & 0.37\\
\hline
\end{tabular}
\end{center}
\end{table}
\begin{table}[!ht]
\scriptsize
\caption{Observable dependence on purpose for dyads with average age in the 170-180 cm range. Lengths in millimetres, times in seconds.}
\label{table1e2}
\begin{center}
\begin{tabular} {|c|c|c|c|c|c|}
\hline
Purpose &  $N^k_g$ &  $V$ &    $r$ & $x$  & $|y|$ \\
\hline
Leisure & 188 & 1138 $\pm$ 12   ($\sigma$=168) &801 $\pm$ 15  ($\sigma$=212) &628 $\pm$ 12 ($\sigma$=159) &366 $\pm$ 21 ($\sigma$=291)    \\ 
\hline
Work & 233 & 1291 $\pm$ 10   ($\sigma$=157) &850 $\pm$ 15  ($\sigma$=230) &730 $\pm$ 10 ($\sigma$=153) &322 $\pm$ 18 ($\sigma$=273)    \\ 
\hline
$F_{1,419}$ & & 92.2 & 5.22 & 44 & 2.53\\
\hline
$p$ & & $<10^{-8}$ & 0.0228 & $<10^{-8}$ & 0.113\\
\hline
$R^2$ & & 0.18 & 0.0123 & 0.0951 & 0.006\\
\hline
$\delta$ & & 0.944 & 0.225 & 0.652 & 0.156\\
\hline
\end{tabular}
\end{center}
\end{table}
\begin{table}[!ht]
\scriptsize
\caption{$p$ values for purpose corresponding to velocity and distance observables at different average age ranges.}
\label{table1e3}
\begin{center}
\begin{tabular} {|c|c|c|c|c|}
\hline
Average height &  $V$ &    $r$ & $x$  & $|y|$ \\
\hline
150-160 cm&  0.596 & 0.194 & 0.96 & 0.268\\
\hline
160-170 cm&  $<10^{-8}$ & 0.0557 & 0.00126 & 0.953\\
\hline
170-180 cm&  $<10^{-8}$ & 0.0228 & $<10^{-8}$ & 0.113\\
\hline
$>$ 180 cm&  0.773 & 0.959 & 0.289 & 0.522\\
\hline
\end{tabular}
\end{center}
\end{table}
\begin{table}[!ht]
\scriptsize
\caption{$\delta$ values for purpose corresponding to velocity and distance observables at different average age ranges.}
\label{table1e4}
\begin{center}
\begin{tabular} {|c|c|c|c|c|}
\hline
Average height &  $V$ &    $r$ & $x$  & $|y|$ \\
\hline
150-160 cm &  0.177 & 0.434 & 0.0166 & 0.37\\
\hline
160-170 cm &  0.696 & 0.217 & 0.368 & 0.00667\\
\hline
170-180 cm &  0.944 & 0.225 & 0.652 & 0.156\\
\hline
$>$ 180 cm &  0.0998 & 0.0177 & 0.368 & 0.22\\
\hline
\end{tabular}
\end{center}
\end{table}

\clearpage
\subsection{Secondary effects and relation}
\label{furtherrel}
\subsubsection{Density}
As discussed above, work-oriented (and thus colleagues) dyads are more present during working days, in which the environment presents a lower density.
Tables  \ref{table2b1} and  \ref{table2b2} show the observables dependence for fixed density ranges ($0 \leq \rho <0.05$ ped/m$^2$ and $0.15 \leq \rho <0.2$ ped/m$^2$, respectively).
The major trends exposed in the main text are present at any density, as confirmed also by tables \ref{table2b3} and  \ref{table2b4}, reporting
$p$ and $\delta$ values, respectively, for all density ranges.

\begin{table}[!ht]
\scriptsize
\caption{Observable dependence on relation for dyads in the $0 \leq \rho <0.05$ ped/m$^2$ range. Lengths in millimetres, times in seconds.}
\label{table2b1}
\begin{center}
\begin{tabular} {|c|c|c|c|c|c|}
\hline
Relation&  $N^k_g$ &  $V$ &    $r$ & $x$  & $|y|$ \\
\hline
Colleagues & 202 & 1276 $\pm$ 12   ($\sigma$=169) &934 $\pm$ 21  ($\sigma$=298) &751 $\pm$ 15 ($\sigma$=207) &409 $\pm$ 26 ($\sigma$=376)    \\ 
\hline
Couples & 62 & 1103 $\pm$ 25   ($\sigma$=193) &760 $\pm$ 38  ($\sigma$=297) &600 $\pm$ 22 ($\sigma$=177) &359 $\pm$ 40 ($\sigma$=314)    \\ 
\hline
Families & 125 & 1084 $\pm$ 19   ($\sigma$=208) &894 $\pm$ 30  ($\sigma$=331) &617 $\pm$ 16 ($\sigma$=175) &512 $\pm$ 37 ($\sigma$=413)    \\ 
\hline
Friends & 193 & 1130 $\pm$ 17   ($\sigma$=230) &830 $\pm$ 19  ($\sigma$=258) &685 $\pm$ 12 ($\sigma$=162) &338 $\pm$ 23 ($\sigma$=326)    \\ 
\hline
$F_{3,578}$ & & 30.5 & 7.59 & 19 & 6.17\\
\hline
$p$ & & $<10^{-8}$ & 5.52$\cdot 10^{-5}$ & $<10^{-8}$ & 0.000396\\
\hline
$R^2$ & & 0.137 & 0.0379 & 0.0896 & 0.031\\
\hline
$\delta$ & & 1.03 & 0.585 & 0.754 & 0.482\\
\hline
\end{tabular}
\end{center}
\end{table}

\begin{table}[!ht]
\scriptsize
\caption{Observable dependence on relation for dyads in the $0.15 \leq \rho <0.2$ ped/m$^2$ range. Lengths in millimetres, times in seconds.}
\label{table2b2}
\begin{center}
\begin{tabular} {|c|c|c|c|c|c|}
\hline
Relation&  $N^k_g$ &  $V$ &    $r$ & $x$  & $|y|$ \\
\hline
Colleagues & 22 & 1229 $\pm$ 27   ($\sigma$=125) &754 $\pm$ 26  ($\sigma$=123) &673 $\pm$ 25 ($\sigma$=117) &237 $\pm$ 40 ($\sigma$=186)    \\ 
\hline
Couples & 19 & 1064 $\pm$ 28   ($\sigma$=124) &663 $\pm$ 31  ($\sigma$=135) &542 $\pm$ 22 ($\sigma$=97.1) &290 $\pm$ 36 ($\sigma$=159)    \\ 
\hline
Families & 68 & 1068 $\pm$ 22   ($\sigma$=180) &802 $\pm$ 30  ($\sigma$=247) &532 $\pm$ 21 ($\sigma$=170) &465 $\pm$ 44 ($\sigma$=362)    \\ 
\hline
Friends & 57 & 1107 $\pm$ 22   ($\sigma$=168) &753 $\pm$ 22  ($\sigma$=164) &603 $\pm$ 20 ($\sigma$=149) &332 $\pm$ 33 ($\sigma$=251)    \\ 
\hline
$F_{3,162}$ & & 5.54 & 2.59 & 5.87 & 4.77\\
\hline
$p$ & & 0.0012 & 0.055 & 0.000794 & 0.00327\\
\hline
$R^2$ & & 0.0931 & 0.0457 & 0.098 & 0.0811\\
\hline
$\delta$ & & 1.32 & 0.613 & 0.888 & 0.694\\
\hline
\end{tabular}
\end{center}
\end{table}
\begin{table}[!ht]
\scriptsize
\caption{$p$ values for relation corresponding to velocity and distance observables at different density ranges.}
\label{table2b3}
\begin{center}
\begin{tabular} {|c|c|c|c|c|}
\hline
Density&   $V$ &    $r$ & $x$  & $|y|$ \\
\hline
0-0.05 ped/m$^2$&  $<10^{-8}$ & 5.52$\cdot 10^{-5}$ & $<10^{-8}$ & 0.000396\\
\hline
0.05-0.1 ped/m$^2$&  $<10^{-8}$ & 0.000158 & $<10^{-8}$ & $<10^{-8}$\\
\hline
0.1-0.15 ped/m$^2$&  $<10^{-8}$ & 2.22$\cdot 10^{-5}$ & $<10^{-8}$ & $<10^{-8}$\\
\hline
0.15-0.2 ped/m$^2$&  0.0012 & 0.055 & 0.000794 & 0.00327\\
\hline
 0.2-0.25 ped/m$^2$&  0.378 & 0.327 & 0.144 & 0.664\\
\hline
\end{tabular}
\end{center}
\end{table}
\begin{table}[!ht]
\scriptsize
\caption{$\delta$ values for relation corresponding to velocity and distance observables at different density ranges.}
\label{table2b4}
\begin{center}
\begin{tabular} {|c|c|c|c|c|}
\hline
Density&   $V$ &    $r$ & $x$  & $|y|$ \\
\hline
0-0.05 ped/m$^2$ &  1.03 & 0.585 & 0.754 & 0.482\\
\hline
0.05-0.1 ped/m$^2$ &  1.07 & 0.476 & 0.732 & 0.538\\
\hline
0.1-0.15 ped/m$^2$ &  1.16 & 0.666 & 0.857 & 0.761\\
\hline
0.15-0.2 ped/m$^2$ &  1.32 & 0.613 & 0.888 & 0.694\\
\hline
 0.2-0.25 ped/m$^2$ &  0.675 & 0.798 & 0.983 & 0.974\\
\hline
\end{tabular}
\end{center}
\end{table}
\subsubsection{Gender}
We now compare the results regarding relation for groups of given gender (two females, mixed and  two males) in tables \ref{table2c1}, \ref{table2c2} and \ref{table2c3}. 
Differences in the distributions (and the corresponding trends) are still significant in fixed gender groups, with the exception of the $r$ female and male distributions, although,
as shown by the relatively high $\delta$ values, this may be due due by the low amount of data in some categories. The patterns analysed in the main text are mostly respected, 
although we may notice some differences such as female friends walking at an higher distance than colleagues,
and two male families walking at a very high speed.
\begin{table}[!ht]
\scriptsize
\caption{Observable dependence on relation for 2 females dyads. Lengths in millimetres, times in seconds.}
\label{table2c1}
\begin{center}
\begin{tabular} {|c|c|c|c|c|c|}
\hline
Relation&  $N^k_g$ &  $V$ &    $r$ & $x$  & $|y|$ \\
\hline
Colleagues & 24 & 1167 $\pm$ 30   ($\sigma$=145) &735 $\pm$ 26  ($\sigma$=128) &664 $\pm$ 20 ($\sigma$=95.9) &238 $\pm$ 34 ($\sigma$=168)    \\ 
\hline
Families & 28 & 1023 $\pm$ 32   ($\sigma$=171) &847 $\pm$ 58  ($\sigma$=305) &565 $\pm$ 27 ($\sigma$=140) &488 $\pm$ 77 ($\sigma$=405)    \\ 
\hline
Friends & 184 & 1105 $\pm$ 15   ($\sigma$=197) &777 $\pm$ 15  ($\sigma$=205) &658 $\pm$ 8.5 ($\sigma$=115) &293 $\pm$ 20 ($\sigma$=274)    \\ 
\hline
$F_{2,233}$ & & 3.85 & 1.9 & 7.85 & 6.48\\
\hline
$p$ & & 0.0227 & 0.153 & 0.000503 & 0.00182\\
\hline
$R^2$ & & 0.032 & 0.016 & 0.0631 & 0.0527\\
\hline
$\delta$ & & 0.902 & 0.465 & 0.817 & 0.783\\
\hline
\end{tabular}
\end{center}
\end{table}

\begin{table}[!ht]
\scriptsize
\caption{Observable dependence on relation for mixed gender dyads. Lengths in millimetres, times in seconds.}
\label{table2c2}
\begin{center}
\begin{tabular} {|c|c|c|c|c|c|}
\hline
Relation&  $N^k_g$ &  $V$ &    $r$ & $x$  & $|y|$ \\
\hline
Colleagues & 35 & 1228 $\pm$ 30   ($\sigma$=175) &923 $\pm$ 55  ($\sigma$=327) &702 $\pm$ 27 ($\sigma$=158) &440 $\pm$ 75 ($\sigma$=445)    \\ 
\hline
Couples & 96 & 1099 $\pm$ 17   ($\sigma$=169) &714 $\pm$ 22  ($\sigma$=219) &600 $\pm$ 15 ($\sigma$=150) &291 $\pm$ 24 ($\sigma$=231)    \\ 
\hline
Families & 183 & 1078 $\pm$ 13   ($\sigma$=182) &860 $\pm$ 21  ($\sigma$=285) &588 $\pm$ 13 ($\sigma$=173) &493 $\pm$ 28 ($\sigma$=372)    \\ 
\hline
Friends & 20 & 1153 $\pm$ 41   ($\sigma$=183) &820 $\pm$ 43  ($\sigma$=192) &616 $\pm$ 43 ($\sigma$=192) &391 $\pm$ 70 ($\sigma$=311)    \\ 
\hline
$F_{3,330}$ & & 7.47 & 7.99 & 4.63 & 7.26\\
\hline
$p$ & & 7.49$\cdot 10^{-5}$ & 3.72$\cdot 10^{-5}$ & 0.00345 & 9.96$\cdot 10^{-5}$\\
\hline
$R^2$ & & 0.0636 & 0.0677 & 0.0404 & 0.0619\\
\hline
$\delta$ & & 0.832 & 0.83 & 0.67 & 0.61\\
\hline
\end{tabular}
\end{center}
\end{table}

\begin{table}[!ht]
\scriptsize
\caption{Observable dependence on relation for 2 males dyads. Lengths in millimetres, times in seconds.}
\label{table2c3}
\begin{center}
\begin{tabular} {|c|c|c|c|c|c|}
\hline
Relation&  $N^k_g$ &  $V$ &    $r$ & $x$  & $|y|$ \\
\hline
Colleagues & 299 & 1287 $\pm$ 8.8   ($\sigma$=152) &852 $\pm$ 13  ($\sigma$=220) &724 $\pm$ 9.3 ($\sigma$=160) &329 $\pm$ 16 ($\sigma$=273)    \\ 
\hline
Families & 35 & 1234 $\pm$ 39   ($\sigma$=229) &891 $\pm$ 63  ($\sigma$=375) &571 $\pm$ 31 ($\sigma$=182) &537 $\pm$ 79 ($\sigma$=467)    \\ 
\hline
Friends & 114 & 1187 $\pm$ 19   ($\sigma$=198) &811 $\pm$ 17  ($\sigma$=186) &676 $\pm$ 14 ($\sigma$=147) &335 $\pm$ 23 ($\sigma$=246)    \\ 
\hline
$F_{2,445}$ & & 14.5 & 2.1 & 16.1 & 8.35\\
\hline
$p$ & & 8.13$\cdot 10^{-7}$ & 0.124 & 1.76$\cdot 10^{-7}$ & 0.000276\\
\hline
$R^2$ & & 0.0611 & 0.00933 & 0.0675 & 0.0362\\
\hline
$\delta$ & & 0.607 & 0.329 & 0.94 & 0.698\\
\hline
\end{tabular}
\end{center}
\end{table}

\subsubsection{Age}
Tables  \ref{table2d1} and  \ref{table2d2} show the observables dependence for fixed minimum age ranges (20-29 and 50-59 years, respectively). 
The major trends exposed in the main text are present even when the age is kept fixed.
We may  notice that $|y|$ assumes a very high value in families even when children are not present\footnote{This may be 
  related to a selection bias in coders, that may have labelled mixed gender dyads as ``couples'' or ``families'' depending on their proximity and ``abreastness''.}. $p$ and $\delta$
values for the relation feature at different minimum age ranges are shown in,
respectively, tables \ref{table2d3} and \ref{table2d4}.

\begin{table}[!ht]
\scriptsize
\caption{Observable dependence on relation for dyads with minimum age in the 20-29 years range. Lengths in millimetres, times in seconds.}
\label{table2d1}
\begin{center}
\begin{tabular} {|c|c|c|c|c|c|}
\hline
Relation&  $N^k_g$ &  $V$ &    $r$ & $x$  & $|y|$ \\
\hline
Colleagues & 86 & 1255 $\pm$ 18   ($\sigma$=165) &813 $\pm$ 20  ($\sigma$=185) &706 $\pm$ 16 ($\sigma$=144) &291 $\pm$ 24 ($\sigma$=219)    \\ 
\hline
Couples & 74 & 1115 $\pm$ 19   ($\sigma$=165) &711 $\pm$ 27  ($\sigma$=229) &600 $\pm$ 18 ($\sigma$=154) &281 $\pm$ 28 ($\sigma$=243)    \\ 
\hline
Families & 23 & 1109 $\pm$ 39   ($\sigma$=187) &877 $\pm$ 58  ($\sigma$=277) &581 $\pm$ 37 ($\sigma$=177) &527 $\pm$ 78 ($\sigma$=373)    \\ 
\hline
Friends & 164 & 1186 $\pm$ 13   ($\sigma$=164) &801 $\pm$ 16  ($\sigma$=208) &683 $\pm$ 10 ($\sigma$=128) &298 $\pm$ 21 ($\sigma$=265)    \\ 
\hline
$F_{3,343}$ & & 10.9 & 5.05 & 11.1 & 5.9\\
\hline
$p$ & & 7.21$\cdot 10^{-7}$ & 0.00195 & 5.89$\cdot 10^{-7}$ & 0.00062\\
\hline
$R^2$ & & 0.0872 & 0.0423 & 0.0883 & 0.0491\\
\hline
$\delta$ & & 0.861 & 0.684 & 0.825 & 0.882\\
\hline
\end{tabular}
\end{center}
\end{table}

\begin{table}[!ht]
\scriptsize
\caption{Observable dependence on relation for dyads with minimum age in the 50-59 years range. Lengths in millimetres, times in seconds.}
\label{table2d2}
\begin{center}
\begin{tabular} {|c|c|c|c|c|c|}
\hline
Relation&  $N^k_g$ &  $V$ &    $r$ & $x$  & $|y|$ \\
\hline
Colleagues & 52 & 1274 $\pm$ 22   ($\sigma$=159) &844 $\pm$ 24  ($\sigma$=172) &700 $\pm$ 20 ($\sigma$=142) &350 $\pm$ 36 ($\sigma$=263)    \\ 
\hline
Families & 28 & 1048 $\pm$ 32   ($\sigma$=169) &846 $\pm$ 55  ($\sigma$=289) &562 $\pm$ 34 ($\sigma$=182) &492 $\pm$ 78 ($\sigma$=410)    \\ 
\hline
Friends & 22 & 1051 $\pm$ 36   ($\sigma$=167) &759 $\pm$ 44  ($\sigma$=208) &637 $\pm$ 24 ($\sigma$=115) &308 $\pm$ 59 ($\sigma$=276)    \\ 
\hline
$F_{2,99}$ & & 23.4 & 1.29 & 7.63 & 2.55\\
\hline
$p$ & & $<10^{-8}$ & 0.28 & 0.000825 & 0.0835\\
\hline
$R^2$ & & 0.321 & 0.0254 & 0.134 & 0.0489\\
\hline
$\delta$ & & 1.39 & 0.339 & 0.878 & 0.514\\
\hline
\end{tabular}
\end{center}
\end{table}

\begin{table}[!ht]
\scriptsize
\caption{$p$ values for relation in different minimum age ranges.}
\label{table2d3}
\begin{center}
\begin{tabular} {|c|c|c|c|c|}
\hline
Minimum age &  $V$ &    $r$ & $x$  & $|y|$ \\
\hline
10-19 years&  0.558 & 0.049 & 0.615 & 0.0313\\
\hline
20-29 years&  7.21$\cdot 10^{-7}$ & 0.00195 & 5.89$\cdot 10^{-7}$ & 0.00062\\
\hline
30-39 years&  $<10^{-8}$ & 0.0128 & $<10^{-8}$ & 0.0513\\
\hline
40-49 years&  1.02$\cdot 10^{-5}$ & 0.39 & 0.000266 & 0.537\\
\hline
50-59 years&  $<10^{-8}$ & 0.28 & 0.000825 & 0.0835\\
\hline
60-69 years&  0.0525 & 0.248 & 0.745 & 0.388\\
\hline
$\geq$ 70 years&  0.385 & 0.251 & 0.198 & 0.237\\
\hline
\end{tabular}
\end{center}
\end{table}

\begin{table}[!ht]
\scriptsize
\caption{$\delta$ values for relation in different minimum age ranges.}
\label{table2d4}
\begin{center}
\begin{tabular} {|c|c|c|c|c|}
\hline
Minimum age&   $V$ &    $r$ & $x$  & $|y|$ \\
\hline
10-19 years &  0.888 & 3.23 & 0.609 & 2.49\\
\hline
20-29 years &  0.861 & 0.684 & 0.825 & 0.882\\
\hline
30-39 years &  1.48 & 0.77 & 1.05 & 0.636\\
\hline
40-49 years &  1.2 & 0.529 & 0.848 & 0.557\\
\hline
50-59 years &  1.39 & 0.339 & 0.878 & 0.514\\
\hline
60-69 years &  1.21 & 1.12 & 0.386 & 0.669\\
\hline
$\geq$ 70 years &  0.612 & 0.844 & 0.937 & 0.865\\
\hline
\end{tabular}
\end{center}
\end{table}

\subsubsection{Height}
Tables  \ref{table2e1} and  \ref{table2e2} show the observables dependence for fixed average height ranges (160-170 and 170-180 cm, respectively), showing that the distributions
are still different in a significant way, and that the major patterns exposed in the main
text are confirmed. Tables \ref{table2e3} and \ref{table2e4} show respectively
the $p$ and $\delta$ values for all height ranges. $\delta$ values are always high showing that
some reduced $p$ values are due to the low number of groups in some ranges.

\begin{table}[!ht]
\scriptsize
\caption{Observable dependence on relation for dyads with average height in the 160-170 cm range. Lengths in millimetres, times in seconds.}
\label{table2e1}
\begin{center}
\begin{tabular} {|c|c|c|c|c|c|}
\hline
Relation&  $N^k_g$ &  $V$ &    $r$ & $x$  & $|y|$ \\
\hline
Colleagues & 89 & 1240 $\pm$ 16   ($\sigma$=149) &862 $\pm$ 26  ($\sigma$=249) &686 $\pm$ 18 ($\sigma$=167) &380 $\pm$ 37 ($\sigma$=350)    \\ 
\hline
Couples & 47 & 1106 $\pm$ 28   ($\sigma$=191) &731 $\pm$ 33  ($\sigma$=226) &622 $\pm$ 28 ($\sigma$=189) &287 $\pm$ 30 ($\sigma$=204)    \\ 
\hline
Families & 121 & 1090 $\pm$ 18   ($\sigma$=196) &854 $\pm$ 27  ($\sigma$=295) &593 $\pm$ 15 ($\sigma$=169) &487 $\pm$ 34 ($\sigma$=371)    \\ 
\hline
Friends & 172 & 1135 $\pm$ 13   ($\sigma$=169) &798 $\pm$ 17  ($\sigma$=221) &659 $\pm$ 10 ($\sigma$=131) &321 $\pm$ 23 ($\sigma$=302)    \\ 
\hline
$F_{3,425}$ & & 13.3 & 3.96 & 7.08 & 7.42\\
\hline
$p$ & & 2.47$\cdot 10^{-8}$ & 0.0084 & 0.000119 & 7.42$\cdot 10^{-5}$\\
\hline
$R^2$ & & 0.086 & 0.0272 & 0.0476 & 0.0498\\
\hline
$\delta$ & & 0.843 & 0.542 & 0.551 & 0.599\\
\hline
\end{tabular}
\end{center}
\end{table}

\begin{table}[!ht]
\scriptsize
\caption{Observable dependence on relation for dyads with average age in the 170-180 cm range. Lengths in millimetres, times in seconds.}
\label{table2e2}
\begin{center}
\begin{tabular} {|c|c|c|c|c|c|}
\hline
Relation&  $N^k_g$ &  $V$ &    $r$ & $x$  & $|y|$ \\
\hline
Colleagues & 231 & 1293 $\pm$ 10   ($\sigma$=157) &859 $\pm$ 15  ($\sigma$=232) &738 $\pm$ 10 ($\sigma$=157) &325 $\pm$ 18 ($\sigma$=274)    \\ 
\hline
Couples & 45 & 1089 $\pm$ 22   ($\sigma$=145) &700 $\pm$ 33  ($\sigma$=219) &576 $\pm$ 14 ($\sigma$=95.4) &300 $\pm$ 39 ($\sigma$=264)    \\ 
\hline
Families & 56 & 1107 $\pm$ 22   ($\sigma$=166) &818 $\pm$ 31  ($\sigma$=234) &557 $\pm$ 20 ($\sigma$=148) &462 $\pm$ 48 ($\sigma$=361)    \\ 
\hline
Friends & 71 & 1162 $\pm$ 20   ($\sigma$=166) &811 $\pm$ 19  ($\sigma$=156) &679 $\pm$ 17 ($\sigma$=145) &328 $\pm$ 26 ($\sigma$=215)    \\ 
\hline
$F_{3,399}$ & & 38.9 & 6.77 & 31.5 & 4.13\\
\hline
$p$ & & $<10^{-8}$ & 0.000183 & $<10^{-8}$ & 0.00672\\
\hline
$R^2$ & & 0.226 & 0.0485 & 0.192 & 0.0301\\
\hline
$\delta$ & & 1.32 & 0.692 & 1.16 & 0.503\\
\hline
\end{tabular}
\end{center}
\end{table}

\begin{table}[!ht]
\scriptsize
\caption{$p$ values for relation at different average height ranges.}
\label{table2e3}
\begin{center}
\begin{tabular} {|c|c|c|c|c|}
\hline
Average height &  $V$ &    $r$ & $x$  & $|y|$ \\
\hline
$<$ 140 cm&  0.362 & 0.108 & 0.61 & 0.0849\\
\hline
140-150 cm&  0.12 & 0.181 & 0.785 & 0.299\\
\hline
150-160 cm&  0.842 & 0.133 & 0.803 & 0.0402\\
\hline
160-170 cm&  2.47$\cdot 10^{-8}$ & 0.0084 & 0.000119 & 7.42$\cdot 10^{-5}$\\
\hline
170-180 cm&  $<10^{-8}$ & 0.000183 & $<10^{-8}$ & 0.00672\\
\hline
$>$ 180 cm&  0.00432 & 0.551 & 0.126 & 0.951\\
\hline
\end{tabular}
\end{center}
\end{table}

\begin{table}[!ht]
\scriptsize
\caption{$\delta$ values for relation at different average height ranges.}
\label{table2e4}
\begin{center}
\begin{tabular} {|c|c|c|c|c|}
\hline
Average height &  $V$ &    $r$ & $x$  & $|y|$ \\
\hline
$<$ 140 cm &  0.767 & 1.41 & 0.423 & 1.53\\
\hline
140-150 cm &  0.977 & 0.798 & 0.159 & 0.612\\
\hline
150-160 cm &  0.405 & 0.658 & 0.521 & 0.594\\
\hline
160-170 cm &  0.843 & 0.542 & 0.551 & 0.599\\
\hline
170-180 cm &  1.32 & 0.692 & 1.16 & 0.503\\
\hline
$>$ 180 cm &  2.78 & 0.972 & 1.41 & 0.331\\
\hline
\end{tabular}
\end{center}
\end{table}

\clearpage
\subsection{Secondary effects and gender}
\label{furthergen}
\subsubsection{Density}
Tables \ref{table3c1} and \ref{table3c2} show the dependence on gender of observables at fixed density ranges ($0 \leq \rho <0.05$ ped/m$^2$ and $0.15 \leq \rho <0.2$ ped/m$^2$, respectively).
We may see that only the $r$ observable loses the statistically significant gender dependence at high density (but still shows it at lower density, when pedestrians may move more freely; furthermore, the effect size is almost not affected by density), while
the other observables preserve it at any density. Tables \ref{table3c3} and \ref{table3c4}
show the dependence of, respectively, the gender $p$ and $\delta$ values at different density
values.

\begin{table}[!ht]
\scriptsize
\caption{Observable dependence on gender in the $0 \leq \rho <0.05$ ped/m$^2$ density range. Lengths in millimetres, times in seconds.}
\label{table3c1}
\begin{center}
\begin{tabular} {|c|c|c|c|c|c|}
\hline
Gender&  $N^k_g$ &  $V$ &    $r$ & $x$  & $|y|$ \\
\hline
Two females & 160 & 1095 $\pm$ 17   ($\sigma$=219) &818 $\pm$ 21  ($\sigma$=267) &669 $\pm$ 11 ($\sigma$=138) &337 $\pm$ 27 ($\sigma$=346)    \\ 
\hline
Mixed & 217 & 1112 $\pm$ 13   ($\sigma$=196) &870 $\pm$ 23  ($\sigma$=340) &642 $\pm$ 13 ($\sigma$=194) &448 $\pm$ 28 ($\sigma$=409)    \\ 
\hline
Two males & 259 & 1254 $\pm$ 12   ($\sigma$=196) &914 $\pm$ 18  ($\sigma$=283) &733 $\pm$ 13 ($\sigma$=217) &404 $\pm$ 22 ($\sigma$=351)    \\ 
\hline
$F_{2,633}$ & & 41.5 & 5.06 & 14.1 & 4.11\\
\hline
$p$ & & $<10^{-8}$ & 0.00658 & 1.04$\cdot 10^{-6}$ & 0.0169\\
\hline
$R^2$ & & 0.116 & 0.0157 & 0.0426 & 0.0128\\
\hline
$\delta$ & & 0.771 & 0.346 & 0.441 & 0.289\\
\hline
\end{tabular}
\end{center}
\end{table}

\begin{table}[!ht]
\scriptsize
\caption{Observable dependence on gender in the $0.15 \leq \rho <0.2$ ped/m$^2$ density range. Lengths in millimetres, times in seconds.}
\label{table3c2}
\begin{center}
\begin{tabular} {|c|c|c|c|c|c|}
\hline
Gender&  $N^k_g$ &  $V$ &    $r$ & $x$  & $|y|$ \\
\hline
Two females & 35 & 1073 $\pm$ 28   ($\sigma$=164) &714 $\pm$ 26  ($\sigma$=152) &572 $\pm$ 18 ($\sigma$=107) &318 $\pm$ 39 ($\sigma$=230)    \\ 
\hline
Mixed & 73 & 1062 $\pm$ 18   ($\sigma$=152) &782 $\pm$ 29  ($\sigma$=247) &521 $\pm$ 20 ($\sigma$=172) &448 $\pm$ 42 ($\sigma$=361)    \\ 
\hline
Two males & 59 & 1171 $\pm$ 23   ($\sigma$=178) &767 $\pm$ 19  ($\sigma$=147) &644 $\pm$ 18 ($\sigma$=136) &304 $\pm$ 28 ($\sigma$=218)    \\ 
\hline
$F_{2,164}$ & & 7.81 & 1.4 & 11.2 & 4.6\\
\hline
$p$ & & 0.000578 & 0.249 & 2.88$\cdot 10^{-5}$ & 0.0114\\
\hline
$R^2$ & & 0.0869 & 0.0168 & 0.12 & 0.0531\\
\hline
$\delta$ & & 0.665 & 0.308 & 0.786 & 0.471\\
\hline
\end{tabular}
\end{center}
\end{table}

\begin{table}[!ht]
\scriptsize
\caption{$p$ values for gender in different density ranges.}
\label{table3c3}
\begin{center}
\begin{tabular} {|c|c|c|c|c|}
\hline
Density&   $V$ &    $r$ & $x$  & $|y|$ \\
\hline
0-0.05 ped/m$^2$&  $<10^{-8}$ & 0.00658 & 1.04$\cdot 10^{-6}$ & 0.0169\\
\hline
0.05-0.1 ped/m$^2$&  $<10^{-8}$ & 0.0448 & $<10^{-8}$ & 0.00164\\
\hline
0.1-0.15 ped/m$^2$&  $<10^{-8}$ & 0.897 & 9.41$\cdot 10^{-8}$ & 0.00478\\
\hline
0.15-0.2 ped/m$^2$&  0.000578 & 0.249 & 2.88$\cdot 10^{-5}$ & 0.0114\\
\hline
 0.2-0.25 ped/m$^2$&  0.0304 & 0.0628 & 0.31 & 0.43\\
\hline
\end{tabular}
\end{center}
\end{table}

\begin{table}[!ht]
\scriptsize
\caption{$\delta$ values for gender in different density ranges.}
\label{table3c4}
\begin{center}
\begin{tabular} {|c|c|c|c|c|}
\hline
Density&   $V$ &    $r$ & $x$  & $|y|$ \\
\hline
0-0.05 ped/m$^2$ &  0.771 & 0.346 & 0.441 & 0.289\\
\hline
0.05-0.1 ped/m$^2$ &  0.771 & 0.235 & 0.537 & 0.291\\
\hline
0.1-0.15 ped/m$^2$ &  0.737 & 0.0554 & 0.582 & 0.35\\
\hline
0.15-0.2 ped/m$^2$ &  0.665 & 0.308 & 0.786 & 0.471\\
\hline
 0.2-0.25 ped/m$^2$ &  1.31 & 1.56 & 0.942 & 0.751\\
\hline
\end{tabular}
\end{center}
\end{table}

\subsubsection{Relation}
Tables \ref{table3b1}, \ref{table3b2} and \ref{table3b3} show the gender dependence of observables in, respectively, colleagues, families and friends (couples are not shown being exclusively
of mixed gender).

\begin{table}[!ht]
\scriptsize
\caption{Observable dependence on gender for colleagues. Lengths in millimetres, times in seconds.}
\label{table3b1}
\begin{center}
\begin{tabular} {|c|c|c|c|c|c|}
\hline
Gender&  $N^k_g$ &  $V$ &    $r$ & $x$  & $|y|$ \\
\hline
Two females & 24 & 1167 $\pm$ 30   ($\sigma$=145) &735 $\pm$ 26  ($\sigma$=128) &664 $\pm$ 20 ($\sigma$=95.9) &238 $\pm$ 34 ($\sigma$=168)    \\ 
\hline
Mixed & 35 & 1228 $\pm$ 30   ($\sigma$=175) &923 $\pm$ 55  ($\sigma$=327) &702 $\pm$ 27 ($\sigma$=158) &440 $\pm$ 75 ($\sigma$=445)    \\ 
\hline
Two males & 299 & 1287 $\pm$ 8.8   ($\sigma$=152) &852 $\pm$ 13  ($\sigma$=220) &724 $\pm$ 9.3 ($\sigma$=160) &329 $\pm$ 16 ($\sigma$=273)    \\ 
\hline
$F_{2,355}$ & & 8.49 & 4.82 & 1.78 & 3.7\\
\hline
$p$ & & 0.00025 & 0.00862 & 0.17 & 0.0256\\
\hline
$R^2$ & & 0.0457 & 0.0264 & 0.00995 & 0.0204\\
\hline
$\delta$ & & 0.798 & 0.709 & 0.38 & 0.561\\
\hline
\end{tabular}
\end{center}
\end{table}

\begin{table}[!ht]
\scriptsize
\caption{Observable dependence on gender for families. Lengths in millimetres, times in seconds.}
\label{table3b2}
\begin{center}
\begin{tabular} {|c|c|c|c|c|c|}
\hline
Gender&  $N^k_g$ &  $V$ &    $r$ & $x$  & $|y|$ \\
\hline
Two females & 28 & 1023 $\pm$ 32   ($\sigma$=171) &847 $\pm$ 58  ($\sigma$=305) &565 $\pm$ 27 ($\sigma$=140) &488 $\pm$ 77 ($\sigma$=405)    \\ 
\hline
Mixed & 183 & 1078 $\pm$ 13   ($\sigma$=182) &860 $\pm$ 21  ($\sigma$=285) &588 $\pm$ 13 ($\sigma$=173) &493 $\pm$ 28 ($\sigma$=372)    \\ 
\hline
Two males & 35 & 1234 $\pm$ 39   ($\sigma$=229) &891 $\pm$ 63  ($\sigma$=375) &571 $\pm$ 31 ($\sigma$=182) &537 $\pm$ 79 ($\sigma$=467)    \\ 
\hline
$F_{2,243}$ & & 12.3 & 0.197 & 0.308 & 0.198\\
\hline
$p$ & & 8.41$\cdot 10^{-6}$ & 0.821 & 0.735 & 0.821\\
\hline
$R^2$ & & 0.0917 & 0.00162 & 0.00253 & 0.00163\\
\hline
$\delta$ & & 1.03 & 0.128 & 0.135 & 0.112\\
\hline
\end{tabular}
\end{center}
\end{table}

\begin{table}[!ht]
\scriptsize
\caption{Observable dependence on gender for friends. Lengths in millimetres, times in seconds.}
\label{table3b3}
\begin{center}
\begin{tabular} {|c|c|c|c|c|c|}
\hline
Gender&  $N^k_g$ &  $V$ &    $r$ & $x$  & $|y|$ \\
\hline
Two females & 184 & 1105 $\pm$ 15   ($\sigma$=197) &777 $\pm$ 15  ($\sigma$=205) &658 $\pm$ 8.5 ($\sigma$=115) &293 $\pm$ 20 ($\sigma$=274)    \\ 
\hline
Mixed & 20 & 1153 $\pm$ 41   ($\sigma$=183) &820 $\pm$ 43  ($\sigma$=192) &616 $\pm$ 43 ($\sigma$=192) &391 $\pm$ 70 ($\sigma$=311)    \\ 
\hline
Two males & 114 & 1187 $\pm$ 19   ($\sigma$=198) &811 $\pm$ 17  ($\sigma$=186) &676 $\pm$ 14 ($\sigma$=147) &335 $\pm$ 23 ($\sigma$=246)    \\ 
\hline
$F_{2,315}$ & & 6.02 & 1.27 & 1.91 & 1.76\\
\hline
$p$ & & 0.00272 & 0.283 & 0.15 & 0.173\\
\hline
$R^2$ & & 0.0368 & 0.00798 & 0.012 & 0.0111\\
\hline
$\delta$ & & 0.412 & 0.213 & 0.388 & 0.354\\
\hline
\end{tabular}
\end{center}
\end{table}

We may see that males dyads are farther and faster than female ones regardless of relation (although the differences in $r$, $x$ and $|y|$ are quite reduced in families and friends).
Mixed dyad behaviour, on the other hand, depends strongly on relation. Mixed dyads are the only ones including couples, and this affects strongly their behaviour, and represent also the largest
part of families. They are very little represented in friends (interestingly, mixed dyads of friends walk much closer, in abreast distance, than same sex dyads, although their absolute distance
is higher than in females). The ``colleagues'' category could represent a fair field for comparing the effect of gender, and in it the mixed behaviour is somehow in between the two sexes (although
the absolute distance $r$ and group depth $|y|$ are very large, suggesting not abreast formations) but in our set colleagues are extremely biased towards males, and thus the analysis in hindered by low female and mixed dyads numbers. Finally we may notice that in families and friends, the effect of gender on distance ($r$, $x$ and $|y|$) is very reduced, but the one on velocity is persistent.
The velocity effect size in families is nevertheless more than two times the one for friends.
\subsubsection{Age}
Tables \ref{table3d1} and \ref{table3d2} show the dependence on gender of observables at fixed average age ranges (20-29 years and 50-59 years, respectively).
Interestingly, the differences between young two females and two males dyads are reduced (and almost absent regarding the distance observables $r$, $x$ and $|y|$), 
while they are very strong in elder groups. Young mixed dyad behaviour is strongly influenced by the presence of couples.
\begin{table}[!ht]
\scriptsize
\caption{Observable dependence on gender in the 20-29 years average age range. Lengths in millimetres, times in seconds.}
\label{table3d1}
\begin{center}
\begin{tabular} {|c|c|c|c|c|c|}
\hline
Gender&  $N^k_g$ &  $V$ &    $r$ & $x$  & $|y|$ \\
\hline
Two females & 111 & 1166 $\pm$ 16   ($\sigma$=170) &791 $\pm$ 21  ($\sigma$=220) &686 $\pm$ 10 ($\sigma$=110) &275 $\pm$ 26 ($\sigma$=271)    \\ 
\hline
Mixed & 125 & 1122 $\pm$ 16   ($\sigma$=175) &784 $\pm$ 23  ($\sigma$=255) &612 $\pm$ 16 ($\sigma$=182) &360 $\pm$ 27 ($\sigma$=307)    \\ 
\hline
Two males & 134 & 1247 $\pm$ 14   ($\sigma$=164) &803 $\pm$ 17  ($\sigma$=201) &689 $\pm$ 13 ($\sigma$=148) &301 $\pm$ 20 ($\sigma$=235)    \\ 
\hline
$F_{2,367}$ & & 18 & 0.235 & 10.5 & 3.03\\
\hline
$p$ & & 3.37$\cdot 10^{-8}$ & 0.791 & 3.77$\cdot 10^{-5}$ & 0.0496\\
\hline
$R^2$ & & 0.0895 & 0.00128 & 0.054 & 0.0162\\
\hline
$\delta$ & & 0.739 & 0.0832 & 0.47 & 0.291\\
\hline
\end{tabular}
\end{center}
\end{table}

\begin{table}[!ht]
\scriptsize
\caption{Observable dependence on gender in the 50-59 years average age range. Lengths in millimetres, times in seconds.}
\label{table3d2}
\begin{center}
\begin{tabular} {|c|c|c|c|c|c|}
\hline
Gender&  $N^k_g$ &  $V$ &    $r$ & $x$  & $|y|$ \\
\hline
Two females & 20 & 1010 $\pm$ 30   ($\sigma$=136) &708 $\pm$ 21  ($\sigma$=95.5) &613 $\pm$ 27 ($\sigma$=121) &254 $\pm$ 34 ($\sigma$=151)    \\ 
\hline
Mixed & 34 & 1071 $\pm$ 29   ($\sigma$=170) &856 $\pm$ 48  ($\sigma$=278) &608 $\pm$ 32 ($\sigma$=189) &462 $\pm$ 66 ($\sigma$=388)    \\ 
\hline
Two males & 60 & 1255 $\pm$ 22   ($\sigma$=168) &847 $\pm$ 25  ($\sigma$=192) &686 $\pm$ 18 ($\sigma$=141) &369 $\pm$ 38 ($\sigma$=298)    \\ 
\hline
$F_{2,111}$ & & 22.8 & 3.69 & 3.37 & 2.81\\
\hline
$p$ & & $<10^{-8}$ & 0.0281 & 0.0379 & 0.0643\\
\hline
$R^2$ & & 0.291 & 0.0623 & 0.0573 & 0.0482\\
\hline
$\delta$ & & 1.52 & 0.646 & 0.486 & 0.646\\
\hline
\end{tabular}
\end{center}
\end{table}
Tables \ref{table3d3} and \ref{table3d4} show the $p$ and $\delta$ values for gender
in different average age ranges. Minimum ages ranges are shown in tables \ref{table3d5} and \ref{table3d6}.
\begin{table}[!ht]
\scriptsize
\caption{$p$ values for gender in different average age ranges.}
\label{table3d3}
\begin{center}
\begin{tabular} {|c|c|c|c|c|}
\hline
Average age &  $V$ &    $r$ & $x$  & $|y|$ \\
\hline
10-19 years&  0.0301 & 0.685 & 0.573 & 0.903\\
\hline
20-29 years&  3.37$\cdot 10^{-8}$ & 0.791 & 3.77$\cdot 10^{-5}$ & 0.0496\\
\hline
30-39 years&  $<10^{-8}$ & 0.0477 & 7.66$\cdot 10^{-8}$ & 0.0433\\
\hline
40-49 years&  $<10^{-8}$ & 0.106 & 0.000167 & 0.856\\
\hline
50-59 years&  $<10^{-8}$ & 0.0281 & 0.0379 & 0.0643\\
\hline
60-69 years&  0.00145 & 0.495 & 0.17 & 0.655\\
\hline
$\geq$ 70 years&  0.245 & 0.564 & 0.543 & 0.598\\
\hline
\end{tabular}
\end{center}
\end{table}
\begin{table}[!ht]
\scriptsize
\caption{$\delta$ values for gender in different average age ranges.}
\label{table3d4}
\begin{center}
\begin{tabular} {|c|c|c|c|c|}
\hline
Average age &  $V$ &    $r$ & $x$  & $|y|$ \\
\hline
10-19 years &  0.769 & 0.241 & 0.309 & 0.14\\
\hline
20-29 years &  0.739 & 0.0832 & 0.47 & 0.291\\
\hline
30-39 years &  0.87 & 0.48 & 0.732 & 0.322\\
\hline
40-49 years &  1.3 & 0.329 & 0.619 & 0.0946\\
\hline
50-59 years &  1.52 & 0.646 & 0.486 & 0.646\\
\hline
60-69 years &  1.49 & 0.457 & 0.666 & 0.27\\
\hline
$\geq$ 70 years &  1.56 & 0.802 & 0.915 & 0.788\\
\hline
\end{tabular}
\end{center}
\end{table}
\begin{table}[!ht]
\scriptsize
\caption{$p$ values for gender in different minimum age ranges.}
\label{table3d5}
\begin{center}
\begin{tabular} {|c|c|c|c|c|}
\hline
Minimum age &  $V$ &    $r$ & $x$  & $|y|$ \\
\hline
0-9 years&  0.0872 & 0.17 & 0.577 & 0.198\\
\hline
10-19 years&  0.00563 & 0.497 & 0.981 & 0.484\\
\hline
20-29 years&  1.67$\cdot 10^{-7}$ & 0.665 & 8.91$\cdot 10^{-5}$ & 0.0654\\
\hline
30-39 years&  $<10^{-8}$ & 0.0904 & 1.99$\cdot 10^{-8}$ & 0.027\\
\hline
40-49 years&  3.02$\cdot 10^{-8}$ & 0.193 & 0.000602 & 0.778\\
\hline
50-59 years&  3.78$\cdot 10^{-8}$ & 0.0458 & 0.0555 & 0.0743\\
\hline
60-69 years&  0.00245 & 0.446 & 0.105 & 0.651\\
\hline
$\geq$ 70 years&  0.245 & 0.564 & 0.543 & 0.598\\
\hline
\end{tabular}
\end{center}
\end{table}
\begin{table}[!ht]
\scriptsize
\caption{$\delta$ values for gender in different minimum age ranges.}
\label{table3d6}
\begin{center}
\begin{tabular} {|c|c|c|c|c|}
\hline
Minimum age &  $V$ &    $r$ & $x$  & $|y|$ \\
\hline
0-9 years &  1.11 & 1.21 & 0.449 & 1.17\\
\hline
10-19 years &  0.949 & 0.473 & 0.0487 & 0.421\\
\hline
20-29 years &  0.715 & 0.111 & 0.475 & 0.271\\
\hline
30-39 years &  0.907 & 0.396 & 0.749 & 0.306\\
\hline
40-49 years &  1.62 & 0.343 & 0.65 & 0.113\\
\hline
50-59 years &  1.47 & 0.616 & 0.487 & 0.659\\
\hline
60-69 years &  1.45 & 0.541 & 0.79 & 0.275\\
\hline
$\geq$ 70 years &  1.56 & 0.802 & 0.915 & 0.788\\
\hline
\end{tabular}
\end{center}
\end{table}
\subsubsection{Height}
Tables \ref{table3e1} and \ref{table3e2} show the dependence on gender of observables at fixed average height ranges (150-160 and 170-180 cm, respectively).
The results (in particular for the shorter height range the effect size,
that helps in dealing with the reduced number of groups) show that differences between the sexes are still present when we consider individuals of similar height.

\begin{table}[!ht]
\scriptsize
\caption{Observable dependence on gender in the 150-160 cm average height range. Lengths in millimetres, times in seconds.}
\label{table3e1}
\begin{center}
\begin{tabular} {|c|c|c|c|c|c|}
\hline
Gender&  $N^k_g$ &  $V$ &    $r$ & $x$  & $|y|$ \\
\hline
Two females & 75 & 1094 $\pm$ 27   ($\sigma$=232) &791 $\pm$ 26  ($\sigma$=225) &627 $\pm$ 15 ($\sigma$=131) &352 $\pm$ 36 ($\sigma$=316)    \\ 
\hline
Mixed & 25 & 1045 $\pm$ 35   ($\sigma$=176) &796 $\pm$ 43  ($\sigma$=217) &603 $\pm$ 33 ($\sigma$=167) &376 $\pm$ 60 ($\sigma$=300)    \\ 
\hline
Two males & 18 & 1272 $\pm$ 82   ($\sigma$=346) &921 $\pm$ 92  ($\sigma$=390) &674 $\pm$ 42 ($\sigma$=180) &493 $\pm$ 110 ($\sigma$=447)    \\ 
\hline
$F_{2,115}$ & & 4.95 & 1.89 & 1.22 & 1.24\\
\hline
$p$ & & 0.00869 & 0.156 & 0.3 & 0.294\\
\hline
$R^2$ & & 0.0792 & 0.0318 & 0.0207 & 0.0211\\
\hline
$\delta$ & & 0.873 & 0.493 & 0.415 & 0.408\\
\hline

\end{tabular}
\end{center}
\end{table}

\begin{table}[!ht]
\scriptsize
\caption{Observable dependence on gender in the 170-180 cm average height range. Lengths in millimetres, times in seconds.}
\label{table3e2}
\begin{center}
\begin{tabular} {|c|c|c|c|c|c|}
\hline
Gender&  $N^k_g$ &  $V$ &    $r$ & $x$  & $|y|$ \\
\hline
Two females & 16 & 1091 $\pm$ 48   ($\sigma$=191) &741 $\pm$ 30  ($\sigma$=119) &662 $\pm$ 22 ($\sigma$=88.5) &238 $\pm$ 32 ($\sigma$=128)    \\ 
\hline
Mixed & 121 & 1127 $\pm$ 16   ($\sigma$=171) &797 $\pm$ 23  ($\sigma$=250) &598 $\pm$ 14 ($\sigma$=149) &396 $\pm$ 31 ($\sigma$=338)    \\ 
\hline
Two males & 284 & 1270 $\pm$ 9.6   ($\sigma$=161) &846 $\pm$ 13  ($\sigma$=213) &723 $\pm$ 9.4 ($\sigma$=159) &324 $\pm$ 15 ($\sigma$=258)    \\ 
\hline
$F_{2,418}$ & & 36.7 & 3.4 & 27.8 & 3.89\\
\hline
$p$ & & $<10^{-8}$ & 0.0344 & $<10^{-8}$ & 0.0212\\
\hline
$R^2$ & & 0.149 & 0.016 & 0.117 & 0.0183\\
\hline
$\delta$ & & 1.1 & 0.502 & 0.799 & 0.492\\
\hline
\end{tabular}
\end{center}
\end{table}
Tables \ref{table3e3} and \ref{table3e4} show, respectively, the gender $p$ and $\delta$ values for different average height ranges.
\begin{table}[!ht]
\scriptsize
\caption{Gender $p$ values for different average height ranges. Lengths in millimetres, times in seconds.}
\label{table3e3}
\begin{center}
\begin{tabular} {|c|c|c|c|c|}
\hline
Average height &  $V$ &    $r$ & $x$  & $|y|$ \\
\hline
$<$ 140 cm&  0.614 & 0.0596 & 0.958 & 0.148\\
\hline
140-150 cm&  0.000737 & 0.372 & 0.0226 & 0.306\\
\hline
150-160 cm&  0.00869 & 0.156 & 0.3 & 0.294\\
\hline
160-170 cm&  1.0 $\cdot 10^{-5}$ & 0.0653 & 0.0212 & 0.000455\\
\hline
170-180 cm&  $<10^{-8}$ & 0.0344 & $<10^{-8}$ & 0.0212\\
\hline
$>$ 180 cm&  0.0241 & 0.191 & 0.137 & 0.647\\
\hline
\end{tabular}
\end{center}
\end{table}
\begin{table}[!ht]
\scriptsize
\caption{Gender $\delta$ values for different average height ranges. Lengths in millimetres, times in seconds.}
\label{table3e4}
\begin{center}
\begin{tabular} {|c|c|c|c|c|}
\hline
Average height &  $V$ &    $r$ & $x$  & $|y|$ \\
\hline
$<$ 140 cm &  0.774 & 1.89 & 0.183 & 1.39\\
\hline
140-150 cm &  2.06 & 0.693 & 1.47 & 0.868\\
\hline
150-160 cm &  0.873 & 0.493 & 0.415 & 0.408\\
\hline
160-170 cm &  0.523 & 0.279 & 0.235 & 0.416\\
\hline
170-180 cm &  1.1 & 0.502 & 0.799 & 0.492\\
\hline
$>$ 180 cm &  1.28 & 0.708 & 0.811 & 0.246\\
\hline
\end{tabular}
\end{center}
\end{table}
\clearpage
\subsection{Secondary effects and age}
\label{furtherage}
\subsubsection{Density}
Tables  \ref{table4fur_dl} and \ref{table4fur_dh} show the age dependence of observables in, respectively, the $0 \leq \rho \leq 0.05$ ped/m$^2$ and $0.15 \leq \rho \leq 0.2$ ped/m$^2$ 
density ranges. Results mostly reflect those of the main text, with high or relatively high $\delta$ values suggesting that some not very good $p$ values may be due to the scarcity
of data in the children and elderly categories (i.e. the categories with the most different behaviour).
A remarkable feature, presented with the caveats related to sensor noise in the tracking of children,  is that while in general velocity decreases with density, this is not true for dyads with children, 
as shown in figure \ref{fastchild} in the main text.

\begin{table}[!ht]
\scriptsize
\caption{Observable dependence on minimum age in the $0 \leq \rho \leq 0.05$ ped/m$^2$ density range. Lengths in millimetres, times in seconds.}
\label{table4fur_dl}
\begin{center}
\begin{tabular} {|c|c|c|c|c|c|}
\hline
Minimum age&  $N^k_g$ &  $V$ &    $r$ & $x$  & $|y|$ \\
\hline
0-9 years & 9 & 1075 $\pm$ 68   ($\sigma$=205) &1078 $\pm$ 97  ($\sigma$=291) &663 $\pm$ 62 ($\sigma$=186) &704 $\pm$ 140 ($\sigma$=414)    \\ 
\hline
10-19 years & 44 & 1175 $\pm$ 43   ($\sigma$=288) &802 $\pm$ 40  ($\sigma$=262) &661 $\pm$ 25 ($\sigma$=167) &337 $\pm$ 44 ($\sigma$=294)    \\ 
\hline
20-29 years & 184 & 1198 $\pm$ 14   ($\sigma$=196) &853 $\pm$ 23  ($\sigma$=313) &694 $\pm$ 14 ($\sigma$=193) &357 $\pm$ 27 ($\sigma$=372)    \\ 
\hline
30-39 years & 185 & 1196 $\pm$ 16   ($\sigma$=217) &894 $\pm$ 22  ($\sigma$=306) &696 $\pm$ 16 ($\sigma$=223) &418 $\pm$ 27 ($\sigma$=368)    \\ 
\hline
40-49 years & 87 & 1150 $\pm$ 20   ($\sigma$=191) &909 $\pm$ 32  ($\sigma$=297) &683 $\pm$ 22 ($\sigma$=202) &440 $\pm$ 42 ($\sigma$=395)    \\ 
\hline
50-59 years & 71 & 1157 $\pm$ 23   ($\sigma$=198) &844 $\pm$ 27  ($\sigma$=228) &678 $\pm$ 18 ($\sigma$=149) &381 $\pm$ 38 ($\sigma$=320)    \\ 
\hline
60-69 years & 47 & 1022 $\pm$ 25   ($\sigma$=174) &912 $\pm$ 51  ($\sigma$=348) &670 $\pm$ 27 ($\sigma$=182) &481 $\pm$ 62 ($\sigma$=424)    \\ 
\hline
$\geq$ 70 years & 9 & 891 $\pm$ 31   ($\sigma$=92.6) &815 $\pm$ 100  ($\sigma$=307) &605 $\pm$ 19 ($\sigma$=55.8) &411 $\pm$ 130 ($\sigma$=394)    \\ 
\hline
$F_{7,628}$ & & 6.98 & 1.61 & 0.527 & 1.94\\
\hline
$p$ & & 4.97$\cdot 10^{-8}$ & 0.129 & 0.815 & 0.0613\\
\hline
$R^2$ & & 0.0722 & 0.0176 & 0.00584 & 0.0211\\
\hline
$\delta$ & & 1.59 & 1.03 & 0.419 & 1.16\\
\hline
\end{tabular}
\end{center}
\end{table}

\begin{table}[!ht]
\scriptsize
\caption{Observable dependence on minimum age in the $0.15 \leq \rho \leq 0.2$ ped/m$^2$ density range. Lengths in millimetres, times in seconds.}
\label{table4fur_dh}
\begin{center}
\begin{tabular} {|c|c|c|c|c|c|}
\hline
Minimum age&  $N^k_g$ &  $V$ &    $r$ & $x$  & $|y|$ \\
\hline
0-9 years & 6 & 1284 $\pm$ 110   ($\sigma$=258) &693 $\pm$ 51  ($\sigma$=126) &485 $\pm$ 47 ($\sigma$=116) &401 $\pm$ 92 ($\sigma$=225)    \\ 
\hline
10-19 years & 14 & 1146 $\pm$ 47   ($\sigma$=176) &806 $\pm$ 65  ($\sigma$=244) &571 $\pm$ 39 ($\sigma$=147) &426 $\pm$ 91 ($\sigma$=341)    \\ 
\hline
20-29 years & 72 & 1099 $\pm$ 16   ($\sigma$=133) &745 $\pm$ 20  ($\sigma$=167) &598 $\pm$ 17 ($\sigma$=145) &322 $\pm$ 30 ($\sigma$=255)    \\ 
\hline
30-39 years & 39 & 1102 $\pm$ 31   ($\sigma$=192) &766 $\pm$ 33  ($\sigma$=208) &575 $\pm$ 24 ($\sigma$=149) &372 $\pm$ 50 ($\sigma$=313)    \\ 
\hline
40-49 years & 17 & 1121 $\pm$ 32   ($\sigma$=131) &763 $\pm$ 37  ($\sigma$=152) &547 $\pm$ 42 ($\sigma$=172) &403 $\pm$ 68 ($\sigma$=279)    \\ 
\hline
50-59 years & 10 & 1057 $\pm$ 53   ($\sigma$=167) &739 $\pm$ 60  ($\sigma$=190) &485 $\pm$ 58 ($\sigma$=184) &416 $\pm$ 110 ($\sigma$=343)    \\ 
\hline
60-69 years & 7 & 1021 $\pm$ 75   ($\sigma$=199) &967 $\pm$ 130  ($\sigma$=354) &616 $\pm$ 84 ($\sigma$=222) &618 $\pm$ 160 ($\sigma$=423)    \\ 
\hline
$\geq$ 70 years & 2 & 760 $\pm$ 18   ($\sigma$=25.4) &644 $\pm$ 22  ($\sigma$=31) &585 $\pm$ 8.9 ($\sigma$=12.5) &185 $\pm$ 52 ($\sigma$=73.5)    \\ 
\hline
$F_{7,159}$ & & 2.76 & 1.46 & 1.11 & 1.21\\
\hline
$p$ & & 0.00993 & 0.184 & 0.359 & 0.299\\
\hline
$R^2$ & & 0.108 & 0.0606 & 0.0466 & 0.0506\\
\hline
$\delta$ & & 2.22 & 0.987 & 0.652 & 1.1\\
\hline
\end{tabular}
\end{center}
\end{table}
Tables \ref{table4fur_d1} and \ref{table4fur_d2} show, respectively, the minimum age $p$ and $\delta$ values in different density ranges
\begin{table}[!ht]
\scriptsize
\caption{Minimum age $p$ values in different density ranges. Lengths in millimetres, times in seconds.}
\label{table4fur_d1}
\begin{center}
\begin{tabular} {|c|c|c|c|c|}
\hline
Density&   $V$ &    $r$ & $x$  & $|y|$ \\
\hline
0-0.05 ped/m$^2$&  4.97$\cdot 10^{-8}$ & 0.129 & 0.815 & 0.0613\\
\hline
0.05-0.1 ped/m$^2$&  $<10^{-8}$ & 0.00286 & 0.0232 & 1.26$\cdot 10^{-6}$\\
\hline
0.1-0.15 ped/m$^2$&  8.51$\cdot 10^{-8}$ & 0.0207 & 0.00346 & 7.22$\cdot 10^{-6}$\\
\hline
0.15-0.2 ped/m$^2$&  0.00993 & 0.184 & 0.359 & 0.299\\
\hline
 0.2-0.25 ped/m$^2$&  0.651 & 0.504 & 0.118 & 0.328\\
\hline
\end{tabular}
\end{center}
\end{table}
\begin{table}[!ht]
\scriptsize
\caption{Minimum age $p$ values in different density ranges. Lengths in millimetres, times in seconds.}
\label{table4fur_d2}
\begin{center}
\begin{tabular} {|c|c|c|c|c|}
\hline
Density &  $V$ &    $r$ & $x$  & $|y|$ \\
\hline
0-0.05 ped/m$^2$ &  1.59 & 1.03 & 0.419 & 1.16\\
\hline
0.05-0.1 ped/m$^2$ &  1.6 & 0.941 & 0.605 & 1.36\\
\hline
0.1-0.15 ped/m$^2$ &  2.25 & 0.689 & 0.93 & 0.924\\
\hline
0.15-0.2 ped/m$^2$ &  2.22 & 0.987 & 0.652 & 1.1\\
\hline
 0.2-0.25 ped/m$^2$ &  0.513 & 2.14 & 1.15 & 1.14\\
\hline
\end{tabular}
\end{center}
\end{table}
\subsubsection{Relation}
Tables \ref{table4fur_col}, \ref{table4fur_coup}, \ref{table4fur_fam} and \ref{table4fur_fri} show the age dependence of observables in, respectively,
colleagues, couples, families and friends. Observable values almost have no age dependence in the 20-59 years age
(with the exclusion of friend velocity). It is interesting to note that the $|y|$ distribution assumes a larger value for 
families even when only adults are involved. Another interesting, although expectable, result is that while dyads with teenagers are very abreast in the friends category,
they are not abreast in the family one (the $|y|$ values is almost doubled in families). 

\begin{table}[!ht]
\scriptsize
\caption{Observable dependence on minimum age for colleagues. Lengths in millimetres, times in seconds.}
\label{table4fur_col}
\begin{center}
\begin{tabular} {|c|c|c|c|c|c|}
\hline
Minimum age&  $N^k_g$ &  $V$ &    $r$ & $x$  & $|y|$ \\
\hline
20-29 years & 86 & 1255 $\pm$ 18   ($\sigma$=165) &813 $\pm$ 20  ($\sigma$=185) &706 $\pm$ 16 ($\sigma$=144) &291 $\pm$ 24 ($\sigma$=219)    \\ 
\hline
30-39 years & 145 & 1293 $\pm$ 14   ($\sigma$=167) &861 $\pm$ 21  ($\sigma$=257) &734 $\pm$ 14 ($\sigma$=165) &331 $\pm$ 25 ($\sigma$=301)    \\ 
\hline
40-49 years & 71 & 1258 $\pm$ 14   ($\sigma$=119) &870 $\pm$ 28  ($\sigma$=236) &714 $\pm$ 19 ($\sigma$=161) &363 $\pm$ 38 ($\sigma$=324)    \\ 
\hline
50-59 years & 52 & 1274 $\pm$ 22   ($\sigma$=159) &844 $\pm$ 24  ($\sigma$=172) &700 $\pm$ 20 ($\sigma$=142) &350 $\pm$ 36 ($\sigma$=263)    \\ 
\hline
60-69 years & 4 & 1217 $\pm$ 36   ($\sigma$=72) &1075 $\pm$ 220  ($\sigma$=433) &692 $\pm$ 100 ($\sigma$=208) &617 $\pm$ 320 ($\sigma$=632)    \\ 
\hline
$F_{4,353}$ & & 1.17 & 1.72 & 0.702 & 1.62\\
\hline
$p$ & & 0.326 & 0.146 & 0.591 & 0.168\\
\hline
$R^2$ & & 0.013 & 0.0191 & 0.00789 & 0.018\\
\hline
$\delta$ & & 0.461 & 1.32 & 0.252 & 1.33\\
\hline
\end{tabular}
\end{center}
\end{table}

\begin{table}[!ht]
\scriptsize
\caption{Observable dependence on minimum age for couples. Lengths in millimetres, times in seconds.}
\label{table4fur_coup}
\begin{center}
\begin{tabular} {|c|c|c|c|c|c|}
\hline
Minimum age&  $N^k_g$ &  $V$ &    $r$ & $x$  & $|y|$ \\
\hline
10-19 years & 2 & 958 $\pm$ 180   ($\sigma$=253) &919 $\pm$ 53  ($\sigma$=74.7) &725 $\pm$ 58 ($\sigma$=81.7) &480 $\pm$ 180 ($\sigma$=257)    \\ 
\hline
20-29 years & 74 & 1115 $\pm$ 19   ($\sigma$=165) &711 $\pm$ 27  ($\sigma$=229) &600 $\pm$ 18 ($\sigma$=154) &281 $\pm$ 28 ($\sigma$=243)    \\ 
\hline
30-39 years & 17 & 1049 $\pm$ 37   ($\sigma$=151) &670 $\pm$ 35  ($\sigma$=143) &572 $\pm$ 32 ($\sigma$=130) &274 $\pm$ 30 ($\sigma$=124)    \\ 
\hline
40-49 years & 3 & 1091 $\pm$ 110   ($\sigma$=187) &897 $\pm$ 110  ($\sigma$=198) &684 $\pm$ 66 ($\sigma$=115) &501 $\pm$ 130 ($\sigma$=223)    \\ 
\hline
$F_{3,92}$ & & 1.2 & 1.53 & 0.966 & 1.36\\
\hline
$p$ & & 0.315 & 0.211 & 0.412 & 0.261\\
\hline
$R^2$ & & 0.0376 & 0.0476 & 0.0306 & 0.0424\\
\hline
$\delta$ & & 0.946 & 1.78 & 1.19 & 1.65\\
\hline
\end{tabular}
\end{center}
\end{table}

\begin{table}[!ht]
\scriptsize
\caption{Observable dependence on minimum age for families. Lengths in millimetres, times in seconds.}
\label{table4fur_fam}
\begin{center}
\begin{tabular} {|c|c|c|c|c|c|}
\hline
Minimum age&  $N^k_g$ &  $V$ &    $r$ & $x$  & $|y|$ \\
\hline
0-9 years & 31 & 1143 $\pm$ 42   ($\sigma$=235) &995 $\pm$ 69  ($\sigma$=383) &529 $\pm$ 34 ($\sigma$=189) &701 $\pm$ 87 ($\sigma$=485)    \\ 
\hline
10-19 years & 36 & 1163 $\pm$ 38   ($\sigma$=230) &831 $\pm$ 49  ($\sigma$=296) &617 $\pm$ 30 ($\sigma$=179) &415 $\pm$ 58 ($\sigma$=347)    \\ 
\hline
20-29 years & 23 & 1109 $\pm$ 39   ($\sigma$=187) &877 $\pm$ 58  ($\sigma$=277) &581 $\pm$ 37 ($\sigma$=177) &527 $\pm$ 78 ($\sigma$=373)    \\ 
\hline
30-39 years & 46 & 1078 $\pm$ 23   ($\sigma$=159) &814 $\pm$ 33  ($\sigma$=225) &561 $\pm$ 24 ($\sigma$=163) &458 $\pm$ 49 ($\sigma$=330)    \\ 
\hline
40-49 years & 41 & 1116 $\pm$ 31   ($\sigma$=199) &801 $\pm$ 28  ($\sigma$=181) &582 $\pm$ 23 ($\sigma$=149) &431 $\pm$ 40 ($\sigma$=256)    \\ 
\hline
50-59 years & 28 & 1048 $\pm$ 32   ($\sigma$=169) &846 $\pm$ 55  ($\sigma$=289) &562 $\pm$ 34 ($\sigma$=182) &492 $\pm$ 78 ($\sigma$=410)    \\ 
\hline
60-69 years & 37 & 1030 $\pm$ 24   ($\sigma$=145) &911 $\pm$ 63  ($\sigma$=382) &642 $\pm$ 25 ($\sigma$=154) &512 $\pm$ 75 ($\sigma$=456)    \\ 
\hline
$\geq$ 70 years & 4 & 847 $\pm$ 52   ($\sigma$=104) &926 $\pm$ 190  ($\sigma$=384) &550 $\pm$ 38 ($\sigma$=75.6) &575 $\pm$ 240 ($\sigma$=477)    \\ 
\hline
$F_{7,238}$ & & 2.83 & 1.52 & 1.46 & 1.74\\
\hline
$p$ & & 0.00758 & 0.162 & 0.182 & 0.101\\
\hline
$R^2$ & & 0.0767 & 0.0427 & 0.0412 & 0.0486\\
\hline
$\delta$ & & 1.42 & 0.679 & 0.659 & 0.686\\
\hline
\end{tabular}
\end{center}
\end{table}

\begin{table}[!ht]
\scriptsize
\caption{Observable dependence on minimum age for friends. Lengths in millimetres, times in seconds.}
\label{table4fur_fri}
\begin{center}
\begin{tabular} {|c|c|c|c|c|c|}
\hline
Minimum age&  $N^k_g$ &  $V$ &    $r$ & $x$  & $|y|$ \\
\hline
10-19 years & 23 & 1143 $\pm$ 61   ($\sigma$=292) &681 $\pm$ 15  ($\sigma$=73.9) &621 $\pm$ 16 ($\sigma$=78.5) &217 $\pm$ 19 ($\sigma$=93.2)    \\ 
\hline
20-29 years & 164 & 1186 $\pm$ 13   ($\sigma$=164) &801 $\pm$ 16  ($\sigma$=208) &683 $\pm$ 10 ($\sigma$=128) &298 $\pm$ 21 ($\sigma$=265)    \\ 
\hline
30-39 years & 56 & 1143 $\pm$ 28   ($\sigma$=206) &817 $\pm$ 24  ($\sigma$=178) &644 $\pm$ 22 ($\sigma$=162) &366 $\pm$ 38 ($\sigma$=286)    \\ 
\hline
40-49 years & 19 & 1089 $\pm$ 47   ($\sigma$=206) &819 $\pm$ 49  ($\sigma$=213) &682 $\pm$ 21 ($\sigma$=92.3) &341 $\pm$ 68 ($\sigma$=295)    \\ 
\hline
50-59 years & 22 & 1051 $\pm$ 36   ($\sigma$=167) &759 $\pm$ 44  ($\sigma$=208) &637 $\pm$ 24 ($\sigma$=115) &308 $\pm$ 59 ($\sigma$=276)    \\ 
\hline
60-69 years & 26 & 996 $\pm$ 38   ($\sigma$=192) &808 $\pm$ 40  ($\sigma$=202) &625 $\pm$ 33 ($\sigma$=169) &383 $\pm$ 59 ($\sigma$=299)    \\ 
\hline
$\geq$ 70 years & 8 & 906 $\pm$ 32   ($\sigma$=91.4) &716 $\pm$ 56  ($\sigma$=159) &606 $\pm$ 18 ($\sigma$=52.2) &290 $\pm$ 85 ($\sigma$=239)    \\ 
\hline
$F_{6,311}$ & & 7.17 & 1.82 & 1.98 & 1.29\\
\hline
$p$ & & 3.58$\cdot 10^{-7}$ & 0.0947 & 0.0678 & 0.262\\
\hline
$R^2$ & & 0.121 & 0.0339 & 0.0368 & 0.0242\\
\hline
$\delta$ & & 1.73 & 0.904 & 0.608 & 0.731\\
\hline
\end{tabular}
\end{center}
\end{table}
\subsubsection{Gender}
Tables \ref{table4fur_0o}, \ref{table4fur_1o} and \ref{table4fur_2o}  show the age dependence of observables in, respectively, dyads with two females, mixed dyads and two males.
The results are similar to those shown in the main text. Although based on an extremely reduced number of data, it is interesting to note the large difference in velocity between two males and two 
females dyads with children (mixed dyads show a value in between, probably due to the fact that they include male and female parents), 
and the very large $|y|$ (non-abreast formation) value assumed in two females dyads (mixed dyads on the opposite are more abreast).
This values are based on very few groups, but differences are nevertheless larger than standard errors, and could reflect differences in the relation that children have with fathers and mothers (at
least in the observed cultural environment). 
\begin{table}[!ht]
\scriptsize
\caption{Observable dependence on minimum age for two females dyads. Lengths in millimetres, times in seconds.}
\label{table4fur_0o}
\begin{center}
\begin{tabular} {|c|c|c|c|c|c|}
\hline
Minimum age&  $N^k_g$ &  $V$ &    $r$ & $x$  & $|y|$ \\
\hline
0-9 years & 6 & 985 $\pm$ 88   ($\sigma$=215) &1252 $\pm$ 150  ($\sigma$=378) &525 $\pm$ 78 ($\sigma$=192) &993 $\pm$ 220 ($\sigma$=535)    \\ 
\hline
10-19 years & 20 & 1075 $\pm$ 58   ($\sigma$=258) &738 $\pm$ 48  ($\sigma$=216) &621 $\pm$ 23 ($\sigma$=103) &291 $\pm$ 64 ($\sigma$=288)    \\ 
\hline
20-29 years & 110 & 1169 $\pm$ 16   ($\sigma$=167) &789 $\pm$ 21  ($\sigma$=221) &684 $\pm$ 10 ($\sigma$=107) &273 $\pm$ 26 ($\sigma$=277)    \\ 
\hline
30-39 years & 55 & 1108 $\pm$ 25   ($\sigma$=188) &777 $\pm$ 23  ($\sigma$=171) &639 $\pm$ 16 ($\sigma$=118) &330 $\pm$ 33 ($\sigma$=246)    \\ 
\hline
40-49 years & 24 & 1040 $\pm$ 33   ($\sigma$=163) &827 $\pm$ 54  ($\sigma$=266) &622 $\pm$ 25 ($\sigma$=123) &404 $\pm$ 77 ($\sigma$=379)    \\ 
\hline
50-59 years & 17 & 1015 $\pm$ 35   ($\sigma$=143) &704 $\pm$ 24  ($\sigma$=97.3) &623 $\pm$ 29 ($\sigma$=120) &240 $\pm$ 32 ($\sigma$=133)    \\ 
\hline
60-69 years & 17 & 923 $\pm$ 31   ($\sigma$=130) &791 $\pm$ 52  ($\sigma$=213) &580 $\pm$ 36 ($\sigma$=149) &390 $\pm$ 85 ($\sigma$=349)    \\ 
\hline
$\geq$ 70 years & 3 & 958 $\pm$ 14   ($\sigma$=23.6) &629 $\pm$ 31  ($\sigma$=53.5) &587 $\pm$ 40 ($\sigma$=69) &186 $\pm$ 22 ($\sigma$=37.4)    \\ 
\hline
$F_{7,244}$ & & 6.27 & 4.83 & 3.74 & 5.64\\
\hline
$p$ & & 8.87$\cdot 10^{-7}$ & 4.06$\cdot 10^{-5}$ & 0.000721 & 4.77$\cdot 10^{-6}$\\
\hline
$R^2$ & & 0.153 & 0.122 & 0.0969 & 0.139\\
\hline
$\delta$ & & 1.51 & 1.94 & 1.42 & 1.78\\
\hline
\end{tabular}
\end{center}
\end{table}

\begin{table}[!ht]
\scriptsize
\caption{Observable dependence on minimum age for mixed dyads. Lengths in millimetres, times in seconds.}
\label{table4fur_1o}
\begin{center}
\begin{tabular} {|c|c|c|c|c|c|}
\hline
Minimum age&  $N^k_g$ &  $V$ &    $r$ & $x$  & $|y|$ \\
\hline
0-9 years & 12 & 1119 $\pm$ 56   ($\sigma$=193) &888 $\pm$ 75  ($\sigma$=259) &573 $\pm$ 61 ($\sigma$=212) &547 $\pm$ 83 ($\sigma$=287)    \\ 
\hline
10-19 years & 16 & 1060 $\pm$ 45   ($\sigma$=181) &840 $\pm$ 54  ($\sigma$=214) &620 $\pm$ 36 ($\sigma$=145) &417 $\pm$ 78 ($\sigma$=313)    \\ 
\hline
20-29 years & 120 & 1123 $\pm$ 16   ($\sigma$=175) &782 $\pm$ 23  ($\sigma$=251) &619 $\pm$ 17 ($\sigma$=182) &352 $\pm$ 27 ($\sigma$=301)    \\ 
\hline
30-39 years & 93 & 1143 $\pm$ 19   ($\sigma$=180) &834 $\pm$ 30  ($\sigma$=286) &601 $\pm$ 19 ($\sigma$=179) &435 $\pm$ 40 ($\sigma$=386)    \\ 
\hline
40-49 years & 53 & 1141 $\pm$ 26   ($\sigma$=191) &802 $\pm$ 25  ($\sigma$=178) &614 $\pm$ 21 ($\sigma$=150) &400 $\pm$ 34 ($\sigma$=245)    \\ 
\hline
50-59 years & 34 & 1078 $\pm$ 29   ($\sigma$=168) &848 $\pm$ 47  ($\sigma$=277) &604 $\pm$ 32 ($\sigma$=188) &455 $\pm$ 66 ($\sigma$=387)    \\ 
\hline
60-69 years & 38 & 1042 $\pm$ 26   ($\sigma$=160) &905 $\pm$ 61  ($\sigma$=378) &642 $\pm$ 25 ($\sigma$=152) &506 $\pm$ 73 ($\sigma$=451)    \\ 
\hline
$\geq$ 70 years & 5 & 831 $\pm$ 44   ($\sigma$=99) &868 $\pm$ 160  ($\sigma$=363) &563 $\pm$ 33 ($\sigma$=72.8) &484 $\pm$ 210 ($\sigma$=463)    \\ 
\hline
$F_{7,363}$ & & 3.64 & 1.11 & 0.387 & 1.32\\
\hline
$p$ & & 0.000822 & 0.358 & 0.91 & 0.24\\
\hline
$R^2$ & & 0.0656 & 0.0209 & 0.0074 & 0.0248\\
\hline
$\delta$ & & 1.76 & 0.431 & 0.536 & 0.652\\
\hline
\end{tabular}
\end{center}
\end{table}

\begin{table}[!ht]
\scriptsize
\caption{Observable dependence on minimum age for two males dyads. Lengths in millimetres, times in seconds.}
\label{table4fur_2o}
\begin{center}
\begin{tabular} {|c|c|c|c|c|c|}
\hline
Minimum age&  $N^k_g$ &  $V$ &    $r$ & $x$  & $|y|$ \\
\hline
0-9 years & 13 & 1237 $\pm$ 65   ($\sigma$=233) &975 $\pm$ 120  ($\sigma$=425) &491 $\pm$ 42 ($\sigma$=153) &709 $\pm$ 150 ($\sigma$=540)    \\ 
\hline
10-19 years & 27 & 1277 $\pm$ 48   ($\sigma$=252) &803 $\pm$ 58  ($\sigma$=303) &628 $\pm$ 34 ($\sigma$=175) &376 $\pm$ 65 ($\sigma$=337)    \\ 
\hline
20-29 years & 134 & 1241 $\pm$ 14   ($\sigma$=156) &806 $\pm$ 16  ($\sigma$=180) &697 $\pm$ 13 ($\sigma$=148) &296 $\pm$ 18 ($\sigma$=208)    \\ 
\hline
30-39 years & 144 & 1280 $\pm$ 16   ($\sigma$=190) &860 $\pm$ 18  ($\sigma$=222) &732 $\pm$ 14 ($\sigma$=173) &331 $\pm$ 22 ($\sigma$=259)    \\ 
\hline
40-49 years & 72 & 1257 $\pm$ 14   ($\sigma$=122) &875 $\pm$ 27  ($\sigma$=233) &715 $\pm$ 19 ($\sigma$=159) &365 $\pm$ 39 ($\sigma$=328)    \\ 
\hline
50-59 years & 60 & 1254 $\pm$ 22   ($\sigma$=168) &846 $\pm$ 25  ($\sigma$=193) &683 $\pm$ 19 ($\sigma$=144) &373 $\pm$ 38 ($\sigma$=297)    \\ 
\hline
60-69 years & 12 & 1134 $\pm$ 48   ($\sigma$=167) &930 $\pm$ 89  ($\sigma$=307) &711 $\pm$ 54 ($\sigma$=189) &462 $\pm$ 120 ($\sigma$=406)    \\ 
\hline
$\geq$ 70 years & 4 & 902 $\pm$ 48   ($\sigma$=95.7) &802 $\pm$ 92  ($\sigma$=183) &619 $\pm$ 19 ($\sigma$=37.6) &410 $\pm$ 140 ($\sigma$=289)    \\ 
\hline
$F_{7,458}$ & & 3.77 & 1.82 & 5.09 & 4.07\\
\hline
$p$ & & 0.000553 & 0.081 & 1.39$\cdot 10^{-5}$ & 0.000239\\
\hline
$R^2$ & & 0.0544 & 0.0271 & 0.0722 & 0.0586\\
\hline
$\delta$ & & 2.01 & 0.445 & 1.41 & 1.64\\
\hline
\end{tabular}
\end{center}
\end{table}
\subsubsection{Height}
Tables  \ref{table4fur_150} and \ref{table4fur_170} show the age dependence of observables in, respectively, the 150-160 cm  and 170-180 cm minimum height
ranges. These data, which respect the patterns highlighted in the main text, present a sufficient number of groups in each category and are thus reliable.
In some situation, a noisy tracking of a child may cause to have a category with very poor and not reliable representation (e.g. groups with children but with a tall minimum height)
causing some irregular behaviour in the $p$ and $\delta$ values of tables \ref{table4fur_h1} and \ref{table4fur_h2}.

\begin{table}[!ht]
\scriptsize
\caption{Observable dependence on minimum age in the 150-160 cm minimum height range. Lengths in millimetres, times in seconds.}
\label{table4fur_150}
\begin{center}
\begin{tabular} {|c|c|c|c|c|c|}
\hline
Minimum age&  $N^k_g$ &  $V$ &    $r$ & $x$  & $|y|$ \\
\hline
10-19 years & 21 & 1124 $\pm$ 54   ($\sigma$=246) &783 $\pm$ 47  ($\sigma$=215) &606 $\pm$ 31 ($\sigma$=143) &352 $\pm$ 73 ($\sigma$=333)    \\ 
\hline
20-29 years & 75 & 1157 $\pm$ 21   ($\sigma$=184) &800 $\pm$ 27  ($\sigma$=232) &668 $\pm$ 14 ($\sigma$=122) &311 $\pm$ 35 ($\sigma$=307)    \\ 
\hline
30-39 years & 48 & 1109 $\pm$ 24   ($\sigma$=168) &804 $\pm$ 32  ($\sigma$=223) &624 $\pm$ 20 ($\sigma$=139) &392 $\pm$ 44 ($\sigma$=305)    \\ 
\hline
40-49 years & 32 & 1108 $\pm$ 39   ($\sigma$=218) &828 $\pm$ 37  ($\sigma$=211) &589 $\pm$ 27 ($\sigma$=151) &452 $\pm$ 58 ($\sigma$=328)    \\ 
\hline
50-59 years & 19 & 1067 $\pm$ 39   ($\sigma$=171) &757 $\pm$ 34  ($\sigma$=146) &611 $\pm$ 36 ($\sigma$=156) &334 $\pm$ 52 ($\sigma$=228)    \\ 
\hline
60-69 years & 33 & 1008 $\pm$ 28   ($\sigma$=163) &808 $\pm$ 65  ($\sigma$=375) &641 $\pm$ 20 ($\sigma$=112) &365 $\pm$ 75 ($\sigma$=429)    \\ 
\hline
$\geq$ 70 years & 5 & 883 $\pm$ 52   ($\sigma$=117) &666 $\pm$ 48  ($\sigma$=108) &538 $\pm$ 30 ($\sigma$=67.2) &272 $\pm$ 88 ($\sigma$=197)    \\ 
\hline
$F_{6,226}$ & & 3.65 & 0.427 & 2.13 & 0.849\\
\hline
$p$ & & 0.00177 & 0.86 & 0.0506 & 0.533\\
\hline
$R^2$ & & 0.0883 & 0.0112 & 0.0536 & 0.022\\
\hline
$\delta$ & & 1.52 & 0.802 & 1.09 & 0.569\\
\hline
\end{tabular}
\end{center}
\end{table}

\begin{table}[!ht]
\scriptsize
\caption{Observable dependence on minimum age in the 170-180 cm minimum height range. Lengths in millimetres, times in seconds.}
\label{table4fur_170}
\begin{center}
\begin{tabular} {|c|c|c|c|c|c|}
\hline
Minimum age&  $N^k_g$ &  $V$ &    $r$ & $x$  & $|y|$ \\
\hline
20-29 years & 95 & 1209 $\pm$ 16   ($\sigma$=152) &808 $\pm$ 18  ($\sigma$=173) &688 $\pm$ 16 ($\sigma$=156) &309 $\pm$ 22 ($\sigma$=219)    \\ 
\hline
30-39 years & 90 & 1269 $\pm$ 21   ($\sigma$=203) &838 $\pm$ 24  ($\sigma$=225) &729 $\pm$ 16 ($\sigma$=155) &300 $\pm$ 26 ($\sigma$=246)    \\ 
\hline
40-49 years & 45 & 1265 $\pm$ 18   ($\sigma$=120) &820 $\pm$ 20  ($\sigma$=137) &720 $\pm$ 20 ($\sigma$=131) &298 $\pm$ 26 ($\sigma$=176)    \\ 
\hline
50-59 years & 30 & 1241 $\pm$ 33   ($\sigma$=182) &862 $\pm$ 44  ($\sigma$=238) &635 $\pm$ 27 ($\sigma$=148) &436 $\pm$ 68 ($\sigma$=371)    \\ 
\hline
$F_{3,256}$ & & 2.21 & 0.709 & 3.36 & 2.61\\
\hline
$p$ & & 0.0873 & 0.548 & 0.0194 & 0.0517\\
\hline
$R^2$ & & 0.0253 & 0.00823 & 0.0379 & 0.0297\\
\hline
$\delta$ & & 0.339 & 0.282 & 0.613 & 0.51\\
\hline
\end{tabular}
\end{center}
\end{table}

\begin{table}[!ht]
\scriptsize
\caption{Minimum age $p$ values in different minimum height ranges. Lengths in millimetres, times in seconds.}
\label{table4fur_h1}
\begin{center}
\begin{tabular} {|c|c|c|c|c|}
\hline
Minimum height &  $V$ &    $r$ & $x$  & $|y|$ \\
\hline
$<$ 140 cm&  0.0333 & 0.137 & 0.0326 & 0.0184\\
\hline
140-150 cm&  0.0129 & 0.65 & 0.858 & 0.615\\
\hline
150-160 cm&  0.00177 & 0.86 & 0.0506 & 0.533\\
\hline
160-170 cm&  0.000643 & 0.00561 & 0.807 & 0.00456\\
\hline
170-180 cm&  0.0873 & 0.548 & 0.0194 & 0.0517\\
\hline
$>$ 180 cm&  0.98 & 0.292 & 0.56 & 0.0386\\
\hline
\end{tabular}
\end{center}
\end{table}

\begin{table}[!ht]
\scriptsize
\caption{Minimum age $\delta$ values in different minimum height ranges. Lengths in millimetres, times in seconds.}
\label{table4fur_h2}
\begin{center}
\begin{tabular} {|c|c|c|c|c|}
\hline
Minimum height&   $V$ &    $r$ & $x$  & $|y|$ \\
\hline
$<$ 140 cm &  0.829 & 0.566 & 0.83 & 0.919\\
\hline
140-150 cm &  7.42 & 1.27 & 0.946 & 1.69\\
\hline
150-160 cm &  1.52 & 0.802 & 1.09 & 0.569\\
\hline
160-170 cm &  1.83 & 1.09 & 0.926 & 0.818\\
\hline
170-180 cm &  0.339 & 0.282 & 0.613 & 0.51\\
\hline
$>$ 180 cm &  0.181 & 1.2 & 1.53 & 1.26\\
\hline
\end{tabular}
\end{center}
\end{table}
\clearpage
\subsection{Secondary effects and height}
\label{furtherhei}
\subsubsection{Density}
Tables \ref{table5ca} and \ref{table5cb} show the dependence of observables on minimum height in the $0 \leq \rho \leq 0.05$ and $0.15 \leq \rho \leq 0.2$ ped/m$^{2}$ ranges, respectively.
The trends discussed in the main text are still present. We notice again a tendency of short people (most probably children) to walk faster at higher density.
$p$ and $\delta$ values for minimum height at different densities are shown in tables \ref{table5cex1} and \ref{table5cex2}
\begin{table}[!ht]
\scriptsize
\caption{Observable dependence on minimum height for dyads in the $0 \leq \rho \leq 0.05$ ped/m$^{2}$ range. Lengths in millimetres, times in seconds.}
\label{table5ca}
\begin{center}
\begin{tabular} {|c|c|c|c|c|c|}
\hline
Minimum height&  $N^k_g$ &  $V$ &    $r$ & $x$  & $|y|$ \\
\hline
$<$ 140 cm & 21 & 1138 $\pm$ 63   ($\sigma$=288) &1034 $\pm$ 80  ($\sigma$=368) &693 $\pm$ 44 ($\sigma$=201) &616 $\pm$ 92 ($\sigma$=421)    \\ 
\hline
140-150 cm & 29 & 1067 $\pm$ 57   ($\sigma$=304) &876 $\pm$ 51  ($\sigma$=275) &671 $\pm$ 40 ($\sigma$=218) &420 $\pm$ 64 ($\sigma$=346)    \\ 
\hline
150-160 cm & 148 & 1104 $\pm$ 18   ($\sigma$=224) &837 $\pm$ 25  ($\sigma$=304) &648 $\pm$ 11 ($\sigma$=128) &395 $\pm$ 32 ($\sigma$=390)    \\ 
\hline
160-170 cm & 290 & 1162 $\pm$ 11   ($\sigma$=187) &880 $\pm$ 19  ($\sigma$=318) &688 $\pm$ 13 ($\sigma$=217) &409 $\pm$ 22 ($\sigma$=379)    \\ 
\hline
170-180 cm & 141 & 1259 $\pm$ 16   ($\sigma$=188) &878 $\pm$ 21  ($\sigma$=253) &718 $\pm$ 17 ($\sigma$=198) &364 $\pm$ 28 ($\sigma$=331)    \\ 
\hline
$>$ 180 cm & 7 & 1242 $\pm$ 69   ($\sigma$=182) &929 $\pm$ 73  ($\sigma$=194) &793 $\pm$ 45 ($\sigma$=120) &316 $\pm$ 100 ($\sigma$=270)    \\ 
\hline
$F_{5,630}$ & & 9.97 & 1.72 & 2.32 & 1.8\\
\hline
$p$ & & $<10^{-8}$ & 0.128 & 0.0422 & 0.11\\
\hline
$R^2$ & & 0.0733 & 0.0135 & 0.0181 & 0.0141\\
\hline
$\delta$ & & 0.906 & 0.634 & 1.13 & 0.767\\
\hline
\end{tabular}
\end{center}
\end{table}
\begin{table}[!ht]
\scriptsize
\caption{Observable dependence on minimum height for dyads in the $0.15 \leq \rho \leq 0.2$ ped/m$^{2}$ range. Lengths in millimetres, times in seconds.}
\label{table5cb}
\begin{center}
\begin{tabular} {|c|c|c|c|c|c|}
\hline
Minimum height&  $N^k_g$ &  $V$ &    $r$ & $x$  & $|y|$ \\
\hline
$<$ 140 cm & 8 & 1185 $\pm$ 57   ($\sigma$=162) &872 $\pm$ 150  ($\sigma$=416) &512 $\pm$ 68 ($\sigma$=193) &555 $\pm$ 190 ($\sigma$=543)    \\ 
\hline
140-150 cm & 6 & 1166 $\pm$ 130   ($\sigma$=315) &965 $\pm$ 170  ($\sigma$=409) &604 $\pm$ 73 ($\sigma$=179) &590 $\pm$ 190 ($\sigma$=457)    \\ 
\hline
150-160 cm & 39 & 1068 $\pm$ 23   ($\sigma$=146) &754 $\pm$ 28  ($\sigma$=177) &518 $\pm$ 27 ($\sigma$=169) &408 $\pm$ 52 ($\sigma$=327)    \\ 
\hline
160-170 cm & 72 & 1093 $\pm$ 20   ($\sigma$=170) &722 $\pm$ 16  ($\sigma$=136) &586 $\pm$ 14 ($\sigma$=121) &321 $\pm$ 24 ($\sigma$=203)    \\ 
\hline
170-180 cm & 42 & 1127 $\pm$ 24   ($\sigma$=158) &792 $\pm$ 26  ($\sigma$=170) &618 $\pm$ 26 ($\sigma$=171) &352 $\pm$ 44 ($\sigma$=284)    \\ 
\hline
$F_{4,162}$ & & 1.34 & 3.29 & 2.62 & 2.31\\
\hline
$p$ & & 0.259 & 0.0127 & 0.0368 & 0.0597\\
\hline
$R^2$ & & 0.0319 & 0.0751 & 0.0608 & 0.054\\
\hline
$\delta$ & & 0.787 & 1.44 & 0.606 & 1.18\\
\hline
\end{tabular}
\end{center}
\end{table}
\begin{table}[!ht]
\scriptsize
\caption{$p$ values for minimum height in different density ranges.}
\label{table5cex1}
\begin{center}
\begin{tabular} {|c|c|c|c|c|}
\hline
Density&   $V$ &    $r$ & $x$  & $|y|$ \\
\hline
0-0.05 ped/m$^2$&  $<10^{-8}$ & 0.128 & 0.0422 & 0.11\\
\hline
0.05-0.1 ped/m$^2$&  $<10^{-8}$ & 0.000607 & 0.000112 & 4.09$\cdot 10^{-6}$\\
\hline
0.1-0.15 ped/m$^2$&  1.84$\cdot 10^{-5}$ & $<10^{-8}$ & 3.34$\cdot 10^{-5}$ & $<10^{-8}$\\
\hline
0.15-0.2 ped/m$^2$&  0.259 & 0.0127 & 0.0368 & 0.0597\\
\hline
 0.2-0.25 ped/m$^2$&  0.303 & 0.602 & 0.603 & 0.765\\
\hline
\end{tabular}
\end{center}
\end{table}
\begin{table}[!ht]
\scriptsize
\caption{$\delta$ values for minimum height in different density ranges.}
\label{table5cex2}
\begin{center}
\begin{tabular} {|c|c|c|c|c|}
\hline
Density&   $V$ &    $r$ & $x$  & $|y|$ \\
\hline
0-0.05 ped/m$^2$ &  0.906 & 0.634 & 1.13 & 0.767\\
\hline
0.05-0.1 ped/m$^2$ &  1.17 & 0.664 & 0.63 & 0.973\\
\hline
0.1-0.15 ped/m$^2$ &  0.856 & 1.32 & 1.12 & 1.29\\
\hline
0.15-0.2 ped/m$^2$ &  0.787 & 1.44 & 0.606 & 1.18\\
\hline
 0.2-0.25 ped/m$^2$ &  0.886 & 0.578 & 0.54 & 0.422\\
\hline
\end{tabular}
\end{center}
\end{table}
\subsubsection{Relation}
Tables \ref{table5cc}, \ref{table5cd}, \ref{table5ce} and \ref{table5cf} show the dependence of observables on minimum height for colleagues, couples, families and friends, respectively.
The dependence of observables on height appears to be attenuated when analysed for groups with a fixed relation (and in particular for couples), as shown by the higher $p$ values, and, to a lesser extent,
lower $\delta$ values.
\begin{table}[!ht]
\scriptsize
\caption{Observable dependence on minimum height for colleague dyads. Lengths in millimetres, times in seconds.}
\label{table5cc}
\begin{center}
\begin{tabular} {|c|c|c|c|c|c|}
\hline
Minimum height&  $N^k_g$ &  $V$ &    $r$ & $x$  & $|y|$ \\
\hline
150-160 cm & 15 & 1135 $\pm$ 36   ($\sigma$=141) &732 $\pm$ 25  ($\sigma$=98.2) &652 $\pm$ 22 ($\sigma$=85.9) &265 $\pm$ 31 ($\sigma$=120)    \\ 
\hline
160-170 cm & 159 & 1276 $\pm$ 12   ($\sigma$=157) &874 $\pm$ 21  ($\sigma$=265) &712 $\pm$ 13 ($\sigma$=168) &369 $\pm$ 28 ($\sigma$=351)    \\ 
\hline
170-180 cm & 170 & 1288 $\pm$ 12   ($\sigma$=155) &846 $\pm$ 15  ($\sigma$=202) &731 $\pm$ 12 ($\sigma$=153) &312 $\pm$ 18 ($\sigma$=236)    \\ 
\hline
$>$ 180 cm & 13 & 1220 $\pm$ 36   ($\sigma$=128) &789 $\pm$ 59  ($\sigma$=214) &689 $\pm$ 33 ($\sigma$=120) &263 $\pm$ 67 ($\sigma$=241)    \\ 
\hline
$F_{3,353}$ & & 5.03 & 2.2 & 1.49 & 1.63\\
\hline
$p$ & & 0.002 & 0.0881 & 0.217 & 0.182\\
\hline
$R^2$ & & 0.041 & 0.0183 & 0.0125 & 0.0137\\
\hline
$\delta$ & & 0.996 & 0.556 & 0.53 & 0.307\\
\hline
\end{tabular}
\end{center}
\end{table}

\begin{table}[!ht]
\scriptsize
\caption{Observable dependence on minimum height for couples. Lengths in millimetres, times in seconds.}
\label{table5cd}
\begin{center}
\begin{tabular} {|c|c|c|c|c|c|}
\hline
Minimum height&  $N^k_g$ &  $V$ &    $r$ & $x$  & $|y|$ \\
\hline
150-160 cm & 20 & 1060 $\pm$ 43   ($\sigma$=193) &736 $\pm$ 32  ($\sigma$=143) &631 $\pm$ 25 ($\sigma$=111) &288 $\pm$ 39 ($\sigma$=175)    \\ 
\hline
160-170 cm & 60 & 1114 $\pm$ 21   ($\sigma$=160) &716 $\pm$ 33  ($\sigma$=254) &591 $\pm$ 22 ($\sigma$=171) &305 $\pm$ 34 ($\sigma$=261)    \\ 
\hline
170-180 cm & 15 & 1092 $\pm$ 42   ($\sigma$=162) &678 $\pm$ 35  ($\sigma$=137) &592 $\pm$ 24 ($\sigma$=93.1) &245 $\pm$ 39 ($\sigma$=151)    \\ 
\hline
$F_{2,92}$ & & 0.773 & 0.296 & 0.528 & 0.408\\
\hline
$p$ & & 0.464 & 0.745 & 0.591 & 0.666\\
\hline
$R^2$ & & 0.0165 & 0.00639 & 0.0114 & 0.00878\\
\hline
$\delta$ & & 0.321 & 0.414 & 0.249 & 0.249\\
\hline
\end{tabular}
\end{center}
\end{table}

\begin{table}[!ht]
\scriptsize
\caption{Observable dependence on minimum height for families. Lengths in millimetres, times in seconds.}
\label{table5ce}
\begin{center}
\begin{tabular} {|c|c|c|c|c|c|}
\hline
Minimum height&  $N^k_g$ &  $V$ &    $r$ & $x$  & $|y|$ \\
\hline
$<$ 140 cm & 33 & 1117 $\pm$ 38   ($\sigma$=216) &1062 $\pm$ 72  ($\sigma$=411) &570 $\pm$ 39 ($\sigma$=222) &746 $\pm$ 89 ($\sigma$=509)    \\ 
\hline
140-150 cm & 19 & 1122 $\pm$ 62   ($\sigma$=270) &832 $\pm$ 60  ($\sigma$=262) &636 $\pm$ 32 ($\sigma$=139) &410 $\pm$ 65 ($\sigma$=285)    \\ 
\hline
150-160 cm & 77 & 1107 $\pm$ 23   ($\sigma$=204) &835 $\pm$ 34  ($\sigma$=302) &583 $\pm$ 19 ($\sigma$=163) &466 $\pm$ 44 ($\sigma$=387)    \\ 
\hline
160-170 cm & 99 & 1080 $\pm$ 17   ($\sigma$=170) &831 $\pm$ 24  ($\sigma$=241) &580 $\pm$ 17 ($\sigma$=166) &467 $\pm$ 33 ($\sigma$=332)    \\ 
\hline
170-180 cm & 17 & 1053 $\pm$ 36   ($\sigma$=149) &833 $\pm$ 65  ($\sigma$=268) &566 $\pm$ 37 ($\sigma$=153) &458 $\pm$ 98 ($\sigma$=403)    \\ 
\hline
$F_{4,240}$ & & 0.602 & 4.3 & 0.542 & 4.04\\
\hline
$p$ & & 0.661 & 0.00224 & 0.705 & 0.00344\\
\hline
$R^2$ & & 0.00993 & 0.0668 & 0.00896 & 0.0631\\
\hline
$\delta$ & & 0.312 & 0.788 & 0.478 & 0.76\\
\hline
\end{tabular}
\end{center}
\end{table}

\begin{table}[!ht]
\scriptsize
\caption{Observable dependence on minimum height for friends. Lengths in millimetres, times in seconds.}
\label{table5cf}
\begin{center}
\begin{tabular} {|c|c|c|c|c|c|}
\hline
Minimum height&  $N^k_g$ &  $V$ &    $r$ & $x$  & $|y|$ \\
\hline
$<$ 140 cm & 4 & 1129 $\pm$ 55   ($\sigma$=109) &611 $\pm$ 20  ($\sigma$=39.7) &518 $\pm$ 31 ($\sigma$=61.8) &246 $\pm$ 68 ($\sigma$=136)    \\ 
\hline
140-150 cm & 16 & 1115 $\pm$ 88   ($\sigma$=354) &816 $\pm$ 44  ($\sigma$=175) &628 $\pm$ 43 ($\sigma$=172) &382 $\pm$ 80 ($\sigma$=319)    \\ 
\hline
150-160 cm & 101 & 1100 $\pm$ 20   ($\sigma$=202) &770 $\pm$ 21  ($\sigma$=209) &659 $\pm$ 10 ($\sigma$=104) &287 $\pm$ 27 ($\sigma$=269)    \\ 
\hline
160-170 cm & 142 & 1138 $\pm$ 15   ($\sigma$=174) &802 $\pm$ 18  ($\sigma$=211) &665 $\pm$ 12 ($\sigma$=146) &324 $\pm$ 23 ($\sigma$=276)    \\ 
\hline
170-180 cm & 53 & 1199 $\pm$ 23   ($\sigma$=168) &806 $\pm$ 19  ($\sigma$=138) &673 $\pm$ 18 ($\sigma$=132) &323 $\pm$ 32 ($\sigma$=233)    \\ 
\hline
$>$ 180 cm & 2 & 1606 $\pm$ 23   ($\sigma$=32.1) &928 $\pm$ 61  ($\sigma$=86.6) &805 $\pm$ 120 ($\sigma$=171) &329 $\pm$ 72 ($\sigma$=102)    \\ 
\hline
$F_{5,312}$ & & 4.13 & 1.27 & 1.68 & 0.515\\
\hline
$p$ & & 0.0012 & 0.276 & 0.138 & 0.765\\
\hline
$R^2$ & & 0.0621 & 0.02 & 0.0263 & 0.00819\\
\hline
$\delta$ & & 2.52 & 5.74 & 2.83 & 0.461\\
\hline
\end{tabular}
\end{center}
\end{table}

\subsubsection{Gender}
Tables \ref{table5cg}, \ref{table5ch} and \ref{table5ci} show the dependence of observables on minimum height for two females, mixed and two males dyads, respectively. 
As discussed in the main
text and shown in figure \ref{fastchild2} in the main text, there is a loss  of linearity in $x$, but the patterns described in the main text are still present, although partially attenuated,
when gender is kept fixed.
\begin{table}[!ht]
\scriptsize
\caption{Observable dependence on minimum height for 2 female dyads. Lengths in millimetres, times in seconds.}
\label{table5cg}
\begin{center}
\begin{tabular} {|c|c|c|c|c|c|}
\hline
Minimum height&  $N^k_g$ &  $V$ &    $r$ & $x$  & $|y|$ \\
\hline
$<$ 140 cm & 7 & 956 $\pm$ 74   ($\sigma$=195) &1186 $\pm$ 150  ($\sigma$=385) &529 $\pm$ 68 ($\sigma$=180) &935 $\pm$ 190 ($\sigma$=509)    \\ 
\hline
140-150 cm & 21 & 1022 $\pm$ 57   ($\sigma$=262) &841 $\pm$ 66  ($\sigma$=300) &591 $\pm$ 36 ($\sigma$=166) &439 $\pm$ 96 ($\sigma$=442)    \\ 
\hline
150-160 cm & 114 & 1098 $\pm$ 18   ($\sigma$=197) &780 $\pm$ 20  ($\sigma$=214) &656 $\pm$ 11 ($\sigma$=112) &306 $\pm$ 26 ($\sigma$=279)    \\ 
\hline
160-170 cm & 104 & 1131 $\pm$ 16   ($\sigma$=159) &768 $\pm$ 18  ($\sigma$=185) &658 $\pm$ 11 ($\sigma$=115) &278 $\pm$ 24 ($\sigma$=245)    \\ 
\hline
170-180 cm & 6 & 1123 $\pm$ 77   ($\sigma$=188) &706 $\pm$ 34  ($\sigma$=83.3) &644 $\pm$ 26 ($\sigma$=64.8) &213 $\pm$ 59 ($\sigma$=144)    \\ 
\hline
$F_{4,247}$ & & 2.58 & 6.59 & 3.11 & 9.37\\
\hline
$p$ & & 0.0381 & 4.66$\cdot 10^{-5}$ & 0.0159 & 4.55$\cdot 10^{-7}$\\
\hline
$R^2$ & & 0.0401 & 0.0965 & 0.048 & 0.132\\
\hline
$\delta$ & & 1.08 & 1.66 & 1.08 & 1.86\\
\hline
\end{tabular}
\end{center}
\end{table}

\begin{table}[!ht]
\scriptsize
\caption{Observable dependence on minimum height for mixed gender dyads. Lengths in millimetres, times in seconds.}
\label{table5ch}
\begin{center}
\begin{tabular} {|c|c|c|c|c|c|}
\hline
Minimum height&  $N^k_g$ &  $V$ &    $r$ & $x$  & $|y|$ \\
\hline
$<$ 140 cm & 14 & 1100 $\pm$ 42   ($\sigma$=159) &947 $\pm$ 82  ($\sigma$=307) &590 $\pm$ 53 ($\sigma$=199) &593 $\pm$ 100 ($\sigma$=373)    \\ 
\hline
140-150 cm & 9 & 1107 $\pm$ 66   ($\sigma$=199) &967 $\pm$ 91  ($\sigma$=273) &664 $\pm$ 42 ($\sigma$=127) &552 $\pm$ 120 ($\sigma$=346)    \\ 
\hline
150-160 cm & 99 & 1092 $\pm$ 20   ($\sigma$=195) &829 $\pm$ 29  ($\sigma$=286) &609 $\pm$ 16 ($\sigma$=160) &429 $\pm$ 38 ($\sigma$=376)    \\ 
\hline
160-170 cm & 210 & 1128 $\pm$ 12   ($\sigma$=176) &811 $\pm$ 18  ($\sigma$=262) &618 $\pm$ 13 ($\sigma$=184) &395 $\pm$ 23 ($\sigma$=328)    \\ 
\hline
170-180 cm & 37 & 1083 $\pm$ 28   ($\sigma$=172) &806 $\pm$ 44  ($\sigma$=267) &589 $\pm$ 24 ($\sigma$=143) &404 $\pm$ 61 ($\sigma$=372)    \\ 
\hline
$>$ 180 cm & 2 & 937 $\pm$ 140   ($\sigma$=204) &687 $\pm$ 3.2  ($\sigma$=4.56) &573 $\pm$ 48 ($\sigma$=67.8) &253 $\pm$ 55 ($\sigma$=77.2)    \\ 
\hline
$F_{5,365}$ & & 1.14 & 1.3 & 0.418 & 1.27\\
\hline
$p$ & & 0.34 & 0.262 & 0.836 & 0.277\\
\hline
$R^2$ & & 0.0153 & 0.0175 & 0.00569 & 0.0171\\
\hline
$\delta$ & & 1.08 & 1.09 & 0.751 & 0.945\\
\hline
\end{tabular}
\end{center}
\end{table}

\begin{table}[!ht]
\scriptsize
\caption{Observable dependence on minimum height for male dyads. Lengths in millimetres, times in seconds.}
\label{table5ci}
\begin{center}
\begin{tabular} {|c|c|c|c|c|c|}
\hline
Minimum height&  $N^k_g$ &  $V$ &    $r$ & $x$  & $|y|$ \\
\hline
$<$ 140 cm & 18 & 1221 $\pm$ 48   ($\sigma$=203) &977 $\pm$ 110  ($\sigma$=455) &577 $\pm$ 53 ($\sigma$=226) &631 $\pm$ 130 ($\sigma$=549)    \\ 
\hline
140-150 cm & 9 & 1301 $\pm$ 140   ($\sigma$=405) &861 $\pm$ 85  ($\sigma$=255) &640 $\pm$ 47 ($\sigma$=142) &456 $\pm$ 110 ($\sigma$=343)    \\ 
\hline
150-160 cm & 21 & 1196 $\pm$ 39   ($\sigma$=180) &739 $\pm$ 36  ($\sigma$=164) &600 $\pm$ 22 ($\sigma$=103) &320 $\pm$ 57 ($\sigma$=261)    \\ 
\hline
160-170 cm & 184 & 1238 $\pm$ 13   ($\sigma$=179) &863 $\pm$ 18  ($\sigma$=241) &700 $\pm$ 13 ($\sigma$=174) &372 $\pm$ 23 ($\sigma$=315)    \\ 
\hline
170-180 cm & 219 & 1272 $\pm$ 11   ($\sigma$=156) &834 $\pm$ 12  ($\sigma$=184) &719 $\pm$ 10 ($\sigma$=151) &310 $\pm$ 15 ($\sigma$=224)    \\ 
\hline
$>$ 180 cm & 15 & 1271 $\pm$ 46   ($\sigma$=178) &808 $\pm$ 53  ($\sigma$=207) &705 $\pm$ 35 ($\sigma$=134) &272 $\pm$ 59 ($\sigma$=229)    \\ 
\hline
$F_{5,460}$ & & 1.46 & 2.56 & 4.51 & 5.03\\
\hline
$p$ & & 0.202 & 0.0267 & 0.000499 & 0.00017\\
\hline
$R^2$ & & 0.0156 & 0.0271 & 0.0468 & 0.0518\\
\hline
$\delta$ & & 0.394 & 0.721 & 0.903 & 0.826\\
\hline
\end{tabular}
\end{center}
\end{table}

\subsubsection{Age}
Tables \ref{table5cm} and \ref{table5cn} show the dependence on minimum height of all observables for dyads with minimum age in the 20-29 and 50-59 year ranges, respectively,
showing the effect of removing children from the population.
Finally, tables \ref{table5cex3} and \ref{table5cex4} show the dependence of, respectively, minimum height $p$ and $\delta$ values on minimum age ranges.

\begin{table}[!ht]
\scriptsize
\caption{Observable dependence on minimum height for dyads with minimum age in the 20-29 year range. Lengths in millimetres, times in seconds.}
\label{table5cm}
\begin{center}
\begin{tabular} {|c|c|c|c|c|c|}
\hline
Minimum height&  $N^k_g$ &  $V$ &    $r$ & $x$  & $|y|$ \\
\hline
140-150 cm & 2 & 1161 $\pm$ 180   ($\sigma$=254) &748 $\pm$ 62  ($\sigma$=87.2) &613 $\pm$ 41 ($\sigma$=58.5) &398 $\pm$ 54 ($\sigma$=75.7)    \\ 
\hline
150-160 cm & 75 & 1157 $\pm$ 21   ($\sigma$=184) &800 $\pm$ 27  ($\sigma$=232) &668 $\pm$ 14 ($\sigma$=122) &311 $\pm$ 35 ($\sigma$=307)    \\ 
\hline
160-170 cm & 188 & 1175 $\pm$ 13   ($\sigma$=176) &781 $\pm$ 17  ($\sigma$=231) &658 $\pm$ 12 ($\sigma$=165) &300 $\pm$ 19 ($\sigma$=265)    \\ 
\hline
170-180 cm & 95 & 1209 $\pm$ 16   ($\sigma$=152) &808 $\pm$ 18  ($\sigma$=173) &688 $\pm$ 16 ($\sigma$=156) &309 $\pm$ 22 ($\sigma$=219)    \\ 
\hline
$>$ 180 cm & 3 & 1262 $\pm$ 130   ($\sigma$=223) &974 $\pm$ 170  ($\sigma$=300) &625 $\pm$ 9.5 ($\sigma$=16.4) &556 $\pm$ 220 ($\sigma$=373)    \\ 
\hline
$F_{4,358}$ & & 1.19 & 0.818 & 0.684 & 0.756\\
\hline
$p$ & & 0.315 & 0.514 & 0.603 & 0.554\\
\hline
$R^2$ & & 0.0131 & 0.00906 & 0.00759 & 0.00838\\
\hline
$\delta$ & & 0.567 & 0.906 & 0.482 & 0.962\\
\hline
\end{tabular}
\end{center}
\end{table}

\begin{table}[!ht]
\scriptsize
\caption{Observable dependence on minimum height for dyads with average age in the 50-59 year range. Lengths in millimetres, times in seconds.}
\label{table5cn}
\begin{center}
\begin{tabular} {|c|c|c|c|c|c|}
\hline
Minimum height&  $N^k_g$ &  $V$ &    $r$ & $x$  & $|y|$ \\
\hline
140-150 cm & 3 & 1043 $\pm$ 41   ($\sigma$=70.2) &784 $\pm$ 56  ($\sigma$=97) &746 $\pm$ 63 ($\sigma$=109) &190 $\pm$ 16 ($\sigma$=28)    \\ 
\hline
150-160 cm & 19 & 1067 $\pm$ 39   ($\sigma$=171) &757 $\pm$ 34  ($\sigma$=146) &611 $\pm$ 36 ($\sigma$=156) &334 $\pm$ 52 ($\sigma$=228)    \\ 
\hline
160-170 cm & 59 & 1162 $\pm$ 25   ($\sigma$=191) &830 $\pm$ 29  ($\sigma$=226) &664 $\pm$ 22 ($\sigma$=165) &371 $\pm$ 41 ($\sigma$=315)    \\ 
\hline
170-180 cm & 30 & 1241 $\pm$ 33   ($\sigma$=182) &862 $\pm$ 44  ($\sigma$=238) &635 $\pm$ 27 ($\sigma$=148) &436 $\pm$ 68 ($\sigma$=371)    \\ 
\hline
$F_{3,107}$ & & 3.84 & 0.928 & 0.974 & 0.806\\
\hline
$p$ & & 0.0118 & 0.43 & 0.408 & 0.493\\
\hline
$R^2$ & & 0.0972 & 0.0254 & 0.0266 & 0.0221\\
\hline
$\delta$ & & 1.12 & 0.502 & 0.886 & 0.687\\
\hline
\end{tabular}
\end{center}
\end{table}
\begin{table}[!ht]
\scriptsize
\caption{Minimum height $p$ values in different minimum age ranges. Lengths in millimetres, times in seconds.}
\label{table5cex3}
\begin{center}
\begin{tabular} {|c|c|c|c|c|}
\hline
Minimum age &  $V$ &    $r$ & $x$  & $|y|$ \\
\hline
0-9 years&  0.000332 & 0.143 & 0.662 & 0.127\\
\hline
10-19 years&  0.311 & 0.822 & 0.478 & 0.926\\
\hline
20-29 years&  0.315 & 0.514 & 0.603 & 0.554\\
\hline
30-39 years&  5$\cdot 10^{-7}$ & 0.595 & 0.00388 & 0.0423\\
\hline
40-49 years&  0.000142 & 0.489 & 0.00545 & 0.0649\\
\hline
50-59 years&  0.0118 & 0.43 & 0.408 & 0.493\\
\hline
60-69 years&  0.091 & 0.23 & 0.24 & 0.182\\
\hline
$\geq$ 70 years&  0.627 & 0.0424 & 0.0506 & 0.107\\
\hline
\end{tabular}
\end{center}
\end{table}
\begin{table}[!ht]
\scriptsize
\caption{Minimum height $\delta$ values in different minimum age ranges. Lengths in millimetres, times in seconds.}
\label{table5cex4}
\begin{center}
\begin{tabular} {|c|c|c|c|c|}
\hline
Minimum age &  $V$ &    $r$ & $x$  & $|y|$ \\
\hline
0-9 years &  3.42 & 1.11 & 2 & 1.04\\
\hline
10-19 years &  0.883 & 0.346 & 0.519 & 0.275\\
\hline
20-29 years &  0.567 & 0.906 & 0.482 & 0.962\\
\hline
30-39 years &  2.32 & 0.943 & 0.702 & 1.28\\
\hline
40-49 years &  2.48 & 0.671 & 0.993 & 0.944\\
\hline
50-59 years &  1.12 & 0.502 & 0.886 & 0.687\\
\hline
60-69 years &  0.89 & 0.436 & 0.645 & 0.597\\
\hline
$\geq$ 70 years & 1.08 & 2.05 & 2.01 & 1.7\\
\hline
\end{tabular}
\end{center}
\end{table}
\clearpage
\section{Coder reliability}
\label{coderel}
\subsection{Analysis of coder agreement}
\label{coderag}
We consider a few possible statistical indicators of agreement between coders.
\subsubsection{Cohen's $\kappa$}
Cohen's $\kappa$ \cite{cohen} is a very popular indicator to compare the agreement between two coders, based on the equation
\begin{equation}
\label{kappa}
\kappa=(p-p_r)/(1-p_r),
\end{equation}
where $p$ stands for the agreement rate between coders and $p_r$ for the probability of random agreement.
The agreement between pairs of coders according to this statistics is shown in table \ref{tablecohen}.

\begin{table}[!ht]
\scriptsize
\caption{Agreement between pairs of coders according to Cohen's $\kappa$ statistics. $C_i - C_j$ stands for agreement between coder $i$ and $j$.}
\label{tablecohen}
\begin{center}
\begin{tabular} {|c|c|c|c|c|c|c|}
\hline
Pair & Purpose & Gender & Relation & Min Age & Avg Age & Max Age \\
	\hline
$C_1 - C_2$ &  0.815 & 0.961 & 0.636 & 0.476 & 0.582 & 0.555\\
	\hline
$C_1 - C_3$ &  0.923 & 0.978 & 0.728 & 0.808 & 0.839 & 0.866 \\
	\hline
$C_2 - C_3$ &  0.810 & 0.944 & 0.647 & 0.449 & 0.508 & 0.526\\
	\hline
\end{tabular}
\end{center}
\end{table}

These results show that in general the agreement is higher for gender, followed by purpose and relation. The agreement between coders 1 and 3 is similar also concerning age, while the agreement 
with coder 2 is quite poor in these categories. Although there is no real sound mathematical way to evaluate the absolute value of these numbers, according to popular benchmarks,
an agreement between 0.8 and 1 is considered as ``almost perfect'', an agreement between 0.6 and 0.8 as ``substantial'', while an agreement between 0.4 and 0.6 is only ``moderate'' \cite{cohen2}

 \subsubsection{Fleiss' $\kappa$}
It generalises eq. \ref{kappa} to deal with multiple coders and categories \cite{fleiss}. The corresponding values are shown in table \ref{tablefleiss}

\begin{table}[!ht]
\scriptsize
\caption{Agreement between coders according to Fleiss' $\kappa$ statistics.}
\label{tablefleiss}
\begin{center}
\begin{tabular} {|c|c|c|c|c|c|}
\hline
  Purpose & Gender & Relation & Min Age & Avg Age & Max Age \\
	\hline
  0.849 & 0.961 & 0.669 & 0.289 & 0.332 & 0.300\\
	\hline
\end{tabular}
\end{center}
\end{table}

We see that, in relative terms, agreement is higher for gender, followed by purpose and relation, and lowest for age. In absolute terms, according to the benchmarks,
we have almost perfect agreement in gender and purpose, substantial in relation and ``fair'' (i.e., worst than ``moderate'') for age indicators, due to the effect of the different coding
by coder 2.

\begin{figure}[ht!]
\begin{center}
\includegraphics[width=1\linewidth]{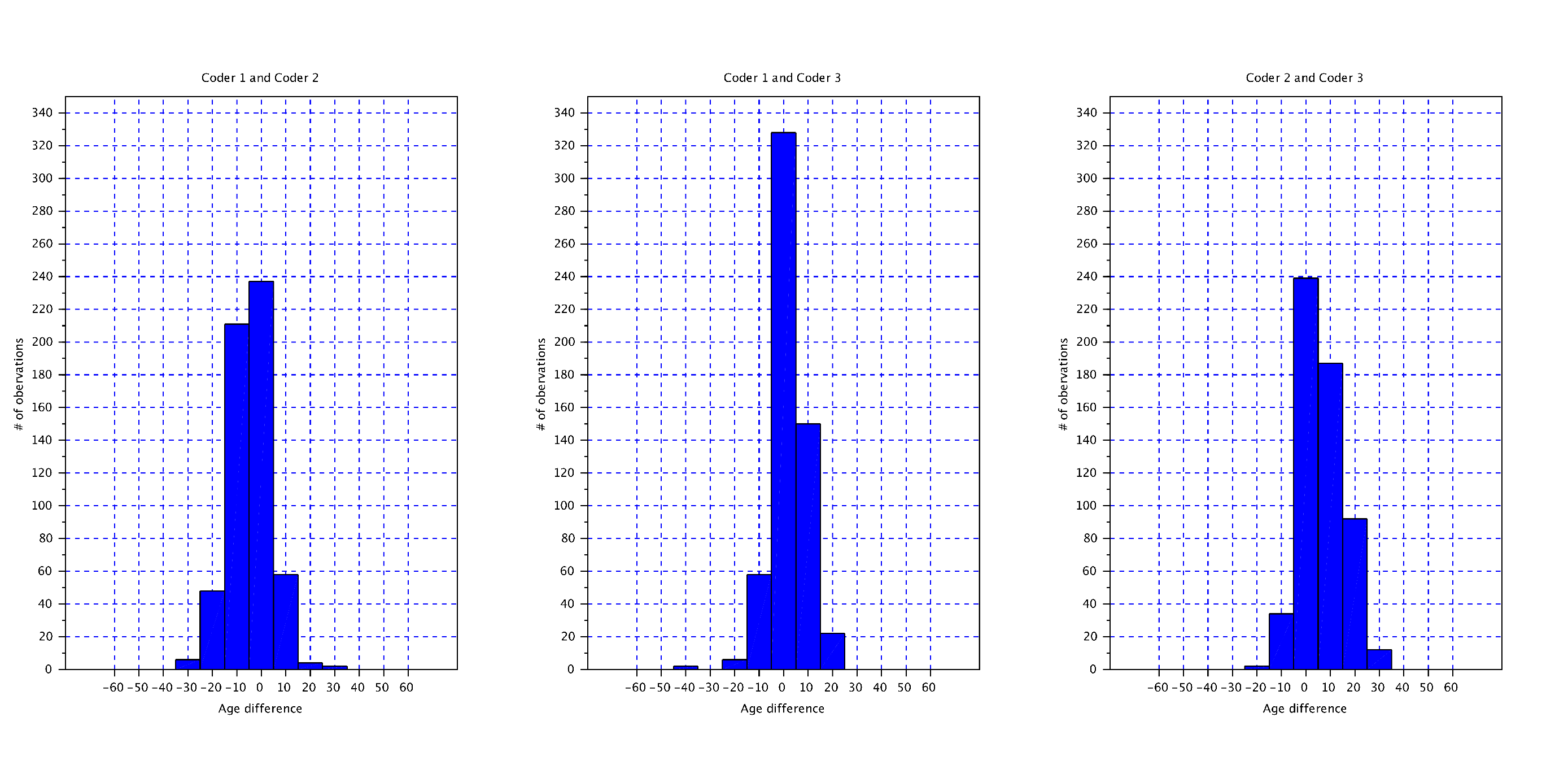} 
\caption{Histograms of age differences between coders.}
\label{fzey}
\end{center}
\end{figure}

Anyway, if we try to plot the age difference between coders, as in figure \ref{fzey}, we see that although disagreement with coder 2 is substantial, it is almost completely limited to
a tendency of coder 2 to put pedestrians in a slightly younger category, i.e. the difference in age between the codings is limited. Nevertheless the Fleiss indicator does not take
in account the magnitude of difference, and is thus not completely adequate to deal with ordered data.

\subsubsection{Krippendorff's $\alpha$}
The Krippendorff $\alpha$ statistics \cite{Krip}, that allows for consideration of quantitative differences between coding results, gives the  results shown in table \ref{tablekrip}.
\begin{table}[!ht]
\scriptsize
\caption{Agreement between coders according to Krippendorff's $\alpha$ statistics. Purpose, gender and relation are ``nominal'' data, age is on an ``interval'', according to the definition of
$\alpha$ statistics.}
\label{tablekrip}
\begin{center}
\begin{tabular} {|c|c|c|c|c|c|}
\hline
  Purpose & Gender & Relation & Min Age & Avg Age & Max Age \\
	\hline
  0.849 & 0.961 & 0.669 & 0.709 & 0.730 & 0.729\\
	\hline
\end{tabular}
\end{center}
\end{table}

Krippendorff does not provide any ``magic number'' but suggests to use data with at least $\alpha>0.667$ (satisfied by all our categories) and require $\alpha>0.8$,
satisfied by purpose and gender,
 for reliable results ($\alpha$ between 0.667 and 0.8 could be used for ``tentative conclusions'').
\subsubsection{Discussion}
Using popular indicators of coder reliability, we have found that, in relative terms, the most reliable coding regards gender, followed by purpose. In absolute terms, according to the
Krippendorff $\alpha$ statistics that can better cope with the nature of our data, we may see that the purpose and gender codings may be considered as enough reliable to
provide sound findings, while the relation and age codings are reliable enough for reporting tentative findings.

The analysis based on these indicators provides an estimate on the reliability of coding of pedestrians in different categories. We may nevertheless use another approach to test the reliability 
of our findings when based on different coding processes. Since for each category we analyse the values of the observables $V$, $r$, $x$ and $|y|$, we may compare these quantitative results between different coders. 

This comparison, which has also the advantage of being based on more mathematically sound statistical indicators (standard errors, ANOVA analysis) is performed in section
\ref{codercomp}, and shows again that for purpose and gender we have an almost perfect quantitative agreement, while for relation and age, although the agreement is less good, the 
major patterns of behaviour are qualitatively observed regardless of coders.
\subsection{Quantitative comparison of results}
\label{codercomp}
\subsubsection{Purpose}
\label{coderpur}
The results (on the common subset of data) for the purpose dependence of all observables between the main coder (coder 1) and the secondary coders are compared in tables \ref{table1f1},
\ref{table1f2} and \ref{table1f3}.

\begin{table}[!ht]
\scriptsize
\caption{Observable dependence on purpose for dyads according to coder 1 (common data set only). Lengths in millimetres, times in seconds.}
\label{table1f1}
\begin{center}
\begin{tabular} {|c|c|c|c|c|c|}
\hline
Purpose &  $N^k_g$ &  $V$ &    $r$ & $x$  & $|y|$ \\
\hline
Leisure & 136 & 1085 $\pm$ 19   ($\sigma$=220) &796 $\pm$ 21  ($\sigma$=248) &636 $\pm$ 13 ($\sigma$=151) &351 $\pm$ 28 ($\sigma$=327)    \\ 
\hline
Work & 132 & 1257 $\pm$ 14   ($\sigma$=157) &829 $\pm$ 17  ($\sigma$=196) &723 $\pm$ 12 ($\sigma$=143) &303 $\pm$ 21 ($\sigma$=241)    \\ 
\hline
$F_{1,266}$ & & 53.1 & 1.41 & 23.5 & 1.88\\
\hline
$p$ & & $<10^{-8}$ & 0.236 & 2.14$\cdot 10^{-6}$ & 0.171\\
\hline
$R^2$ & & 0.166 & 0.00529 & 0.0811 & 0.00703\\
\hline
$\delta$ & & 0.893 & 0.146 & 0.594 & 0.168\\
\hline
\end{tabular}
\end{center}
\end{table}

\begin{table}[!ht]
\scriptsize
\caption{Observable dependence on purpose for dyads according to coder 2 (common data set only). Lengths in millimetres, times in seconds.}
\label{table1f2}
\begin{center}
\begin{tabular} {|c|c|c|c|c|c|}
\hline
Purpose &  $N^k_g$ &  $V$ &    $r$ & $x$  & $|y|$ \\
\hline
Leisure & 151 & 1093 $\pm$ 17   ($\sigma$=212) &793 $\pm$ 20  ($\sigma$=243) &641 $\pm$ 12 ($\sigma$=147) &344 $\pm$ 26 ($\sigma$=318)    \\ 
\hline
Work & 117 & 1269 $\pm$ 15   ($\sigma$=159) &837 $\pm$ 18  ($\sigma$=196) &728 $\pm$ 13 ($\sigma$=146) &306 $\pm$ 22 ($\sigma$=243)    \\ 
\hline
$F_{1,266}$ & & 56.2 & 2.56 & 23.4 & 1.13\\
\hline
$p$ & & $<10^{-8}$ & 0.111 & 2.18$\cdot 10^{-6}$ & 0.289\\
\hline
$R^2$ & & 0.175 & 0.00954 & 0.081 & 0.00422\\
\hline
$\delta$ & & 0.927 & 0.198 & 0.599 & 0.131\\
\hline
\end{tabular}
\end{center}
\end{table}

\begin{table}[!ht]
\scriptsize
\caption{Observable dependence on purpose for dyads according to coder 3 (common data set only). Lengths in millimetres, times in seconds.}
\label{table1f3}
\begin{center}
\begin{tabular} {|c|c|c|c|c|c|}
\hline
Purpose &  $N^k_g$ &  $V$ &    $r$ & $x$  & $|y|$ \\
\hline
Leisure & 133 & 1077 $\pm$ 19   ($\sigma$=217) &789 $\pm$ 22  ($\sigma$=250) &626 $\pm$ 13 ($\sigma$=145) &354 $\pm$ 29 ($\sigma$=330)    \\ 
\hline
Work & 133 & 1262 $\pm$ 14   ($\sigma$=156) &836 $\pm$ 17  ($\sigma$=195) &732 $\pm$ 12 ($\sigma$=144) &302 $\pm$ 21 ($\sigma$=239)    \\ 
\hline
$F_{1,264}$ & & 63.6 & 2.93 & 35.6 & 2.13\\
\hline
$p$ & & $<10^{-8}$ & 0.0881 & $<10^{-8}$ & 0.145\\
\hline
$R^2$ & & 0.194 & 0.011 & 0.119 & 0.00802\\
\hline
$\delta$ & & 0.982 & 0.211 & 0.734 & 0.18\\
\hline
\end{tabular}
\end{center}
\end{table}
The differences between coders are thus always of one standard error or smaller, and the extremely significant statistical differences in the $x$ and $V$ distribution (along with the less
significant $|y|$ and $r$ ones) are reported by all coders.

\subsubsection{Relation}
\label{coderrel}
The results (on the common subset of data) for the relation dependence of all observables between the main coder (coder 1) and the secondary coders are compared in tables \ref{table2f1},
\ref{table2f2} and \ref{table2f3}. While all the major trends exposed in the main text are confirmed, quantitative results between coders may sometimes be different (we refer in particular to
the $|y|$ distribution for couples, extremely narrow according to coder 3).

\begin{table}[!ht]
\scriptsize
\caption{Observable dependence on relation for dyads according to coder 1 (common data set only). Lengths in millimetres, times in seconds.}
\label{table2f1}
\begin{center}
\begin{tabular} {|c|c|c|c|c|c|}
\hline
Relation&  $N^k_g$ &  $V$ &    $r$ & $x$  & $|y|$ \\
\hline
Colleagues & 125 & 1256 $\pm$ 14   ($\sigma$=154) &829 $\pm$ 18  ($\sigma$=196) &725 $\pm$ 13 ($\sigma$=142) &301 $\pm$ 21 ($\sigma$=239)    \\ 
\hline
Couples & 28 & 1087 $\pm$ 37   ($\sigma$=194) &690 $\pm$ 33  ($\sigma$=174) &611 $\pm$ 21 ($\sigma$=112) &248 $\pm$ 37 ($\sigma$=198)    \\ 
\hline
Families & 40 & 1051 $\pm$ 24   ($\sigma$=153) &864 $\pm$ 54  ($\sigma$=341) &594 $\pm$ 21 ($\sigma$=134) &492 $\pm$ 69 ($\sigma$=438)    \\ 
\hline
Friends & 56 & 1121 $\pm$ 36   ($\sigma$=271) &777 $\pm$ 24  ($\sigma$=182) &669 $\pm$ 19 ($\sigma$=145) &286 $\pm$ 32 ($\sigma$=243)    \\ 
\hline
$F_{3,245}$ & & 16.4 & 4.19 & 11.8 & 6.12\\
\hline
$p$ & & $<10^{-8}$ & 0.00651 & 3.06$\cdot 10^{-7}$ & 0.0005\\
\hline
$R^2$ & & 0.167 & 0.0488 & 0.126 & 0.0697\\
\hline
$\delta$ & & 1.33 & 0.612 & 0.934 & 0.678\\
\hline
\end{tabular}
\end{center}
\end{table}

\begin{table}[!ht]
\scriptsize
\caption{Observable dependence on relation for dyads according to coder 2 (common data set only). Lengths in millimetres, times in seconds.}
\label{table2f2}
\begin{center}
\begin{tabular} {|c|c|c|c|c|c|}
\hline
Relation&  $N^k_g$ &  $V$ &    $r$ & $x$  & $|y|$ \\
\hline
Colleagues & 116 & 1267 $\pm$ 14   ($\sigma$=156) &839 $\pm$ 18  ($\sigma$=197) &729 $\pm$ 14 ($\sigma$=147) &308 $\pm$ 23 ($\sigma$=244)    \\ 
\hline
Couples & 44 & 1082 $\pm$ 28   ($\sigma$=184) &703 $\pm$ 21  ($\sigma$=140) &582 $\pm$ 19 ($\sigma$=125) &296 $\pm$ 33 ($\sigma$=221)    \\ 
\hline
Families & 42 & 1054 $\pm$ 25   ($\sigma$=164) &894 $\pm$ 53  ($\sigma$=341) &651 $\pm$ 25 ($\sigma$=163) &451 $\pm$ 70 ($\sigma$=457)    \\ 
\hline
Friends & 66 & 1131 $\pm$ 31   ($\sigma$=254) &786 $\pm$ 23  ($\sigma$=188) &673 $\pm$ 17 ($\sigma$=136) &304 $\pm$ 29 ($\sigma$=238)    \\ 
\hline
$F_{3,264}$ & & 19 & 6.55 & 11.9 & 3.13\\
\hline
$p$ & & $<10^{-8}$ & 0.000276 & 2.54$\cdot 10^{-7}$ & 0.0262\\
\hline
$R^2$ & & 0.178 & 0.0692 & 0.119 & 0.0344\\
\hline
$\delta$ & & 1.35 & 0.74 & 1.04 & 0.437\\
\hline
\end{tabular}
\end{center}
\end{table}

\begin{table}[!ht]
\scriptsize
\caption{Observable dependence on relation for dyads according to coder 3 (common data set only) Lengths in millimetres, times in seconds..}
\label{table2f3}
\begin{center}
\begin{tabular} {|c|c|c|c|c|c|}
\hline
Relation&  $N^k_g$ &  $V$ &    $r$ & $x$  & $|y|$ \\
\hline
Colleagues & 136 & 1259 $\pm$ 14   ($\sigma$=158) &834 $\pm$ 17  ($\sigma$=194) &727 $\pm$ 13 ($\sigma$=147) &304 $\pm$ 21 ($\sigma$=242)    \\ 
\hline
Couples & 23 & 1070 $\pm$ 42   ($\sigma$=204) &624 $\pm$ 20  ($\sigma$=96.4) &578 $\pm$ 20 ($\sigma$=95.1) &182 $\pm$ 17 ($\sigma$=81.2)    \\ 
\hline
Families & 50 & 1053 $\pm$ 24   ($\sigma$=172) &867 $\pm$ 44  ($\sigma$=312) &612 $\pm$ 20 ($\sigma$=140) &478 $\pm$ 59 ($\sigma$=416)    \\ 
\hline
Friends & 54 & 1084 $\pm$ 32   ($\sigma$=235) &780 $\pm$ 27  ($\sigma$=196) &663 $\pm$ 22 ($\sigma$=159) &298 $\pm$ 33 ($\sigma$=245)    \\ 
\hline
$F_{3,259}$ & & 23.4 & 7.57 & 12.4 & 7.52\\
\hline
$p$ & & $<10^{-8}$ & 7.11$\cdot 10^{-5}$ & 1.36$\cdot 10^{-7}$ & 7.61$\cdot 10^{-5}$\\
\hline
$R^2$ & & 0.213 & 0.0807 & 0.125 & 0.0801\\
\hline
$\delta$ & & 1.27 & 0.915 & 1.06 & 0.849\\
\hline
\end{tabular}
\end{center}
\end{table}

\subsubsection{Gender}
\label{codergen}
The results (on the common subset of data) for the gender dependence of all observables between the main coder (coder 1) and the secondary coders are compared in tables \ref{table3f1},
\ref{table3f2} and \ref{table3f3}, showing that there is basically no difference in the coding of gender. 

\begin{table}[!ht]
\scriptsize
\caption{Observable dependence on gender for dyads according to coder 1 (common data set only). Lengths in millimetres, times in seconds.}
\label{table3f1}
\begin{center}
\begin{tabular} {|c|c|c|c|c|c|}
\hline
Gender&  $N^k_g$ &  $V$ &    $r$ & $x$  & $|y|$ \\
\hline
Two females & 55 & 1076 $\pm$ 32   ($\sigma$=240) &745 $\pm$ 21  ($\sigma$=155) &629 $\pm$ 15 ($\sigma$=112) &290 $\pm$ 33 ($\sigma$=242)    \\ 
\hline
Mixed & 86 & 1095 $\pm$ 19   ($\sigma$=173) &820 $\pm$ 31  ($\sigma$=287) &641 $\pm$ 17 ($\sigma$=159) &384 $\pm$ 39 ($\sigma$=360)    \\ 
\hline
Two males & 127 & 1261 $\pm$ 16   ($\sigma$=178) &836 $\pm$ 17  ($\sigma$=195) &727 $\pm$ 13 ($\sigma$=150) &305 $\pm$ 22 ($\sigma$=243)    \\ 
\hline
$F_{2,265}$ & & 27.2 & 3.25 & 12.8 & 2.48\\
\hline
$p$ & & $<10^{-8}$ & 0.0404 & 5.09$\cdot 10^{-6}$ & 0.0855\\
\hline
$R^2$ & & 0.171 & 0.0239 & 0.0879 & 0.0184\\
\hline
$\delta$ & & 0.93 & 0.494 & 0.699 & 0.292\\
\hline
\end{tabular}
\end{center}
\end{table}

\begin{table}[!ht]
\scriptsize
\caption{Observable dependence on gender for dyads according to coder 2 (common data set only). Lengths in millimetres, times in seconds.}
\label{table3f2}
\begin{center}
\begin{tabular} {|c|c|c|c|c|c|}
\hline
Gender&  $N^k_g$ &  $V$ &    $r$ & $x$  & $|y|$ \\
\hline
Two females & 53 & 1078 $\pm$ 33   ($\sigma$=241) &747 $\pm$ 22  ($\sigma$=158) &637 $\pm$ 15 ($\sigma$=106) &286 $\pm$ 32 ($\sigma$=233)    \\ 
\hline
Mixed & 89 & 1093 $\pm$ 18   ($\sigma$=173) &814 $\pm$ 30  ($\sigma$=283) &635 $\pm$ 17 ($\sigma$=159) &382 $\pm$ 38 ($\sigma$=360)    \\ 
\hline
Two males & 126 & 1263 $\pm$ 16   ($\sigma$=177) &838 $\pm$ 17  ($\sigma$=194) &728 $\pm$ 13 ($\sigma$=150) &306 $\pm$ 22 ($\sigma$=244)    \\ 
\hline
$F_{2,265}$ & & 28.2 & 3.12 & 13.3 & 2.48\\
\hline
$p$ & & $<10^{-8}$ & 0.0459 & 3.22$\cdot 10^{-6}$ & 0.0853\\
\hline
$R^2$ & & 0.176 & 0.023 & 0.091 & 0.0184\\
\hline
$\delta$ & & 0.935 & 0.494 & 0.604 & 0.3\\
\hline
\end{tabular}
\end{center}
\end{table}

\begin{table}[!ht]
\scriptsize
\caption{Observable dependence on gender for dyads according to coder 3 (common data set only). Lengths in millimetres, times in seconds.}
\label{table3f3}
\begin{center}
\begin{tabular} {|c|c|c|c|c|c|}
\hline
Gender&  $N^k_g$ &  $V$ &    $r$ & $x$  & $|y|$ \\
\hline
Two females & 55 & 1074 $\pm$ 32   ($\sigma$=239) &742 $\pm$ 21  ($\sigma$=153) &636 $\pm$ 14 ($\sigma$=103) &281 $\pm$ 31 ($\sigma$=230)    \\ 
\hline
Mixed & 89 & 1093 $\pm$ 19   ($\sigma$=175) &824 $\pm$ 31  ($\sigma$=288) &634 $\pm$ 17 ($\sigma$=161) &397 $\pm$ 39 ($\sigma$=368)    \\ 
\hline
Two males & 124 & 1267 $\pm$ 16   ($\sigma$=173) &834 $\pm$ 17  ($\sigma$=190) &730 $\pm$ 13 ($\sigma$=150) &298 $\pm$ 21 ($\sigma$=232)    \\ 
\hline
$F_{2,265}$ & & 30.4 & 3.44 & 14.2 & 3.99\\
\hline
$p$ & & $<10^{-8}$ & 0.0336 & 1.36$\cdot 10^{-6}$ & 0.0196\\
\hline
$R^2$ & & 0.187 & 0.0253 & 0.0969 & 0.0293\\
\hline
$\delta$ & & 0.987 & 0.511 & 0.622 & 0.359\\
\hline
\end{tabular}
\end{center}
\end{table}
\subsubsection{Age}
\label{coderage}
The results (on the common subset of data) for the minimum age dependence of all observables between the main coder (coder 1) and the secondary coders are compared in tables \ref{table4f1},
\ref{table4f2} and \ref{table4f3}. Sadly, almost no groups with children are present in the common set. The drop in velocity with age is, on the other hand, confirmed in 
a statistically significant way by all coders.

\begin{table}[!ht]
\scriptsize
\caption{Observable dependence on minimum age for dyads according to coder 1 (common data set only). Lengths in millimetres, times in seconds.}
\label{table4f1}
\begin{center}
\begin{tabular} {|c|c|c|c|c|c|}
\hline
Minimum age&  $N^k_g$ &  $V$ &    $r$ & $x$  & $|y|$ \\
\hline
10-19 years & 16 & 1157 $\pm$ 86   ($\sigma$=343) &715 $\pm$ 31  ($\sigma$=123) &653 $\pm$ 23 ($\sigma$=92.3) &223 $\pm$ 38 ($\sigma$=151)    \\ 
\hline
20-29 years & 58 & 1183 $\pm$ 28   ($\sigma$=215) &765 $\pm$ 28  ($\sigma$=211) &666 $\pm$ 20 ($\sigma$=149) &268 $\pm$ 33 ($\sigma$=252)    \\ 
\hline
30-39 years & 96 & 1186 $\pm$ 21   ($\sigma$=203) &817 $\pm$ 21  ($\sigma$=211) &689 $\pm$ 17 ($\sigma$=166) &327 $\pm$ 27 ($\sigma$=262)    \\ 
\hline
40-49 years & 41 & 1193 $\pm$ 25   ($\sigma$=161) &811 $\pm$ 27  ($\sigma$=173) &684 $\pm$ 22 ($\sigma$=143) &327 $\pm$ 38 ($\sigma$=245)    \\ 
\hline
50-59 years & 31 & 1210 $\pm$ 29   ($\sigma$=160) &880 $\pm$ 46  ($\sigma$=254) &696 $\pm$ 29 ($\sigma$=160) &407 $\pm$ 65 ($\sigma$=360)    \\ 
\hline
60-69 years & 21 & 1017 $\pm$ 35   ($\sigma$=160) &869 $\pm$ 66  ($\sigma$=304) &671 $\pm$ 34 ($\sigma$=156) &401 $\pm$ 85 ($\sigma$=388)    \\ 
\hline
$\geq$ 70 years & 5 & 949 $\pm$ 15   ($\sigma$=34.2) &913 $\pm$ 170  ($\sigma$=379) &608 $\pm$ 28 ($\sigma$=61.7) &551 $\pm$ 210 ($\sigma$=470)    \\ 
\hline
$F_{6,261}$ & & 3.35 & 1.81 & 0.462 & 1.91\\
\hline
$p$ & & 0.00337 & 0.0974 & 0.836 & 0.0789\\
\hline
$R^2$ & & 0.0715 & 0.0399 & 0.0105 & 0.0421\\
\hline
$\delta$ & & 1.73 & 0.964 & 0.578 & 1.29\\
\hline
\end{tabular}
\end{center}
\end{table}

\begin{table}[!ht]
\scriptsize
\caption{Observable dependence on minimum age for dyads according to coder 2 (common data set only). Lengths in millimetres, times in seconds.}
\label{table4f2}
\begin{center}
\begin{tabular} {|c|c|c|c|c|c|}
\hline
Minimum age&  $N^k_g$ &  $V$ &    $r$ & $x$  & $|y|$ \\
\hline
0-9 years & 2 & 1190 $\pm$ 220   ($\sigma$=312) &749 $\pm$ 110  ($\sigma$=152) &700 $\pm$ 87 ($\sigma$=123) &202 $\pm$ 55 ($\sigma$=77.8)    \\ 
\hline
10-19 years & 16 & 1169 $\pm$ 84   ($\sigma$=334) &682 $\pm$ 20  ($\sigma$=80.1) &646 $\pm$ 20 ($\sigma$=78.4) &172 $\pm$ 18 ($\sigma$=73.3)    \\ 
\hline
20-29 years & 107 & 1163 $\pm$ 19   ($\sigma$=196) &765 $\pm$ 16  ($\sigma$=165) &655 $\pm$ 13 ($\sigma$=138) &288 $\pm$ 22 ($\sigma$=231)    \\ 
\hline
30-39 years & 78 & 1217 $\pm$ 24   ($\sigma$=209) &869 $\pm$ 28  ($\sigma$=244) &727 $\pm$ 21 ($\sigma$=185) &362 $\pm$ 33 ($\sigma$=290)    \\ 
\hline
40-49 years & 32 & 1181 $\pm$ 31   ($\sigma$=176) &855 $\pm$ 44  ($\sigma$=249) &645 $\pm$ 22 ($\sigma$=124) &418 $\pm$ 66 ($\sigma$=373)    \\ 
\hline
50-59 years & 24 & 1074 $\pm$ 32   ($\sigma$=158) &853 $\pm$ 60  ($\sigma$=293) &706 $\pm$ 29 ($\sigma$=143) &343 $\pm$ 74 ($\sigma$=363)    \\ 
\hline
60-69 years & 9 & 1047 $\pm$ 42   ($\sigma$=127) &861 $\pm$ 100  ($\sigma$=311) &655 $\pm$ 45 ($\sigma$=135) &432 $\pm$ 120 ($\sigma$=375)    \\ 
\hline
$F_{6,261}$ & & 2.1 & 3.09 & 2.3 & 2.13\\
\hline
$p$ & & 0.0533 & 0.0061 & 0.0349 & 0.0505\\
\hline
$R^2$ & & 0.0461 & 0.0663 & 0.0503 & 0.0467\\
\hline
$\delta$ & & 0.842 & 0.833 & 0.482 & 1.14\\
\hline
\end{tabular}
\end{center}
\end{table}

\begin{table}[!ht]
\scriptsize
\caption{Observable dependence on minimum age for dyads according to coder 3 (common data set only). Lengths in millimetres, times in seconds.}
\label{table4f3}
\begin{center}
\begin{tabular} {|c|c|c|c|c|c|}
\hline
Minimum age&  $N^k_g$ &  $V$ &    $r$ & $x$  & $|y|$ \\
\hline
10-19 years & 14 & 1163 $\pm$ 98   ($\sigma$=367) &701 $\pm$ 31  ($\sigma$=117) &623 $\pm$ 35 ($\sigma$=130) &218 $\pm$ 48 ($\sigma$=181)    \\ 
\hline
20-29 years & 64 & 1157 $\pm$ 29   ($\sigma$=236) &758 $\pm$ 24  ($\sigma$=194) &658 $\pm$ 20 ($\sigma$=158) &274 $\pm$ 27 ($\sigma$=220)    \\ 
\hline
30-39 years & 50 & 1197 $\pm$ 27   ($\sigma$=193) &830 $\pm$ 32  ($\sigma$=227) &685 $\pm$ 23 ($\sigma$=162) &349 $\pm$ 43 ($\sigma$=302)    \\ 
\hline
40-49 years & 77 & 1205 $\pm$ 19   ($\sigma$=163) &832 $\pm$ 23  ($\sigma$=205) &684 $\pm$ 16 ($\sigma$=141) &351 $\pm$ 33 ($\sigma$=293)    \\ 
\hline
50-59 years & 36 & 1205 $\pm$ 25   ($\sigma$=152) &820 $\pm$ 35  ($\sigma$=207) &722 $\pm$ 28 ($\sigma$=168) &300 $\pm$ 39 ($\sigma$=233)    \\ 
\hline
60-69 years & 20 & 1025 $\pm$ 40   ($\sigma$=179) &903 $\pm$ 74  ($\sigma$=332) &699 $\pm$ 29 ($\sigma$=129) &418 $\pm$ 97 ($\sigma$=436)    \\ 
\hline
$\geq$ 70 years & 7 & 956 $\pm$ 26   ($\sigma$=69.1) &881 $\pm$ 120  ($\sigma$=326) &605 $\pm$ 36 ($\sigma$=96.4) &503 $\pm$ 140 ($\sigma$=382)    \\ 
\hline
$F_{6,261}$ & & 3.69 & 2.04 & 1.33 & 1.67\\
\hline
$p$ & & 0.00153 & 0.0604 & 0.245 & 0.129\\
\hline
$R^2$ & & 0.0783 & 0.0449 & 0.0296 & 0.0369\\
\hline
$\delta$ & & 1.74 & 0.755 & 0.732 & 1.09\\
\hline
\end{tabular}
\end{center}
\end{table}

\clearpage
\section{Dependence on average and maximum age}
\label{maxavage}
\subsection{Average age}
\begin{figure}[ht!]
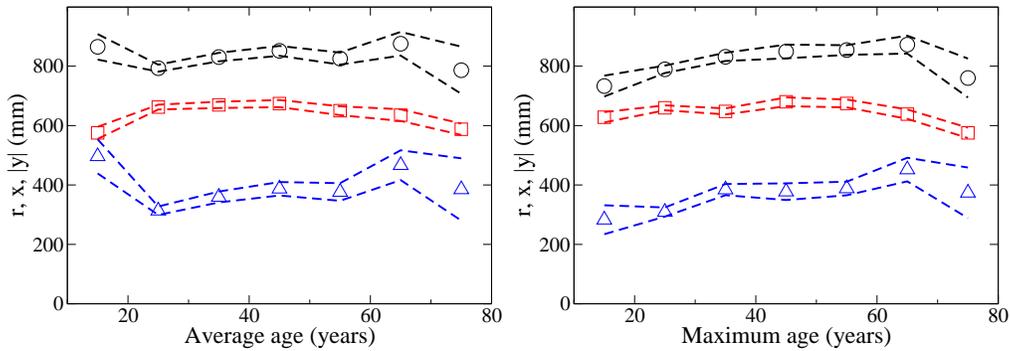

\begin{center}
\includegraphics[width=0.45\linewidth]{f31.eps} \vspace{0.1\linewidth} \includegraphics[width=0.45\linewidth]{f33.eps} 
\caption{$r$ (black and circles), $x$ (red and squares) and $|y|$ (blue and triangles) dependence on average (left) and maximum (right) age. 
Dashed lines provide standard error confidence bars. The point at 75 years corresponds to the ``70 years or more'' slot.}
\label{f4a_av}
\end{center}
\end{figure}

\begin{figure}[ht!]
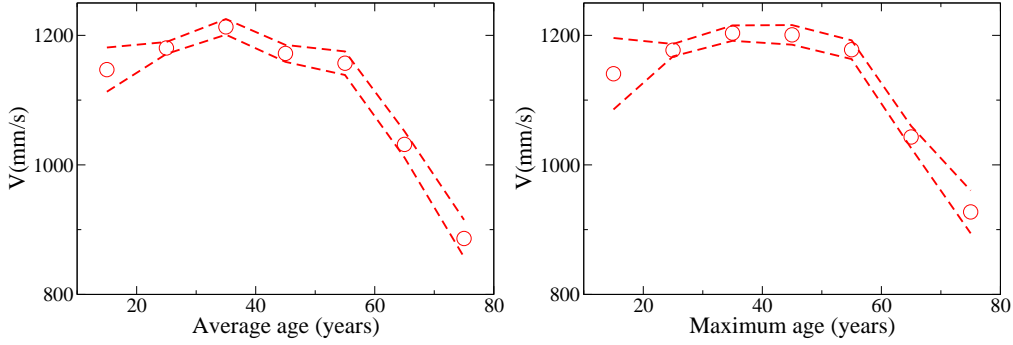

\begin{center}
\includegraphics[width=0.45\linewidth]{f32.eps}  \vspace{0.1\linewidth} \includegraphics[width=0.45\linewidth]{f34.eps} 
\caption{$V$ dependence on average (left) and maximum (right) age. Dashed lines provide standard error confidence bars. The point at 75 years corresponds to the ``70 years or more'' slot.}
\label{f4b_av}
\end{center}
\end{figure}

Table \ref{table4a_av} shows the average age dependence of all observables, and the age dependence of variables $r$, $x$, $|y|$ is also graphically shown in figure
\ref{f4a_av}, while that of $V$ is shown in figure \ref{f4b_av} (left panels).

\begin{table}[!ht]
\scriptsize
\caption{Observable dependence on average age for dyads. Lengths in millimetres, times in seconds.}
\label{table4a_av}
\begin{center}
\begin{tabular} {|c|c|c|c|c|c|}
\hline
Average age&  $N^k_g$ &  $V$ &    $r$ & $x$  & $|y|$ \\
\hline
10-19 years & 60 & 1147 $\pm$ 34   ($\sigma$=264) &865 $\pm$ 43  ($\sigma$=332) &575 $\pm$ 20 ($\sigma$=158) &496 $\pm$ 57 ($\sigma$=445)    \\ 
\hline
20-29 years & 370 & 1181 $\pm$ 9.2   ($\sigma$=178) &793 $\pm$ 12  ($\sigma$=226) &662 $\pm$ 8.1 ($\sigma$=155) &313 $\pm$ 14 ($\sigma$=274)    \\ 
\hline
30-39 years & 269 & 1213 $\pm$ 12   ($\sigma$=199) &831 $\pm$ 14  ($\sigma$=234) &670 $\pm$ 11 ($\sigma$=174) &360 $\pm$ 18 ($\sigma$=302)    \\ 
\hline
40-49 years & 195 & 1172 $\pm$ 13   ($\sigma$=183) &852 $\pm$ 17  ($\sigma$=232) &674 $\pm$ 12 ($\sigma$=167) &387 $\pm$ 23 ($\sigma$=316)    \\ 
\hline
50-59 years & 114 & 1157 $\pm$ 18   ($\sigma$=194) &825 $\pm$ 20  ($\sigma$=217) &650 $\pm$ 15 ($\sigma$=159) &376 $\pm$ 30 ($\sigma$=317)    \\ 
\hline
60-69 years & 69 & 1032 $\pm$ 20   ($\sigma$=168) &875 $\pm$ 40  ($\sigma$=332) &635 $\pm$ 20 ($\sigma$=163) &467 $\pm$ 50 ($\sigma$=416)    \\ 
\hline
$\geq$ 70 years & 12 & 886 $\pm$ 29   ($\sigma$=99.8) &786 $\pm$ 79  ($\sigma$=275) &588 $\pm$ 19 ($\sigma$=66.6) &385 $\pm$ 100 ($\sigma$=363)    \\ 
\hline
$F_{6,1082}$ & & 13.2 & 2.26 & 3.79 & 4.75\\
\hline
$p$ & & $<10^{-8}$ & 0.036 & 0.000955 & 8.72$\cdot 10^{-5}$\\
\hline
$R^2$ & & 0.0681 & 0.0124 & 0.0206 & 0.0257\\
\hline
$\delta$ & & 1.67 & 0.275 & 0.598 & 0.603\\
\hline
\end{tabular}
\end{center}
\end{table}

\subsection{Maximum age}
Table \ref{table4a_max} shows the average age dependence of all observables, and the age dependence of variables $r$, $x$, $|y|$ is also graphically shown in figure
\ref{f4a_av}, while that of $V$ is shown in figure \ref{f4b_av} (right panels).

\begin{table}[!ht]
\scriptsize
\caption{Observable dependence on maximum age for dyads. Lengths in millimetres, times in seconds.}
\label{table4a_max}
\begin{center}
\begin{tabular} {|c|c|c|c|c|c|}
\hline
Maximum age&  $N^k_g$ &  $V$ &    $r$ & $x$  & $|y|$ \\
\hline
10-19 years & 28 & 1141 $\pm$ 55   ($\sigma$=292) &733 $\pm$ 35  ($\sigma$=186) &628 $\pm$ 17 ($\sigma$=92.5) &283 $\pm$ 49 ($\sigma$=258)    \\ 
\hline
20-29 years & 327 & 1177 $\pm$ 9.6   ($\sigma$=174) &789 $\pm$ 12  ($\sigma$=225) &660 $\pm$ 8.2 ($\sigma$=149) &309 $\pm$ 15 ($\sigma$=278)    \\ 
\hline
30-39 years & 292 & 1203 $\pm$ 12   ($\sigma$=204) &831 $\pm$ 14  ($\sigma$=238) &648 $\pm$ 10 ($\sigma$=172) &384 $\pm$ 19 ($\sigma$=321)    \\ 
\hline
40-49 years & 143 & 1201 $\pm$ 15   ($\sigma$=181) &849 $\pm$ 23  ($\sigma$=275) &680 $\pm$ 15 ($\sigma$=176) &377 $\pm$ 28 ($\sigma$=336)    \\ 
\hline
50-59 years & 179 & 1178 $\pm$ 14   ($\sigma$=193) &854 $\pm$ 16  ($\sigma$=217) &674 $\pm$ 13 ($\sigma$=178) &388 $\pm$ 23 ($\sigma$=306)    \\ 
\hline
60-69 years & 105 & 1043 $\pm$ 17   ($\sigma$=174) &872 $\pm$ 30  ($\sigma$=310) &638 $\pm$ 16 ($\sigma$=162) &452 $\pm$ 40 ($\sigma$=407)    \\ 
\hline
$\geq$ 70 years & 15 & 927 $\pm$ 33   ($\sigma$=128) &760 $\pm$ 65  ($\sigma$=254) &575 $\pm$ 18 ($\sigma$=67.9) &374 $\pm$ 85 ($\sigma$=330)    \\ 
\hline
$F_{6,1082}$ & & 14.2 & 3.37 & 1.97 & 3.7\\
\hline
$p$ & & $<10^{-8}$ & 0.0027 & 0.0668 & 0.00122\\
\hline
$R^2$ & & 0.0731 & 0.0183 & 0.0108 & 0.0201\\
\hline
$\delta$ & & 1.38 & 0.484 & 0.619 & 0.443\\
\hline
\end{tabular}
\end{center}
\end{table}
\subsection{Discussion}
It may be seen that the results concerning minimum (section \ref{ageef}), maximum and average age are quite similar above 20 years. Nevertheless,
using minimum age allows us to spot the presence of children below 10 years of age
and verify their peculiar behaviour.
\section{Comparison between minimum, average and maximum height}
\label{heicomp}
\begin{figure}[ht!]
\begin{center}
\includegraphics[width=0.7\linewidth]{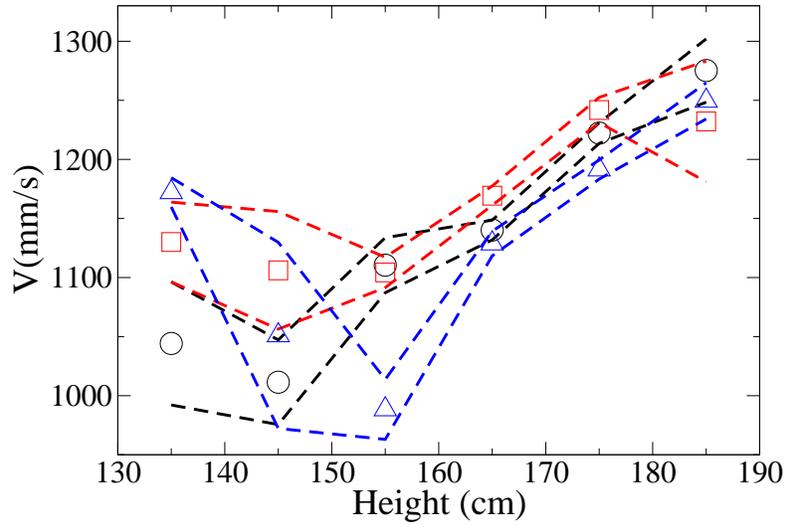} 
\caption{$V$ dependence on average (black and circles), minimum (red and squares) and maximum (blue and triangles) height.
 Dashed lines provide standard error confidence bars.
The points at 135 and 185 cm correspond to the ``less than 140'' and ``more than 180'' cm slots.}
\label{comphfigv}
\end{center}
\end{figure}

\begin{figure}[ht!]
\begin{center}
\includegraphics[width=0.7\linewidth]{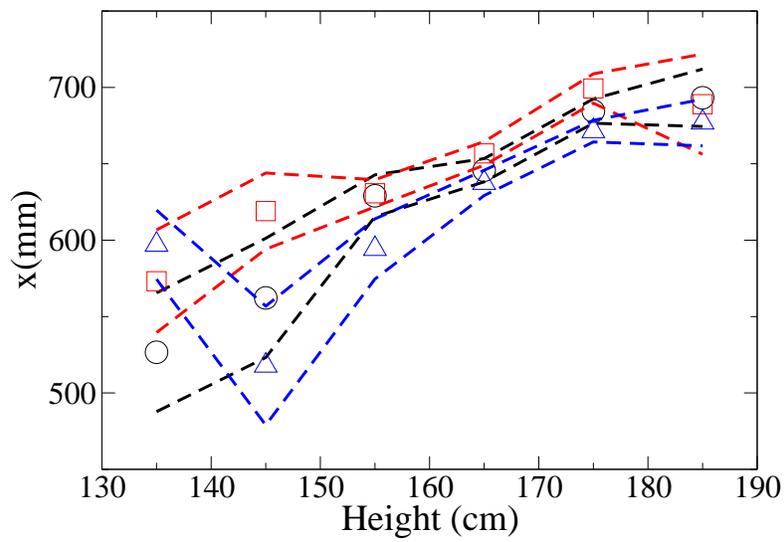} 
\caption{$x$ dependence on average (black and circles), minimum (red and squares) and maximum (blue and triangles) height.
 Dashed lines provide standard error confidence bars.
The points at 135 and 185 cm correspond to the ``less than 140'' and ``more than 180'' cm slots.}
\label{comphfigvbis}
\end{center}
\end{figure}

Tables  \ref{table_hc_av} and \ref{table_hc_max} show the dependence on, respectively, average and maximum height of all observables. Figure
Figures
\ref{comphfigv} and \ref{comphfigvbis} provide, on the other hand, a graphical comparison for the $V$ and $x$ observables. We show these two figures since these observables are mostly growing
with height, and so their analysis is easier. As the figures show, the ``average'' curves are somehow in between the other two curves, with the ``minimum'' curve on the top and the maximum on the 
bottom, as expected for an observable that grows with height\footnote{A group whose tallest person is in, e.g, the 160-170 cm range is expected to have a shorter average height than a group whose
shortest person is in the same range.}. This suggests that dyads have, at least regarding height, a behaviour that is an average of the individual ones, and all three indicators should
be basically equivalent. In the main text we choose to use the ``minimum'' height indicator for two reasons: it allows to better identify dyads with children, and it has a sufficient number
of events in all occupied height slots.

\begin{table}[!ht]
\scriptsize
\caption{Observable dependence on average height for dyads. Lengths in millimetres, times in seconds.}
\label{table_hc_av}
\begin{center}
\begin{tabular} {|c|c|c|c|c|c|}
\hline
Average height&  $N^k_g$ &  $V$ &    $r$ & $x$  & $|y|$ \\
\hline
$<$ 140 cm & 14 & 1044 $\pm$ 52   ($\sigma$=195) &983 $\pm$ 98  ($\sigma$=365) &527 $\pm$ 39 ($\sigma$=145) &672 $\pm$ 130 ($\sigma$=492)    \\ 
\hline
140-150 cm & 22 & 1011 $\pm$ 36   ($\sigma$=168) &910 $\pm$ 81  ($\sigma$=382) &562 $\pm$ 39 ($\sigma$=183) &570 $\pm$ 100 ($\sigma$=476)    \\ 
\hline
150-160 cm & 118 & 1110 $\pm$ 23   ($\sigma$=253) &812 $\pm$ 24  ($\sigma$=260) &629 $\pm$ 14 ($\sigma$=149) &379 $\pm$ 31 ($\sigma$=340)    \\ 
\hline
160-170 cm & 472 & 1140 $\pm$ 8.3   ($\sigma$=181) &821 $\pm$ 12  ($\sigma$=250) &646 $\pm$ 7.6 ($\sigma$=165) &372 $\pm$ 15 ($\sigma$=329)    \\ 
\hline
170-180 cm & 421 & 1222 $\pm$ 8.7   ($\sigma$=179) &828 $\pm$ 11  ($\sigma$=224) &685 $\pm$ 8 ($\sigma$=164) &342 $\pm$ 14 ($\sigma$=282)    \\ 
\hline
$>$ 180 cm & 42 & 1275 $\pm$ 27   ($\sigma$=174) &793 $\pm$ 26  ($\sigma$=171) &693 $\pm$ 19 ($\sigma$=122) &274 $\pm$ 34 ($\sigma$=220)    \\ 
\hline
$F_{5,1083}$ & & 17.9 & 1.95 & 7.31 & 5.74\\
\hline
$p$ & & $<10^{-8}$ & 0.0829 & 9.42$\cdot 10^{-7}$ & 3.1$\cdot 10^{-5}$\\
\hline
$R^2$ & & 0.0764 & 0.00894 & 0.0327 & 0.0258\\
\hline
$\delta$ & & 1.54 & 0.816 & 1.3 & 1.29\\
\hline
\end{tabular}
\end{center}
\end{table}

\begin{table}[!ht]
\scriptsize
\caption{Observable dependence on maximum height for dyads. Lengths in millimetres, times in seconds.}
\label{table_hc_max}
\begin{center}
\begin{tabular} {|c|c|c|c|c|c|}
\hline
Maximum height&  $N^k_g$ &  $V$ &    $r$ & $x$  & $|y|$ \\
\hline
$<$ 140 cm & 3 & 1172 $\pm$ 12   ($\sigma$=21.4) &650 $\pm$ 49  ($\sigma$=84.4) &597 $\pm$ 23 ($\sigma$=39.1) &189 $\pm$ 60 ($\sigma$=104)    \\ 
\hline
140-150 cm & 3 & 1051 $\pm$ 79   ($\sigma$=136) &611 $\pm$ 11  ($\sigma$=19.6) &518 $\pm$ 39 ($\sigma$=67.5) &242 $\pm$ 33 ($\sigma$=57.2)    \\ 
\hline
150-160 cm & 49 & 988 $\pm$ 25   ($\sigma$=178) &782 $\pm$ 34  ($\sigma$=240) &594 $\pm$ 20 ($\sigma$=138) &366 $\pm$ 49 ($\sigma$=346)    \\ 
\hline
160-170 cm & 336 & 1129 $\pm$ 10   ($\sigma$=191) &820 $\pm$ 14  ($\sigma$=256) &637 $\pm$ 8.3 ($\sigma$=151) &378 $\pm$ 19 ($\sigma$=343)    \\ 
\hline
170-180 cm & 556 & 1191 $\pm$ 8.1   ($\sigma$=191) &830 $\pm$ 10  ($\sigma$=237) &671 $\pm$ 7.1 ($\sigma$=167) &359 $\pm$ 13 ($\sigma$=306)    \\ 
\hline
$>$ 180 cm & 142 & 1250 $\pm$ 15   ($\sigma$=183) &843 $\pm$ 21  ($\sigma$=252) &677 $\pm$ 15 ($\sigma$=182) &365 $\pm$ 26 ($\sigma$=313)    \\ 
\hline
$F_{5,1083}$ & & 18.8 & 1.31 & 4.21 & 0.427\\
\hline
$p$ & & $<10^{-8}$ & 0.257 & 0.000849 & 0.83\\
\hline
$R^2$ & & 0.08 & 0.00601 & 0.0191 & 0.00197\\
\hline
$\delta$ & & 1.44 & 0.929 & 0.881 & 0.555\\
\hline
\end{tabular}
\end{center}
\end{table}

\end{document}